\documentclass[times,twocolumn,final,longtitle]{elsarticle}
\usepackage{jasr}
\usepackage{framed,multirow}
\usepackage{threeparttable}

\usepackage{lineno}

\usepackage{amssymb}
\usepackage{latexsym,textcomp}

\usepackage{hyperref}
\hypersetup{
    colorlinks=true,
    linkcolor=blue,
    filecolor=magenta,      
    urlcolor=cyan,
    }

\usepackage{tabularx}
\usepackage[anticlockwise]{rotating}

\usepackage{xcolor}
\definecolor{newcolor}{rgb}{.8,.349,.1}

\usepackage{csquotes}
\usepackage{subfiles} 

\usepackage{aas_macros}
\def\natco{Nature Communications}

\journal{Advances in Space Research}

\begin{document}


\verso{Temmer, M., Scolini, C., Richardson, I.G. \textit{et al}}

\begin{frontmatter}

\title{CME Propagation Through the Heliosphere:\\ Status and Future of Observations and Model Development}%


\author[1]{Manuela~\snm{Temmer}\corref{cor1}}
\cortext[cor1]{Corresponding author: manuela.temmer@uni-graz.at}

\author[2]{Camilla~\snm{Scolini}}
\author[4,5]{Ian~G.~\snm{Richardson}}
\author[6,7]{Stephan~G.~\snm{Heinemann}}
\author[8,9]{Evangelos~\snm{Paouris}}
\author[9]{Angelos~\snm{Vourlidas}}
\author[10]{Mario~M.~\snm{Bisi}}

\author[2]{and writing teams: 
N.~Al-Haddad$^{\rm b}$, 
T.~Amerstorfer$^{\rm j}$, 
L.~Barnard$^{\rm k}$,
D.~Bure\v{s}ov{\'a}$^{\rm l}$,
S.~J.~Hofmeister$^{\rm m}$,
K.~Iwai$^{\rm n}$,
B.~V.~Jackson$^{\rm o}$,
R.~Jarolim$^{\rm a}$, 
L.~K.~Jian$^{\rm c}$,
J.~A.~Linker$^{\rm p}$,
N.~Lugaz$^{\rm b}$,
P.~K.~Manoharan$^{\rm q,r}$,
M.~L.~Mays$^{\rm c}$,
W.~Mishra$^{\rm s}$,
M.~J.~Owens$^{\rm k}$,
E.~Palmerio$^{\rm p}$,
B.~Perri$^{\rm t}$,
J.~Pomoell$^{\rm f}$,
R.~F.~Pinto$^{\rm u}$,
E.~Samara$^{\rm v,w}$,
T.~Singh$^{\rm x}$,
D.~Sur$^{\rm y,z}$,
C.~Verbeke$^{\rm v}$,
A.~M.~Veronig$^{\rm a}$,
B.~Zhuang}

\address[1]{Institute of Physics, University of Graz, 8010 Graz, Austria}
\address[2]{Institute for the Study of Earth, Oceans, and Space, University of New Hampshire, Durham, NH 03824, USA}
\address[4]{Heliophysics Science Division, NASA Goddard Space Flight Center, Greenbelt, MD 20771, USA}
\address[5]{Department of Astronomy, University of Maryland, College Park, MD 20742, USA}
\address[6]{Max-Planck-Institut f\"ur Sonnensystemforschung, 37077 G{\"o}ttingen, Germany}
\address[7]{Department of Physics, University of Helsinki, FI-00014 Helsinki, Finland}
\address[8]{Department of Physics and Astronomy, George Mason University, Fairfax, VA 22030, USA}
\address[9]{Johns Hopkins University Applied Physics Laboratory, Laurel, MD 20723, USA}
\address[10]{RAL Space, United Kingdom Research and Innovation -- Science \& Technology Facilities Council, Harwell Campus, Oxfordshire, OX11 0QX, UK}
\address[11]{Austrian Space Weather Office, GeoSphere Austria, 8020 Graz, Austria}
\address[12]{Department of Meteorology, University of Reading, Reading RG6 6BB, UK}
\address[13]{Institute of Atmospheric Physics, 14100 Prague 4, Czech Republic}
\address[14]{Leibniz-Institut for Astrophysics Potsdam (AIP), 14482 Potsdam, Germany}
\address[15]{Institute for Space-Earth Environmental Research, Nagoya University, Nagoya 464-8601, Japan}
\address[16]{Center for Astrophysics and Space Sciences, University of California San Diego, La Jolla, CA 92093, USA}
\address[17]{Predictive Science Inc., San Diego, CA 92121, USA}
\address[18]{Radio Astronomy Centre, National Centre for Radio Astrophysics, Tata Institute of Fundamental Research, Tamil Nadu 643001, India}
\address[19]{Arecibo Observatory, University of Central Florida, Arecibo, PR 00612, USA}
\address[20]{Indian Institute of Astrophysics, Bengaluru 560034, India}
\address[21]{Universit{\'e} Paris-Saclay, Universit{\'e} Paris Cit{\'e}, CEA, CNRS, AIM, 91191, Gif-sur-Yvette, France}
\address[22]{IRAP, Universit{\'e} Toulouse III --- Paul Sabatier, CNRS, CNES, 31028 Toulouse, France}
\address[23]{Centre for mathematical Plasma Astrophysics (CmPA), KU Leuven, 3001 Leuven, Belgium}
\address[24]{Solar--Terrestrial Centre of Excellence---SIDC, Royal Observatory of Belgium, 1180 Brussels, Belgium}
\address[25]{Center for Space Plasma and Aeronomic Research, The University of Alabama in Huntsville, AL 35805, USA}
\address[26]{CIRES, University of Colorado at Boulder, Boulder, CO 80309, USA}
\address[27]{Narula Institute of Technology, Kolkata, West Bengal 700109, India}

\received{--}
\accepted{--}

\begin{abstract}
The ISWAT (International Space Weather Action Teams) heliosphere clusters H1 and H2 have a focus on interplanetary space and its characteristics, especially on the large-scale co-rotating and transient structures impacting Earth.  Solar wind stream interaction regions, generated by the interaction between high-speed solar wind originating in large-scale open coronal magnetic fields and slower solar wind from closed magnetic fields, are regions of compressed plasma and magnetic field followed by high-speed streams that recur at the ${\sim}27$~day solar rotation period.  Short-term reconfigurations of the lower coronal magnetic field generate flare emissions and provide the energy to accelerate enormous amounts of magnetised plasma and particles in the form of coronal mass ejections into interplanetary space.   The dynamic interplay between these phenomena changes the configuration of interplanetary space on various temporal and spatial scales which in turn influences the propagation of individual structures.  While considerable efforts have been made to model the solar wind, we outline the limitations arising from the rather large uncertainties in parameters inferred from observations that make reliable predictions of the structures impacting Earth difficult.  Moreover, the increased complexity of interplanetary space as solar activity rises in cycle 25 is likely to pose a challenge to these models.  Combining observational and modeling expertise will extend our knowledge of the relationship between these different phenomena and the underlying physical processes, leading to improved models and scientific understanding and more-reliable space-weather forecasting. The current paper summarizes the efforts and progress achieved in recent years, identifies open questions, and gives an outlook for the next 5--10 years.  It acts as basis for updating the existing COSPAR roadmap by \cite{Schrijver2015}, as well as providing a useful and practical guide for peer-users and the next generation of space weather scientists.
\end{abstract}


\end{frontmatter}




\section{Introduction}
\label{sec:section1}
 
Our Sun is an active star that impacts modern life and society by dynamically generating large-scale structures across the heliosphere consisting of plasma and magnetic field that interact with Earth and other planets.  The study of the influence of the Sun on interplanetary space and solar system bodies is often known as ``space weather'' \citep[e.g.,][]{WrightEtAl1997,Cade2015}. Space weather poses a global threat for Earth, though countries are impacted differently depending on their latitudinal position and infrastructure. The most severe consequences come from the effects of intense geomagnetic storms, i.e., disturbances of the Earth's magnetosphere resulting from the impact of coronal mass ejections (CMEs), on advanced human technologies \citep[e.g.,][]{Eastwood2017}.  These may include induced electric currents with the potential to severely disrupt power grids, and degrade communication networks. Space weather can also affect cutting-edge communication, positioning, and navigation technologies. These require reliable and operational connections between ground- and space-based instrumentation to allow individual users of smartphones and other devices to navigate indoors and out, as well as to protect users of navigation products from errors. Increasing demands on the accuracy and reliability of new technologies require a deeper knowledge and more accurate identification of the effects of space weather, including the ability to distinguish between different sources of space weather effects, such as changes in the ionosphere--thermosphere--magnetosphere coupling during space weather events, and their effects on Earth's upper atmosphere. 

Space agencies (e.g., in Europe (ESA), in US (NASA), in China (CNSA), in Russia (Roscosmos), in India (ISRO) or in Japan (JAXA)), international research unions (e.g., Committee on Space Research (COSPAR), International Space Environment Service (ISES), International Space Weather Initiative (ISWI), or Scientific Committee on Solar-Terrestrial Physics (SCOSTEP)), and the United Nations, run extensive space-weather panels and programmes for enhancing awareness of, and preparedness for, strong solar and geomagnetic activity. Individual countries have invested substantial amounts of money to build forecasting capabilities designed to address their own vulnerability to space weather \citep[e.g.,][]{Hapgood2017, opgenoorthEtAl19}. There are also considerable efforts to  translate results from scientific research into operational models and to train forecasters, passing on the knowledge gained to the next generation of space weather researchers. This paper briefly reviews recent progress made in the topics of interest to the \href{https://www.iswat-cospar.org}{ISWAT (international Space Weather Action Teams)} H1+H2 Clusters and places this progress in the context of the COSPAR Space Weather Roadmap paper by \cite{Schrijver2015}. In the following we briefly explain the ISWAT initiative and structure.

\subsection{Interrelation Between the ISWAT Teams at a Glance}  

\begin{figure*}
\centering
\includegraphics[width=1.\linewidth]{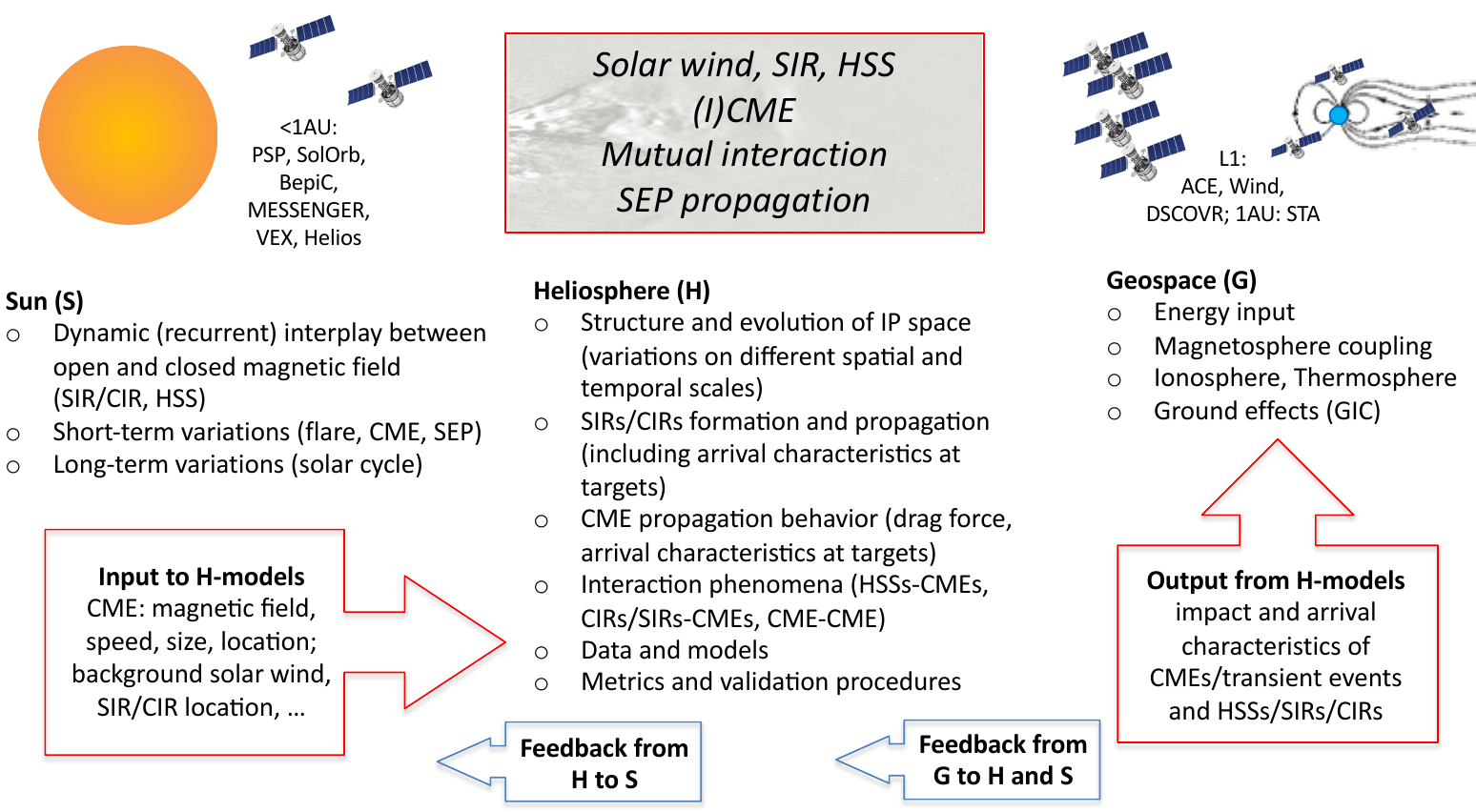}
\caption{Schematic overview of the topics of interest starting from the Sun to interplanetary (IP) space and arrival at Earth (SIR/CIR: stream- and co-rotating interaction region; HSS: high-speed-stream, CME: coronal mass ejection, SEP: solar energetic particles, GIC: ground induced current), which are related to the ISWAT S, H, and G Clusters and the input information required by the H Cluster from the S Cluster together with the output from the H Cluster provided to the G Cluster. This leads to a feedback loop between the Clusters.}
\label{fig:sec1:overview}
\end{figure*}

Space weather, with its many facets, is a highly-interdisciplinary field that requires coordination among research involving different spatial and temporal regimes, starting from the source of events on the Sun (covered by ISWAT Cluster S) through the heliosphere (covered by ISWAT cluster H, and the focus of this paper), to the vicinity of Earth (i.e., Geospace, treated by Cluster G).  The H Clusters' teams focus on research and studies of the background solar wind and propagation of transient events, as well as the mutual interactions between the various large-scale structures, with the aim of improving heliospheric models.  This requires reliable input on the solar perspective from the S Clusters' teams, such as long-term solar activity (S1 Cluster summary, see TI2 paper by \cite{Pevtsov2023}, short-term dynamic changes of the magnetic field on the Sun and the interplay between open and closed magnetic field.  Such input may be used, for example, to model the behavior of the background solar wind \cite[S2 Cluster summary TI2 paper by][]{Arge2023}.  A goal of future heliospheric models is that they will work in real-time, for example by forecasting the geoeffectiveness (i.e., capable of causing a geomagnetic disturbance)  of a CME before the eruption has actually happened on the Sun.  A major challenge for that is that the input parameters for modeling a specific solar eruption (its speed, size, magnetic field, location, etc...) need to be forecast prior to the eruption (see S3 Cluster summary TI2 papers by: \cite{Georgoulis2023} on forecasting; and \cite{Linton2023} on understanding solar eruptions).  In turn, the H Cluster teams provide input in terms of expected impact of CMEs and SIRs (stream interaction regions; and CIRs, i.e., co-rotating interaction region) on the Geospace system for the G Cluster teams (G1 Cluster summary TI2 paper by \cite{Opgenoorth2023} on the geomagnetic environment; G2a Cluster summary TI2 paper by \cite{Bruinsma2023} on atmospheric variability; G2b Cluster summary TI2 paper by \cite{Tsagouri2023} on observational and modeling aspects for the ionospheric variability; and G3 Cluster summary TI2 papers on near-Earth radiation and plasma environment by \cite{Zheng2023,Minow2023,Boyd2023}).  Within the H Cluster, H3 investigates the radiation environment in the heliosphere \citep[solar energetic particles (SEPs) and Galactic cosmic rays (GCRs); see the H3 Cluster summary TI2 paper by \cite{Guo2023} and also the TI1 review paper on SEPs by][]{WHITMAN2022}.  Finally, H4 investigates space weather at other planetary bodies.  These interrelations are also depicted in the schematic overview given in Figure~\ref{fig:sec1:overview}.  As can be seen, the H Clusters act as ``communication link'' between the S and G Clusters.  In combination, the ISWAT Initiative --- with the different Clusters and their respective teams and overarching activities --- provides the best basis for testing theories, developing tools, and evaluating the results (research to operation---R2O; operation to research---O2R).

\subsection{The COSPAR Space Weather Roadmap: Where Do We Stand?}

Extensive research in recent years has enhanced our understanding of the physical processes involved in the interaction between solar wind and transient events, while increased computational power has enabled substantial progress in modeling the solar wind. We have not only developed computationally expensive magnetohydrodynamics (MHD) models as suggested by \cite{Schrijver2015}, but also improved empirical and analytic models. 
Data assimilation (DA) algorithms, combining in situ and remote-sensing-image data, as well as common metrics, have been developed. Together, these advances have enabled improved and more detailed insight into large-scale propagating disturbances and their impact \citep[e.g.,][]{Maysetal2015b,Dumbovicetal2018,Rileyetal2018,Verbekeetal2019a}.  However, a major weakness is a lack of coordination on the validation of models, i.e., determining objectively how well the models perform. For example, validation efforts for individual models may use different choices of events and input data, and there are very few benchmarks that can be used to confront models with each other.  This is partially due to the diversity of modeling approaches, which can make comparisons difficult.  For example, most analytical models give only 1D solutions, while the high computational time and expense makes it challenging for numerical MHD codes to perform the multiple runs required for a full validation. Most models do not provide predictions of the magnetic field.  
Thus, it is still difficult to find the best trade-offs between model accuracy, robustness, and speed, although new numerical techniques are helping to overcome this challenge.  The coupling of different codes into dedicated space-weather frameworks (see Table\ref{tab:swmf}) to model the entire heliosphere demonstrates efforts in the community to combine models and exploit their individual strengths.  In that respect we note the importance of ensemble modeling, where the uncertainties of input parameters for a specific model can be used to derive the probability of a range of outcomes \cite[such as in hurricane track predictions, as recommended by][]{Schrijver2015}, as has already been explored \citep[e.g.,][]{Maysetal2015b,Amerstorferetal2018,Weissetal2021}. 

\begin{table*}[ht]
    \centering
    \begin{tabular}{c|c}
         \href{https://ccmc.gsfc.nasa.gov}{NASA/CCMC} & \cite{Kuznetsova2022} \\
         \href{https://clasp.engin.umich.edu/research/theory-computational-methods/space-weather-modeling-framework/}{SWMF} & \cite{Toth2005, Gombosi2021} \\
        \href{https://esa-vswmc.eu}{ESA/VSWMC} & \cite{Poedts2020a}\\
         \href{https://cidas.isee.nagoya-u.ac.jp/susanoo/}{SUSANOO} & \cite{Shiota2014, Shiota2016}\\
         \href{http://storms-service.irap.omp.eu}{STORMS} & \cite{rouillardetal2020}
    \end{tabular}
    \caption{Examples of Space Weather modeling frameworks in US, Europe and Asia, with links to software download and/or webpage hosting the service (CCMC: Community coordinated modeling center; SWMF: Space Weather Modeling Framework; ESA/VSWMC: European Space Agency/Virtual Space Weather Modeling Center; SUSANOO: Space-Weather-Forecast-Usable System Anchored by Numerical Operations and Observations; STORMS: Solar-Terrestrial Observations and Modeling Service)}
    \label{tab:swmf}
\end{table*}

There are major challenges for further improving our models.  First, current models in ``forecast mode" cannot fully capture the evolution of CME magnetic fields from the eruption on the solar surface to interplanetary space.  Modeling the magnetic field of a CME (often assumed to be a flux rope (FR)) sufficiently reliably to derive the impact at Earth, especially predicting the magnetic field component $B_z$, is the "holy grail” of space weather research (and prediction/forecast).  Within the ISWAT initiative, work on this problem spans expertise in Clusters S3 and H2.  Second, the correct and accurate (i.e., validated) modeling of the background solar wind is still an outstanding issue, which is the topic of Cluster H1.  Besides CME events interacting with the ambient solar wind flow, recent results show that even the quiet solar wind flow itself has a transient component \citep[e.g.,][]{BourouaineEtAl2020}.  We therefore need to better understand the solar wind as time-dependent outflow.  Simulations of CME propagation are only as precise as the accuracy of the background flow allows.   Third, the thorough validation of solar wind models poses a problem since there are only limited locations where solar wind measurements are available to compare with model outputs which usually cover large regions of the heliosphere.  The limited measurements restrict validation procedures and prevent the skill of a model from being reliably quantified.  Fourth, solar activity changes on short-, mid-, and long-term scales (see Cluster S1), requiring dynamic adjustments of model parameters.  For example, the default model parameters derived through statistical studies for cycle 23 need to be adapted when applied to events during solar cycle 24.  A drop in the magnetic field and heliospheric pressure \citep[see e.g.,][]{YermolaevEtAl2022} during the weaker solar cycle 24 led to a cascade of reactions, such as an over-expansion of CMEs in the heliosphere that changed their propagation behavior and the formation of shocks \citep[see e.g.,][]{Gopalswamyetal2015,Lugazetal2017b}.  In addition, cycle 24 revealed a more complex coronal magnetic field leading to more pseudostreamer contributions and, hence, CME trajectories being directed out of the ecliptic \citep[see e.g.,][]{JianEtAl2019}.  As can be seen, there are still many open scientific questions related to advance models for CME and solar wind forecasting.

\textit{There is not a single model or framework currently available that outperforms others, and each model shows strength and weakness on different aspects.}

\subsection{General Methodology}\label{method}
There are a number of different types of space weather models in the heliospheric domain that are designed to provide specific types of predictions.  For example, models may assume different CME structures and use different criteria to assess the impact of the CME at a target such as Earth.  Therefore, caution is advised to ensure that appropriate parameters are considered when comparing model or forecast outputs with actual measurements.

Hit/miss ``categorical" forecasts are concerned with predictions of the arrival or non-arrival of a CME or SIR structure at a given target location.  Additionally, CME propagation models that do not describe the internal magnetic structure of CMEs can be used to predict the time of arrival (ToA) of the CME (most commonly defined as the arrival time of the CME-driven shock, depending on the specifics of the model) but not the arrival of the ejecta or the impact (e.g., geomagnetic) of the CME.  This is true for both the empirical/analytical models and MHD-based cone CME models that are widely employed for forecasting due to their robustness and the relatively-low computational resources required \citep[e.g.,][]{PizzoEtAl2011}.  On the other hand, models describing the CME internal magnetic structure (generally in the form of various magnetic FR or spheromak models) are able to distinguish between the ToA of the CME-driven shock, and the ToA of the ejecta.  In addition, some models also predict the speed on arrival (SoA) and density on arrival (DoA) for both CMEs and SIRs, the major structures contributing to the space weather impact on planetary magnetospheres via compression mechanisms resulting from increased dynamic pressure (see Cluster G).  The prediction can be provided in the form of a single value (e.g., as provided by drag-based and other analytical CME models), or in the form of a time series at a given target \citep[e.g., from MHD models or the OSPREI suite of][]{Kayetal2022}.  Additionally, CME propagation models that differentiate between the shock, sheath, and ejecta components of a CME can provide time series predictions of the magnetic-field components, including the $B_{z}(t)$ component that is most important for assessing geoeffectiveness as it is mainly responsible for erosion of the magnetospheric field \citep[see e.g.,][]{PalEtAl22}.  Times series of other parameters contributing to the interplanetary evolution of CME structures, such as the plasma beta (requiring estimates of the plasma temperature and density) may also be provided, together with the duration of the perturbation which is important in determining the space weather impact of an interplanetary structure.  Predictions of the shock, sheath, and ejecta durations at a given target typically require the use of magnetised CME MHD models.  Little emphasis has been put on the modeling and prediction of the ejecta wake duration so far, with only exploratory studies based on MHD models having been performed \citep[e.g.,][]{Scolini2021}.

Increasing efforts have been devoted to reducing the computation time of CME and global background solar wind models to less than a day so that they may contribute to daily forecasts. The extensive use of code parallelization allows models to run in parallel on a few tens to hundreds of processing cores on computer clusters of various sizes, thereby speeding up the computation. Specific approaches considered include: coupling between empirical coronal models and MHD heliospheric models \citep[e.g.][]{Odstrcil2003,Poedts2020b}, tomographic methods \citep[which can also be used to propagate the background magnetic field and for driving MHD models without the need for other CME parameterizations, e.g.][and references therein]{Bisietal2015,Jacksonetal2020,Gonzi2021}, grid adaptation techniques such as adaptive mesh refinement (AMR) or r-AMR \citep[][]{Verbekeetal2022}, implicit solvers \citep[][]{Mikicetal2018,Poedts2020b}, or interpolation from multi-1D solvers \citep[MULTI-VP and the Alfvén-wave Driven Solar Wind Model AWSOM-r;][]{Pinto2017,Huangetal2020}.

The ENLIL \citep[][]{OdstrcilPizzo1999} and EUHFORIA \citep[European Heliospheric Forecasting Information Asset;][]{Pomoell2018} models have been the work-horses of the space-weather community due to their adaptability, usability and useful performance (see more details on CME propagation models in Section \ref{CMEpropModel}).

\textit{It is important to note that every model makes assumptions that may differ and uses numerical, analytical, or empirical techniques or inputs that naturally introduce simulated behaviors of varying degrees of physical accuracy.} 

\subsection{Availability of Observational Data}\label{Data Availability}

Observations are crucial in space weather, not just to efficiently monitor the Sun and heliosphere, and detect sudden events, but also to provide statistics to improve our understanding of the underlying physics and to better constrain and improve models. In recent years, a plethora of satellite missions have provided valuable data for space weather research, including SOHO \citep[Solar and Heliospheric Observatory;][]{domingo1995}, Wind \citep{ogilvie1995}, ACE \citep[Advanced Composition Explorer;][]{stone1998}, DSCOVR \citep[Deep Space Climate Observatory;][]{DSCOVR}, GOES, Proba-2 \citep[][]{santandrea2013}, the twin STEREO \citep[Solar Terrestrial Relation Observatory;][]{Howard2006} spacecraft, SDO \citep[Solar Dynamics Observatory;][]{pesnell12}. Promising for enhancing our knowledge are the recently launched Parker Solar Probe \citep[PSP;][]{Fox2016} and Solar Orbiter \citep[SolO;][]{Mueller2020} missions. PSP is providing key in situ data in the inner heliosphere extending down to the solar corona, that will improve our understanding of the evolution of solar wind structures as they move out from the Sun, while SolO, in addition to also providing in situ measurements in the inner heliosphere, will provide images out of the ecliptic that will increase the coverage of magnetograms to polar regions.

In the frame of ESA's \href{https://www.esa.int/Space_Safety}{Space Safety Programme} the future operational space weather mission \href{https://www.esa.int/Space_Safety/Vigil}{\textit{Vigil}} is planned to be launched 2029.  \textit{Vigil} will be located permanently at the Lagrange point L5 and is designed as an operational space weather mission to stream a constant feed of near real-time data on potentially-hazardous solar activity, before it comes into view from Earth.  \textit{Vigil} would help to overcome the drawback that, at present, measurements of solar surface magnetic fields are largely confined to the visible hemisphere, by extending the region of surface magnetic field observations that can be fed into the models (see Section \ref{sec:section4}). \cite{Schrijver2015} explicitly mention that extending solar magnetic field coverage will improve multi-day forecasts of individual space weather events. Synchronic real-time magnetograms as opposed to time-delayed synoptic maps will be key for a better global modeling of the magnetic field \citep{Caplanetal2016,Jeong2020}.  Still, however, there are no plans for farside magnetographs, and we are stuck doing the best we can do with helioseismology and ADAPT (Air Force Data Assimilative Photospheric flux Transport) approaches \citep[][]{Argeetal2010}.  It is important to point out that many highly-used missions (e.g., SOHO, ACE, WIND, SDO, STEREO) are aging and that attention needs to be paid to potential losses of critical parts of our heliospheric observatory. 

Complementary data are also available from ground-based facilities, such as magnetograms from the GONG network, radio observations from the Worldwide Interplanetary Scintillation Station (WIPSS) Network \citep[e.g.,][]{bisietal2016} and modern radio telescopes such as the Low Frequency Array (LOFAR) and the Murchison Widefield Array (MWA) \citep[see also e.g., TI1 papers by][and references therein, as well as the \href{http://lofar4sw.eu/}{LOFAR For Space Weather (LOFAR4SW)} project]{CHASHEI2022,CHHETRI2022,FALLOWS2022,Iwaietal2022}, white-light coronagraph data for the low corona from the Mauna Loa Solar Observatory (MLSO), and  high-resolution solar images and spectropolarimetry from the Daniel K. Inouye Solar Telescope \citep[DKIST;][]{Rimmele2020} telescope.  Radio and interplanetary scintillation (IPS) is a promising approach for the future \citep[see also][]{ShaifullahEtAl2020}, in particular for probing latitudinal variations of the solar wind \citep[e.g.,][and references therein]{Sokol2015,PorowskiEtAl2022}.  The magnetic field of CMEs can also be tracked from radio observations of Faraday rotation \citep[e.g.,][with radio telescope systems such as LOFAR, the Karl G. Jansky Very Large Array (VLA), Greenbank, MWA, and the future Square Kilometre Array Observatory (SKA)]{Jensen2010, Jensen2013, Bisi2016, Woodetal2020, Kooi2022}.  In comparison to space missions, ground-based observatories allow for bigger installations with higher resolution and regular maintenance.

\textit{Ensuring continuing support of ground- and space-based infrastructures for space weather observations is crucial. It facilitates to develop more diverse data-driven codes that include DA as recommended by \cite{Schrijver2015}.}

\subsection{H1+H2 Cluster Activities}
In the following we describe the H1+H2 Cluster activities and related open questions towards improving Space Weather forecasts. An overview on the various large-scale interplanetary structures driving Space Weather are given in Section~\ref{sec:section2}. SIRs and CIRs, the main contributors to moderate-to-strong space-weather disturbances at Earth, are not fully understood and have many open questions that are presented in Section~\ref{sec:section3}. CMEs as main source of strong-to-severe space weather disturbances and modeling efforts are described in more detail in Section~\ref{sec:section4}.  Various interaction scenarios between these different types of large-scale solar wind structures are given in Section~\ref{sec:section5}. A recent NASA sponsored \href{https://science.nasa.gov/science-pink/s3fs-public/atoms/files/GapAnalysisReport_full_final.pdf}{Gap Analysis} from the Johns Hopkins Applied Physics Laboratory led by A.~Vourlidas, provides valuable future prospects for model development and improvement, that is presented in Section~\ref{sec:section6}.  In Section~\ref{sec:section7} we give some closing thoughts and point out paths forward.

\section{Structure of Interplanetary Space throughout the Heliosphere}
\label{sec:section2}

\begin{figure}[ht]
\centering
\includegraphics[width=0.8\linewidth]{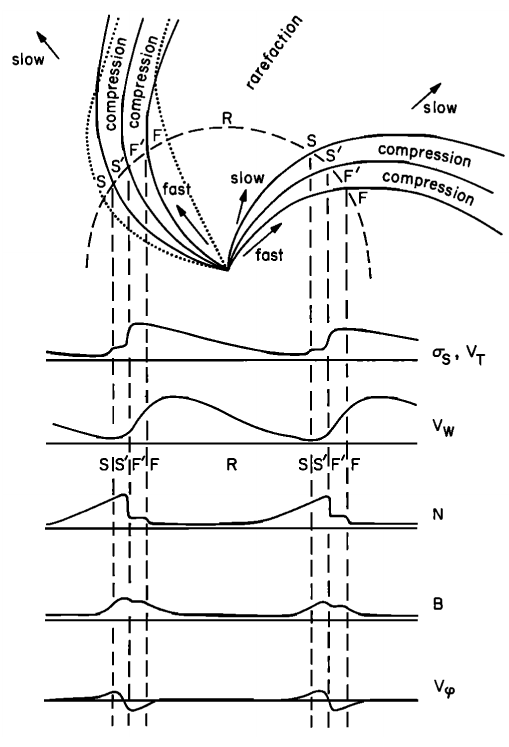}
\caption{Schematic of two high-speed streams co-rotating with the Sun and the associated variations in several plasma parameters at 1~AU: Thermal temperature ($V_T$), magnetic field fluctuation level ($\sigma_s$); solar wind speed ($V_W$); density ($N$); magnetic field intensity ($B$); and transverse component of the solar wind velocity ($V_\phi$). The regions indicated are: the unperturbed slow solar wind (S), compressed, accelerated slow solar wind (S'), compressed, decelerated fast solar wind (F'), unperturbed fast solar wind (F), and a rarefaction (R). S and F form the interaction region, and the stream interface is at the S'–F' boundary. Dotted lines indicate magnetic field lines in the slow and fast solar wind that thread into the interaction region beyond 1~AU \citep{Belcher1971}. }
\label{fig:sec2:bdstreams}
\end{figure}

In this section, we introduce the basic properties of the solar wind and its various large-scale structures including transient CMEs, out to approximately the orbit of Mars. It is not intended to be a comprehensive review of this topic \cite[for such reviews see e.g.,][]{Cranmer2017,Cranmer2019, Richardson2018,Owens2020,LuhmannEtAl2020,Zhang2021,Temmer2021b,Gopalswamy2022}.  

\subsection{Large Scale Structures in the Solar Wind} 
The solar wind, formed from the supersonic expansion of the solar corona \citep[e.g.,][]{Parker1958, Cranmer2019}, is a plasma consisting predominantly of electrons and protons with smaller contributions from helium and heavier ions \citep[e.g., ][]{Vonsteiger2000}.  The solar wind flowing nearly radially away from the Sun drags out coronal magnetic field lines that, because of solar rotation at their footpoints, form an approximately Archimedean spiral configuration in which the interplanetary magnetic field (IMF) is more (less) tightly wound in slower- (higher-) speed solar wind \citep[e.g.,][and references therein]{Owens2013b}. The coronal source of solar wind rotates approximately
every 27 days as seen by an Earth observer, while it takes about 4 days for the radially flowing solar wind plasma to reach 1 AU. This combination produces a local IMF oriented part of a global heliospheric field configuration in a spiral shape with 45 degrees from the radial direction \citep[][]{Parker1961}. The earliest observations of the solar wind \citep{Snyder1963} revealed that the large-scale solar wind is structured into streams of higher-speed solar wind associated with open field lines originating in coronal holes \citep[e.g.,][ see also Section~\ref{sec:section3}]{Krieger1973} interspersed with intervals of slower, denser wind.  The origin of the slow solar wind is still unclear but it is probably of mixed origin in predominantly closed coronal magnetic structures that tend to lie below streamers at the Sun including the streamer belt mapping to the heliospheric current sheet (HCS). Recent PSP observations show that a highly structured slow solar wind can also emerge from within coronal holes \citep{Bale2019}. There is still a debate about the actual origin of the open flux and why some of the open flux at the Sun appears to be ``missing" compared to estimates based on in situ observations in the solar wind \citep[e.g.,][]{Linker2017,Linker2021}.

Typical properties of high-speed stream (HSS) coronal hole flows at 1~AU based on spacecraft observations \citep[e.g.,][]{Ebert2009,Owens2020} include speeds of $\sim$500--800~km~s$^{-1}$, densities of $\sim$2--4~cm$^{-3}$, magnetic field strengths of $\sim$3--4~nT, and proton temperatures of $\sim$2--3$\times 10^5$~K. In slow streamer-belt solar wind, the corresponding values are: speeds of $\sim$300--400~km~s$^{-1}$, densities of $\sim$5--10~cm$^{-3}$, magnetic field strengths of $\sim$4--8~nT and proton temperatures of $\sim$0.5--1$\times 10^5$~K \citep[][]{Schwenn2006,YermolaevEtAl2009}.  The solar wind speed is relatively independent of the heliocentric distance, but the other parameters depend inversely on some power of it.  Ulysses observations indicate that the magnetic field strength appears to be latitude-independent \citep{Smith1995}, suggesting that significant non-radial expansion of the solar wind occurs. We also note that ``typical'' properties might vary, especially when considering different solar cycles and different epochs of a solar cycle (see more details in Section \ref{sec2:solarCycle}).

The interaction between a HSS and the preceding slower solar wind  forms a region of compressed plasma at the leading edge of the HSS that corotates with the Sun (Figure~\ref{fig:sec2:bdstreams}). Such structures are  termed ''co-rotating interaction regions" (CIRs), though the term ``stream interaction region" (SIR) has also been introduced to indicate an interaction region that is only observed on one rotation \cite[e.g.,][]{Jian2006}.  However, the terms are also used interchangeably. Figure~\ref{fig:sec2:bdstreams} shows the typical variations in the solar wind parameters at $\sim$1~AU associated with CIRs including enhancements in the plasma density, magnetic field intensity and proton temperature and a deflection in the solar wind flow direction. CIRs will be discussed in more detail in Section \ref{sec:section3}. See also \cite{Richardson2018} for a recent review about solar wind stream interaction regions throughout the heliosphere.

Transient structures associated with CMEs at the Sun form the other major component of the solar wind. CMEs are identified as bright, outwardly-propagating structures in white-light coronagraph images. They include an enormous mass of coronal material and carry an embedded magnetic field that is stronger than that in the background solar wind. Due to that, they quickly expand in both lateral and radial direction \citep[e.g.,][]{Scolinietal2020}. This strongly influences their 2-D appearance in white-light image data and, hence, the derivation of propagation speed and width, posing a challenge when thinking of accurate inputs for space weather models. CME-associated eruptions are often evident in other remote sensing observations (e.g., extreme ultra-violet---EUV and X-ray low-coronal signatures, and radio signatures), providing critical complementary information on the erupted structures \citep[e.g.,][]{Hudson2001,Palmerio2017}. Taken together, these signatures, when indicative of a frontside Earth-directed CME can provide at best, a lead time of two to three days for arrival at Earth (see Cluster S3). Forecasting when a solar active region will erupt, and predicting the properties of the resulting CME, from solar surface structures prior to eruption, is itself a major space weather challenge as discussed in TI2 paper by \cite{Georgoulis2023}. Occasionally, ``stealth" or ``stealth-like" CMEs are observed in coronagraphs that have weak or no eruptive signatures in the low corona \citep[][]{Robbrechtetal2009,Palmerio2021a}.  

\begin{figure}[ht]
    \centering
    \includegraphics[width=0.8\linewidth]{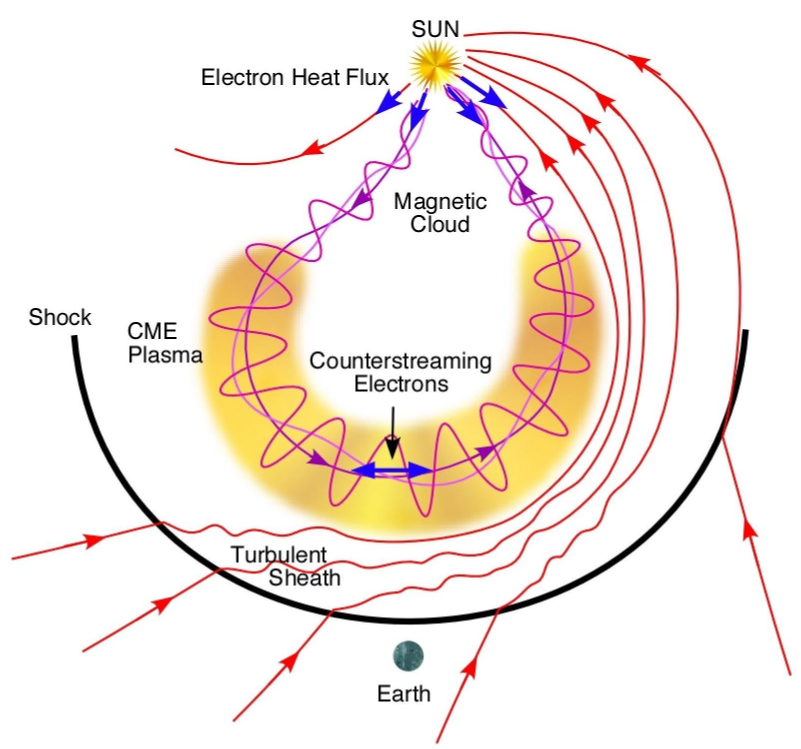}
    \caption{Schematic of the structure of an CME and upstream shock, including a magnetic FR, plasma characteristics (indicated by yellow shading) that differ from those of the ambient solar wind plasma, and counterstreaming suprathermal electron signatures \citep{ZurbuchenRichardson2006}.}
    \label{fig:sec2:cme}
\end{figure}

Figure~\ref{fig:sec2:cme} shows a schematic of a CME propagating out through the solar wind. When observed in situ, a CME is often referred to as an ``interplanetary" CME \citep[ICME; e.g.,][]{Rouillard2011}. Since the link between CMEs at the Sun and ICMEs in the solar wind is now firmly established, for example from STEREO observations \citep[e.g.,][]{Mstl2009}, it is clear that they are the same physical phenomenon, namely a magnetized plasma structure ejected from the Sun. Nevertheless, both CME and ICME are frequently used in the literature to distinguish between CMEs imaged by remote-sensing instruments, such as coronagraphs (revealing global properties) and the related structures observed in situ (revealing local properties). With the differentiation by the observing techniques, the terms CMEs and ICMEs refer to different geometry or scales, and may also refer to various evolutionary stages (but not necessarily, considering heliospheric image data or spacecraft with in situ measurements orbiting close to the Sun). Throughout this paper, we use the term CME for both the imaged and in situ observed cases.

\begin{figure}[ht] 
\centering
\includegraphics[width=0.7\linewidth]{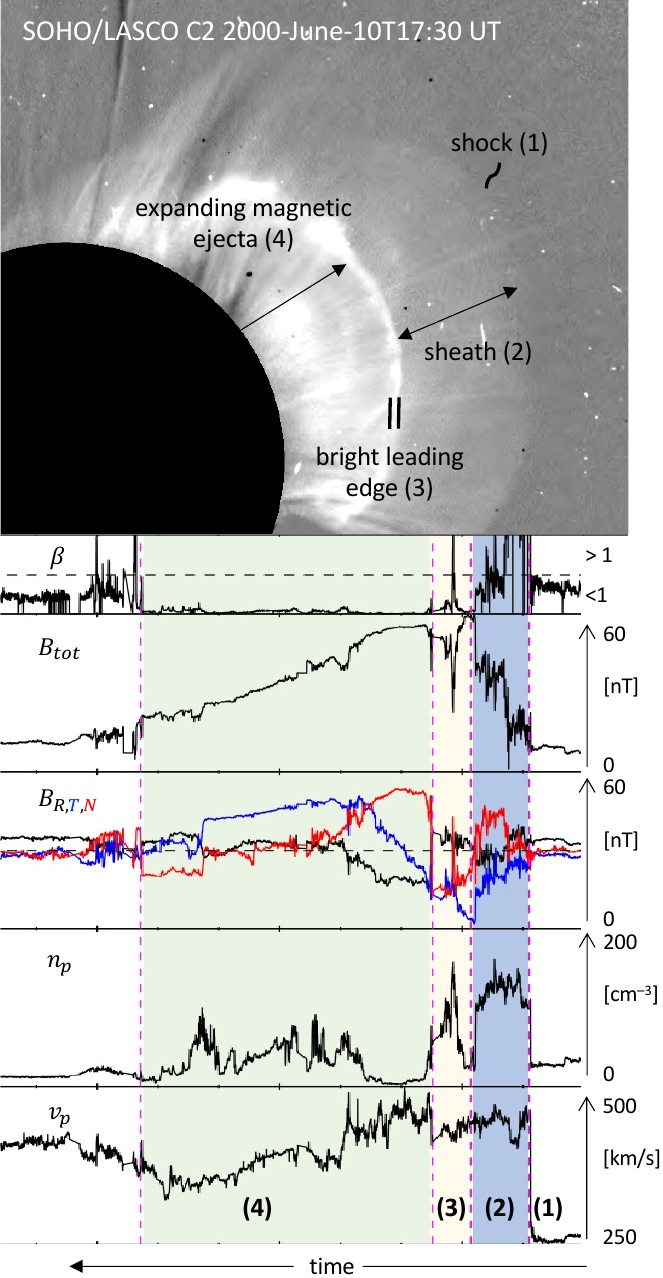}
\caption{Relating CME density structures from white-light image data covering a distance up to about 0.03~AU to in situ plasma and magnetic field measurements at a distance of 0.53~AU. In both data sets we identify the magnetic ejecta region (4) driving several distinct upstream regions, shock (1), sheath (2), and leading edge (3). The image is adapted from \cite{TemmerBothmer2022}.}
\label{fig:sec2:remote_is}
\end{figure}

Figure~\ref{fig:sec2:remote_is} shows the relation between density structures identified from in situ and white-light information. The in situ signatures of CME passage may include first the detection of a forward shock, if the CME speed is sufficiently high compared to the surrounding wind (see also Section~\ref{sec3.3}). This may be followed by a sheath characterized by a pile-up/compression region, then by another density enhancement region called the leading edge, and the magnetic ejecta (sometimes referred to as the ``driver'' of the preceding shock/compression) that is identified by a number of characteristics that differ from those of the background solar wind, due to its origin in an eruptive event \citep[e.g.,][and references therein]{Wimmer-Schweingruberetal2006,ZurbuchenRichardson2006,Kilpua2017a,TemmerBothmer2022}. These characteristics include unusual solar wind charge states and composition \citep[e.g.,][]{Leprietal2001,Gruesbeck2011, Zurbuchen2016, Owens2018, Rivera2019}, bidirectional suprathermal electron heat fluxes, indicating the presence of looped field lines rooted at the Sun \citep[e.g., ][]{Gosling1987}, a monotonic speed decrease (consistent with expansion), low densities and proton temperatures \cite[e.g.,][]{RichardsonCane1995} relative to the ambient wind, often leading to a low plasma beta indicating a magnetically-dominated structure, and elevated helium abundance. Traditionally, ejecta showing a combination of low density, low temperature, and enhanced, slowly-rotating magnetic fields have been known as ``magnetic clouds"  \citep[MCs;][]{Burlagaetal1981}, while structures exhibiting magnetic field rotations but lacking some of the typical plasma signatures have been called ``MC-like" structures \citep[][]{CaneRichardson2003, Lepping2005}. More recently, the terms ``magnetic ejecta" \citep[ME;][]{Winslowetal2015} and ``magnetic obstacle" \citep[MO;][]{Nievesetal2018} have been introduced to refer to ejecta signatures lacking clear rotations in the magnetic field components, with or without associated solar wind plasma observations. The smooth rotations of the magnetic field components have been often interpreted as indicative of possible magnetic flux-rope (MFR) or magnetic flux-rope-like (MFR-like) structures \citep{BothmerSchwenn1998}.  Such CMEs have received considerable attention because the magnetic field configurations are arguably simpler to model and may be consistent with the helical structures occasionally present in coronagraph observations of CMEs. However, only a fraction of CMEs include in situ MC signatures, and this fraction appears to vary with the solar cycle from the majority of CMEs at solar minimum to as small as $\sim20$\% around solar maximum \cite[][]{RichardsonCane2004}.  In the following, we will use the general term ``ejecta'' to refer to the in situ counterparts of CMEs when not distinguishing among the different ejecta sub-classes.  The final structure that may be encountered in situ is a ``wake" following the ejecta. The features of CMEs will be discussed further in Sections~\ref{sec:section4} and \ref{sec:section5}.

The speeds of CMEs observed in situ cover a wide range. Many CMEs have speeds similar to the ambient solar wind, suggesting that they are carried out with the ambient flow, while a few have speeds exceeding 1000~km~s$^{-1}$ \citep[e.g.,][]{RichardsonCane2010}. There is evidence \citep[e.g.][]{Cane1986,Gopalswamy2000} that, as they move away from the Sun, fast CMEs tend to decelerate, even well beyond 1~AU \citep[e.g.,][]{Richardson2014, Witasse2017}, tending towards the ambient solar wind speed, while slow CMEs are accelerated by the ambient solar wind. This may be accounted for by a so-called ``drag force'' that is incorporated into many analytical CME propagation models (see Section~\ref{sec:section4} for more details). 

As noted above, when traveling faster than the background solar wind speed, a CME can generate a shock wave.  Particles accelerated by CME-driven shocks make a major contribution to SEP events, in addition to particles accelerated by solar flares (see the Cluster H3 and TI2 paper by \cite{Guo2023} for more details about SEPs). Also, the intensity of an SEP event tends to be correlated with the speed of the associated CME observed by coronagraphs, and hence many current SEP prediction models, reviewed by \citet{WHITMAN2022}, require such CME observations as an input.

\begin{figure}
\centering
\includegraphics[width=1.0\linewidth]{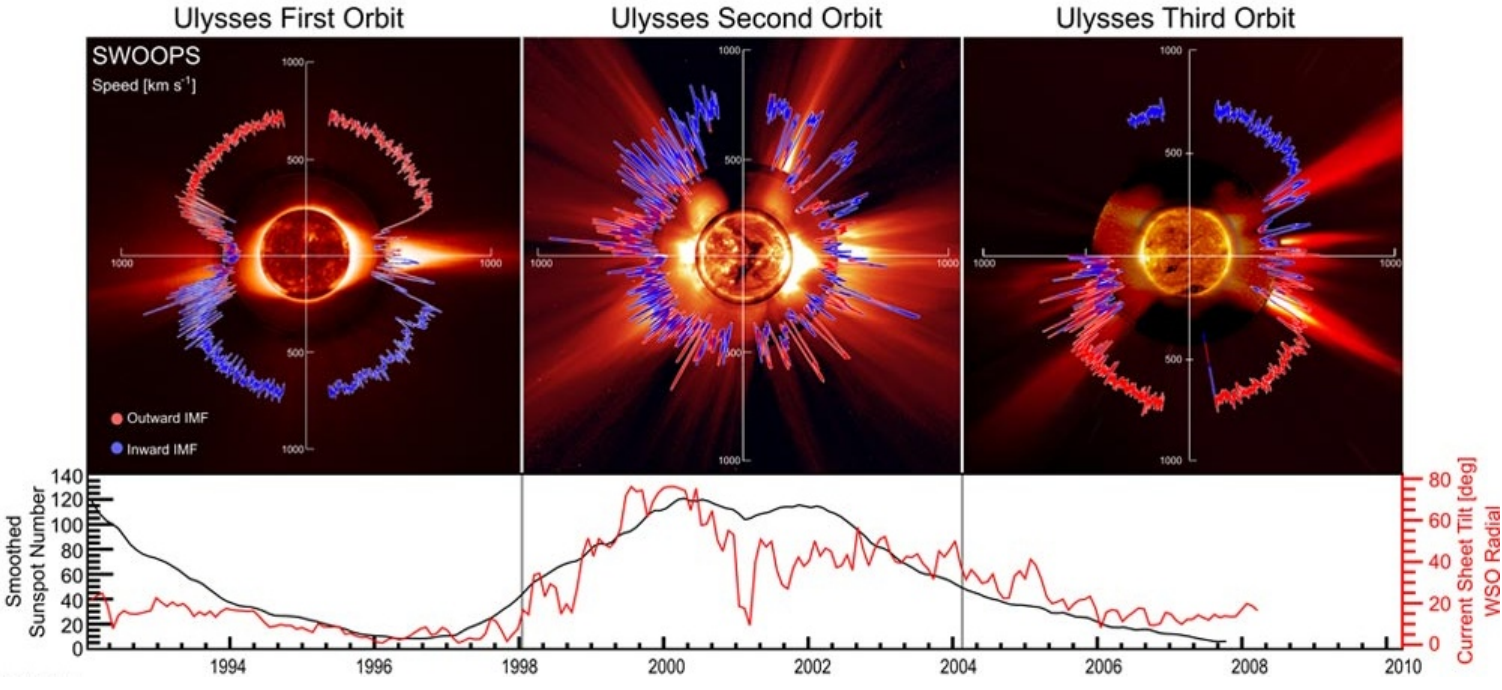}
\caption{Polar plots of the solar wind speed during Ulysses’ three orbits of the Sun showing fast solar wind at high latitudes, slow solar wind at low latitudes, and alternating fast and slow solar wind at mid latitudes, during the first (left) and third (right) orbits around solar minimum. Solar wind speeds are more variable in latitude during the second orbit (centre) around the maximum of solar cycle 23. Red/blue colours represent the IMF direction away from/towards the Sun. Representative observations from SOHO and MLSO illustrate the differences in the streamer belt configuration for each orbit \citep{McComas2008}.}
\label{fig:sec2:uly3orbit}
\end{figure}

\subsection{Solar Cycle Variations}\label{sec2:solarCycle}
The large scale structure of the solar wind is profoundly influenced by the ${\sim}$11~year solar activity cycle. Around the minimum of a solar cycle, coronal holes tend to dominate, expanding from the polar regions to equatorial locations. Solar wind HSSs, originating from coronal holes located near the solar equator, and the associated CIRs formed in front of them then become the source of the recurrent disturbances of Earth's magnetosphere and ionosphere \citep{Verbanac2011}. Figure~\ref{fig:sec2:uly3orbit} \citep[from][]{McComas2008} shows observations from the Ulysses spacecraft, which probed the solar wind up to high latitudes, of the latitudinal structure of the solar wind in a polar plot of the solar wind speed.  Observations from the first Ulysses orbit (left panel) made near solar minimum show high-speed flows at higher latitudes and slower flows at low latitudes above the streamer belt, which is evident in the coronal image from the MLSO.  At low latitudes, there are also intermittent intervals of higher speed flows predominantly associated with equatorward extensions of polar coronal holes or low latitude coronal holes.  Embedded in the streamer belt is a large scale current sheet, the HCS, that separates oppositely-directed magnetic fields from the two polar hemispheres of the Sun \citep[e.g.,][]{Smith2001}; the red/blue color of the speed plot shows the outward/inward magnetic field directions in each hemisphere.  Such a latitudinal organization of solar wind speeds may also be inferred from IPS observations \citep[e.g.,][]{Rickett1991, Manoharan2012, Tokumaru2021}; IPS will be discussed further below.  The middle panel of Figure~\ref{fig:sec2:uly3orbit} shows similar observations from the second Ulysses orbit during a period of high solar activity.  Here, the solar wind speed and magnetic field direction are highly variable in latitude, due to the presence of CMEs propagating away from the Sun over a wide range of latitudes and the higher inclination (tilt angle) of the HCS, resulting from the dominant contribution to the IMF from active regions and the weakening of the polar coronal holes.  The average solar wind speed is also lower than during solar minimum. During the third Ulysses orbit (right panel), again at near solar minimum conditions, the large-scale organization of the solar wind speed with latitude returned, but with the magnetic field polarities in each hemisphere reversed.

 \begin{figure}
\centering
\includegraphics[width=1.0\linewidth]{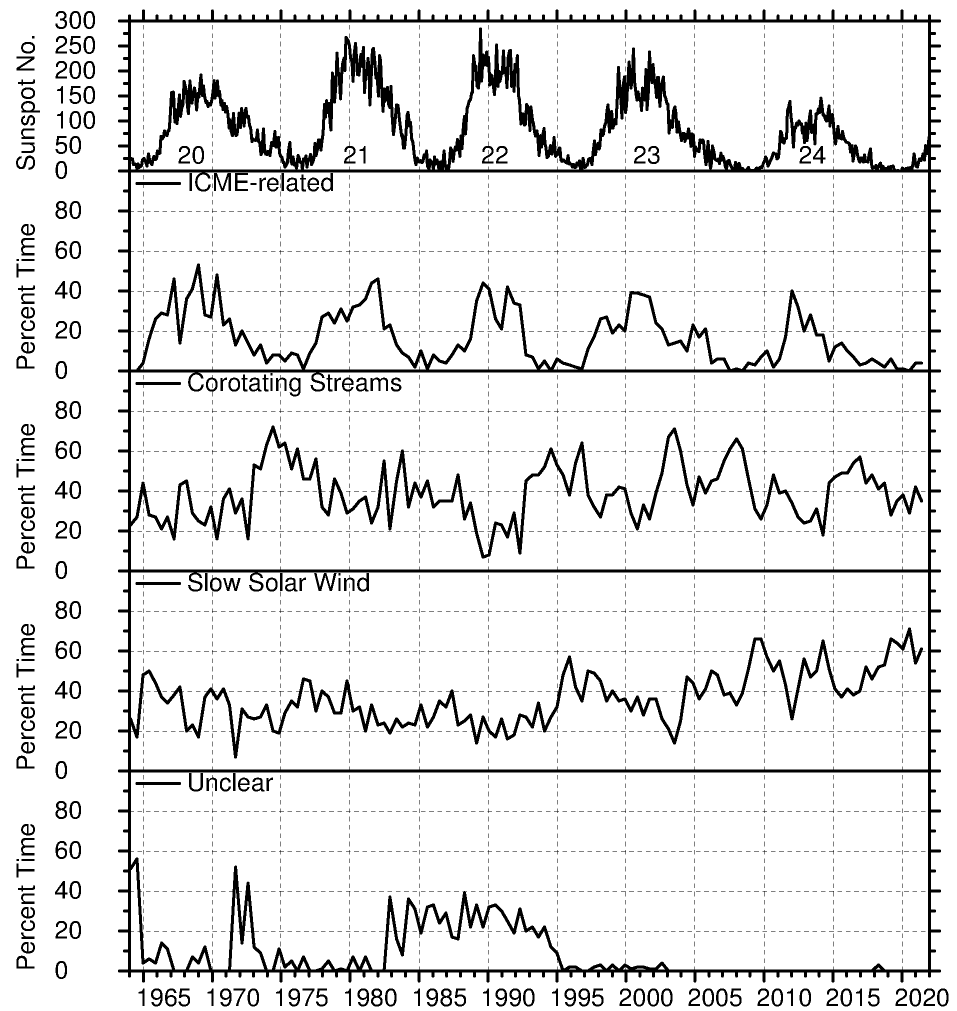}
\caption{Sunspot number (top panel) and the percentage of time the solar wind at Earth is composed of CME-associated structures (e.g., post-shock flows, CMEs), co-rotating HSS, and slow solar wind, for 1964--2021, based on visual examination of OMNI solar wind data and other data sets, as discussed in, and updated from, \cite{RichardsonCane2012}.  The bottom panel shows the time when the solar wind classification could not be determined, predominantly due to data gaps. Note that the occurrence of CME-related flows tends to follow the solar activity cycle, while CIRs are most prominent during the declining and minimum phases of the cycle though are present throughout the cycle.   }
\label{fig:sec2:per}
\end{figure}
 
 \begin{figure}
     \centering
     \includegraphics[width=1.0\linewidth]{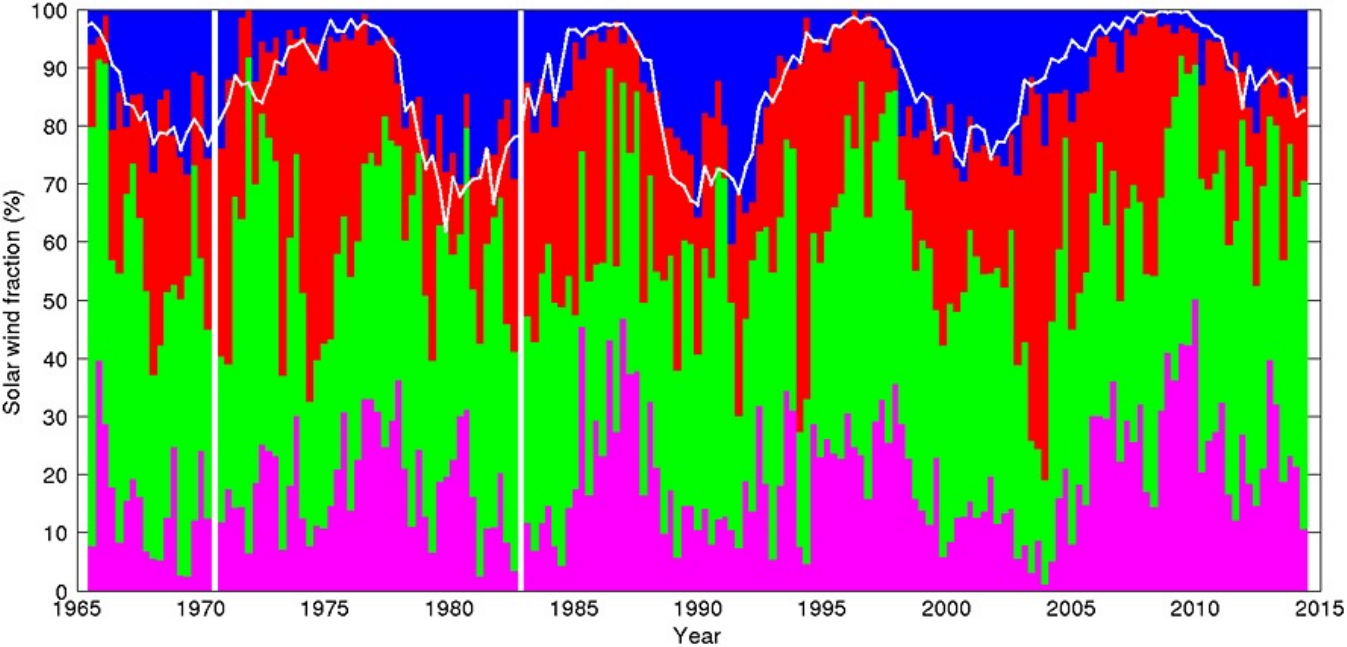}
     \caption{A categorization, based on identifying the characteristic features of solar wind plasma parameters in different types of solar wind, of the OMNI2 data in 1963--2013 into four types of solar wind: ejecta (i.e., CMEs, blue), coronal-hole-origin plasma (red), streamer-belt-origin plasma (green), and sector-reversal-region plasma (purple). The white curve is $100 - 0.2\times$ the sunspot number, i.e., the sunspot number is inverted here compared to Figure~\ref{fig:sec2:per}. White vertical bands are intervals with insufficient data \citep{Xu2015}. The percentage of the time when the classification is judged to be unclear is largely based on data availability, such as in the 1980s-mid 1990s when solar wind data were only available when the measuring spacecraft, IMP8, was in the solar wind.}
     \label{fig:sec2:xub}
 \end{figure}
 
Figure~\ref{fig:sec2:per} shows the relative occurrence time of CME-associated structures, co-rotating HSS, and slow solar wind at Earth in 1964--2021, averaged over six Carrington rotation intervals.   These results are based on a visual inspection of OMNI solar wind observations and other data, and are updated from \cite{Richardson2002} and \cite{RichardsonCane2012}. Figure~\ref{fig:sec2:per} also illustrates the variation in solar wind structure with the solar cycle. The occurrence of CMEs tends, like the CME rate \citep[e.g.,][]{Yashiro2004, Robbrechtetal2009}, to follow solar activity levels. Co-rotating HSS remain present throughout the solar cycle but tend to be predominant during the declining and minimum phases, as does slow solar wind \citep[][]{KamideEtAl98,Verbanacetal11}.  There are also clear cycle-to-cycle variations in Figure~\ref{fig:sec2:per} with a weakening observed for cycle 24. Solar cycle 24 showed a clear drop in all parameters by 20--40\% compared to previous cycles \citep{YermolaevEtAl2021,YermolaevEtAl2022}. Recent studies showed that this might be related to the characteristics of CMEs occurring in different cycles \citep{Bilenko2022}. Especially for modeling, these cycle-to-cycle variations of the solar wind need to be taken into account. Strong variations definitely affect the model performances as the boundary and initial conditions change from epoch to epoch.  

\begin{figure}
    \centering
    \includegraphics[width=1.0\linewidth]{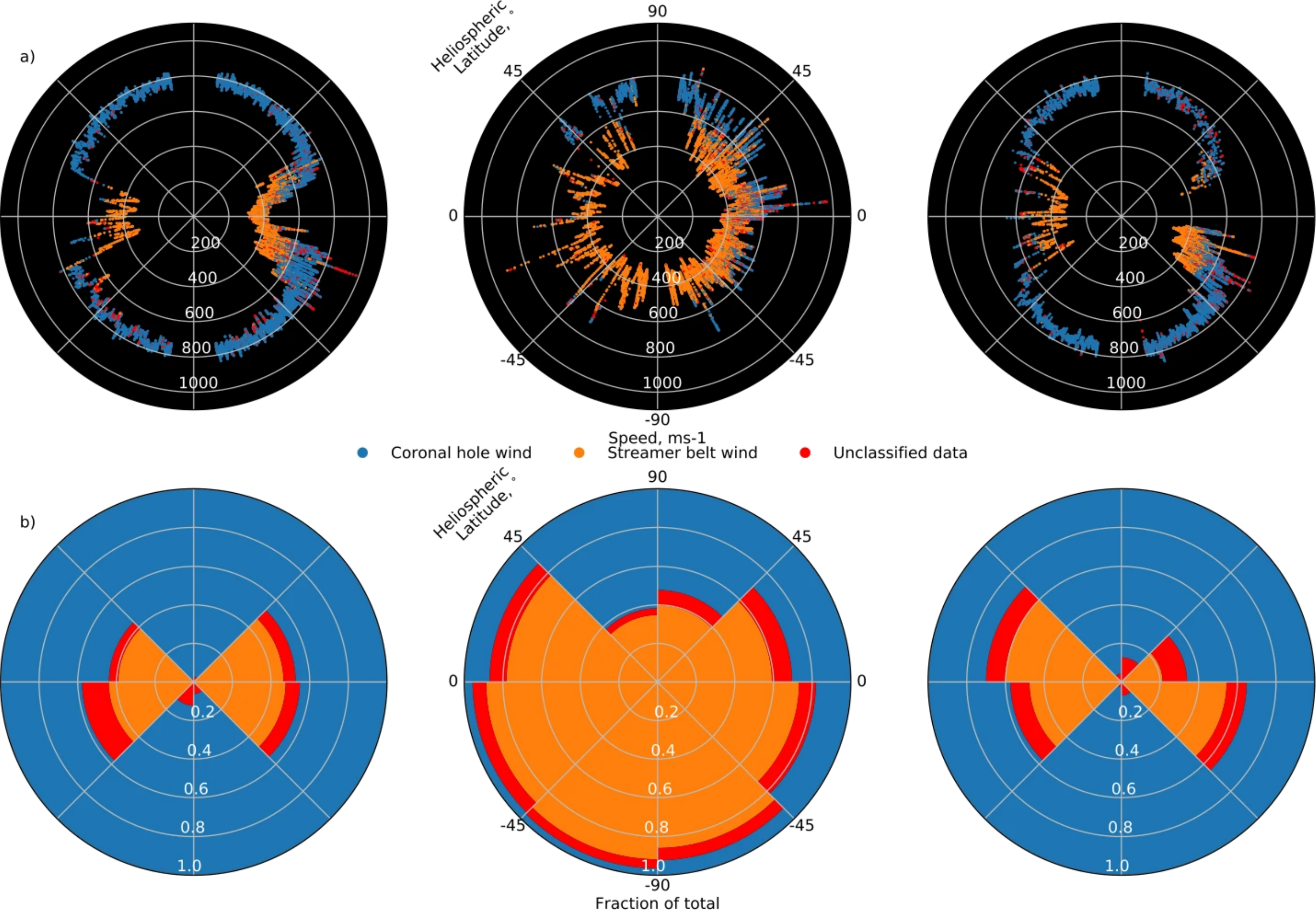}
    \caption{Machine-learning classification of the Ulysses data in Figure~\ref{fig:sec2:uly3orbit} into coronal hole wind (blue), streamer belt wind (orange), and unclassified data (red) \citep{Bloch2020}. The lower plots show the fraction of each type of solar wind as a function of heliolatitude. }
    \label{fig:sec2:bloch}
\end{figure}

Methods of ``automated'' solar wind structure identification, based on combinations of selected solar wind parameters, have also been proposed \citep[e.g.,][]{Neugebauer2003, Zhao2009,Xu2015}, though \cite{Neugebauer2016}  note that the classifications provided by the three schemes they considered were only in agreement 49\% of the time. Figure~\ref{fig:sec2:xub} shows the occurrence of solar wind structures in 1963--2013 obtained by \cite{Xu2015}.  This shows similar patterns to Figure~\ref{fig:sec2:per}, though the slow solar wind is sub-divided into streamer belt flows and sector reversal regions. Recent efforts to classify solar wind structures have utilized machine-learning \citep[ML; e.g.,][]{Camporeale2017,Heidrich2018,Bloch2020,Li2020}. As an example, Figure~\ref{fig:sec2:bloch} \citep[from][]{Bloch2020} shows a ML classification of the solar wind flows at Ulysses shown in Figure~\ref{fig:sec2:uly3orbit}. 

Solar wind structures cover a wide range of size scales \citep{Verscharen2019}, including  small-scale FRs \citep[e.g.,][]{Moldwin2000}, density structures \citep[e.g.,][]{Viall2008}, and features associated with turbulence \citep{Bruno2013}.  Recent movies of heavily-processed coronagraph images offer a tantalizing view of a complex, structured solar wind \citep{DeForest2018}. The small-scale solar wind structuring may have effects on the CME propagation itself as CMEs tend to adjust to the solar wind speed and IMF. This may generate CME frontal deformation, and local measurements from in situ data may influence statistical results. While interesting in their own right, such small-scale structures will not be discussed further in this section.  

Much of our knowledge of the structure and time variation of the solar wind is based on observations from heliospheric spacecraft.  However, these have only probed limited regions of the heliosphere \citep{Verscharen2019}, in some cases only at certain solar activity levels.  Also, with the notable exception of Ulysses, these spacecraft generally remain close to the ecliptic plane, where only a limited ($\sim\pm7^\circ$) sampling of the latitudinal structure of the solar wind is provided by the inclination of the solar equator relative to the ecliptic.  Also, Ulysses was still $\sim$1~AU from the Sun when at the highest latitudes. Hence, there have only been limited studies of the latitudinal structure of the solar wind in the inner heliosphere using in situ observations.  The Helios 1 and 2 spacecraft orbiting at 0.3--1~AU during solar cycle 21 demonstrated clearly that even small changes in spacecraft latitude can significantly affect the solar wind structures observed in situ \citep{Schwenn1978,Burlaga1978}. More recently, similar conclusions were inferred from STEREO measurements when the two spacecraft had a small separation in latitude but observed or missed features associated with large-scale solar wind structures \citep[e.g.,][]{GomezHerreroetal2011}. Since the last Roadmap by \cite{Schrijver2015}, PSP and SolO have been launched to probe the solar wind in the inner heliosphere (see also Section \ref{Data Availability}). PSP has, at the time of writing, already sampled into below 20~Rs from the Sun and detected for the first time the sub-Alfvenic point \citep[][]{Kasperetal2021}, while SolO is commencing a series of maneuvers that will ultimately increase its latitude range to ${\pm}35^\circ$. Both missions will improve our knowledge of the solar wind in the inner heliosphere in the next few years.  Recent planetary spacecraft have provided observations of the solar wind while in their cruise phases and/or in orbit, such as MESSENGER and BepiColombo (Mercury missions), Venus Express (Venus mission), Mars Odyssey, Mars Express, and MAVEN (Mars missions), Rosetta (Comet 67P mission), Juno (Jupiter mission), Cassini (Saturn mission), and New Horizons (Pluto mission), complementing earlier observations of the outer heliosphere from spacecraft such as Pioneers~10 and 11 and Voyagers~1 and 2. \cite{Witasse2017} demonstrated how combined observations from multiple spacecraft may be used to track an CME from the Sun (on October 14, 2014) out to Cassini at 9.9~AU and possibly to New Horizons at 31.6~AU, and Voyager~2 at 110~AU in late March 2016. Other studies of solar wind structures using planetary spacecraft include \cite{Moestl2015}, \cite{Prise2015}, \cite{Janvieretal2019}, \cite{Davies2021}, \cite{Palmerio2021c}, and \cite{Winslowetal2021a}.

\subsection{Modeling the Background Solar Wind}\label{sec:section2:SWmodel}
Models of the solar wind can provide a global view of solar wind structures and help to interpret the structures observed by spacecraft. Several such models are in use in space weather studies, which are described in more detail in Section~\ref{sec3.2:sw_model}. Though differing in details, many based on solving MHD equations on a suitable spatio-temporal grid. (Solar wind models are discussed further in the TI2 paper by \cite{Arge2023}.) Currently, these models generally use as input coronal magnetic field models based on photospheric magnetograms either from the ground (e.g., the GONG network) or spacecraft (e.g., SOHO, SDO).  An example is the Wang--Sheeley--Arge \citep[WSA;][]{Arge2004} model, which is based on an observed anti-correlation between the non-radial expansion of coronal field lines and the solar wind speed \citep[][see also Section~\ref{sec3:fastSW}]{Wang1990}. 
However, differences in slow and fast solar wind composition and charge states \citep{Vonsteiger2000} indicate that expansion alone is not the cause of speed solar wind variations and different solar sources must be involved \citep{Laming2015}. In particular the slow wind is found to have a substantial transient component \citep[e.g.,][]{BourouaineEtAl2020} that in general is not addressed by current modeling (e.g. there is no truly time dependent global solar wind model based on time dependent synoptic maps, although some global models can provide frequent updates based on updating maps such as ADAPT---see also Section~\ref{fig:sec1:overview}).

A major problem with modeling the global solar wind in this way is the absence of photospheric magnetic field observations from the far side of the Sun.  While magnetic fields observed on the front side can be assumed to persist onto the far side, a significant change on the far side, for example due to the emergence of an active region or a change in a coronal hole boundary, which may alter the solar wind structure, will not be detected until it rotates onto the front side.  A recent approach \citep{Jeong2020} uses artificial intelligence (AI) to predict far-side coronal magnetic fields, though this requires far-side EUV observations such as from the STEREO spacecraft \citep[see also][]{HeinemannEtAl2021b}, which will now not be available for several years with STEREO-A returning to the front-side of the Sun in August 2023. Observations of the solar magnetic field from spacecraft at L4 and/or L5, ${\sim}60^\circ$ west/east of the Sun--Earth line \citep{Vourlidas2015,Posner2021, Bemporad2021} will help to reduce, but not remove, this observational gap.  A spacecraft at L5, such as Vigil, would also monitor co-rotating structures around 5 days before they reach Earth \citep[e.g.,][]{Simunac2009}.  In addition, magnetic fields in the polar regions of the Sun are poorly measured from Earth.  SolO moving to higher latitudes in coming years will help to improve our view of the poles. 

 \begin{figure}
\centering
\includegraphics[width=1.0\linewidth]{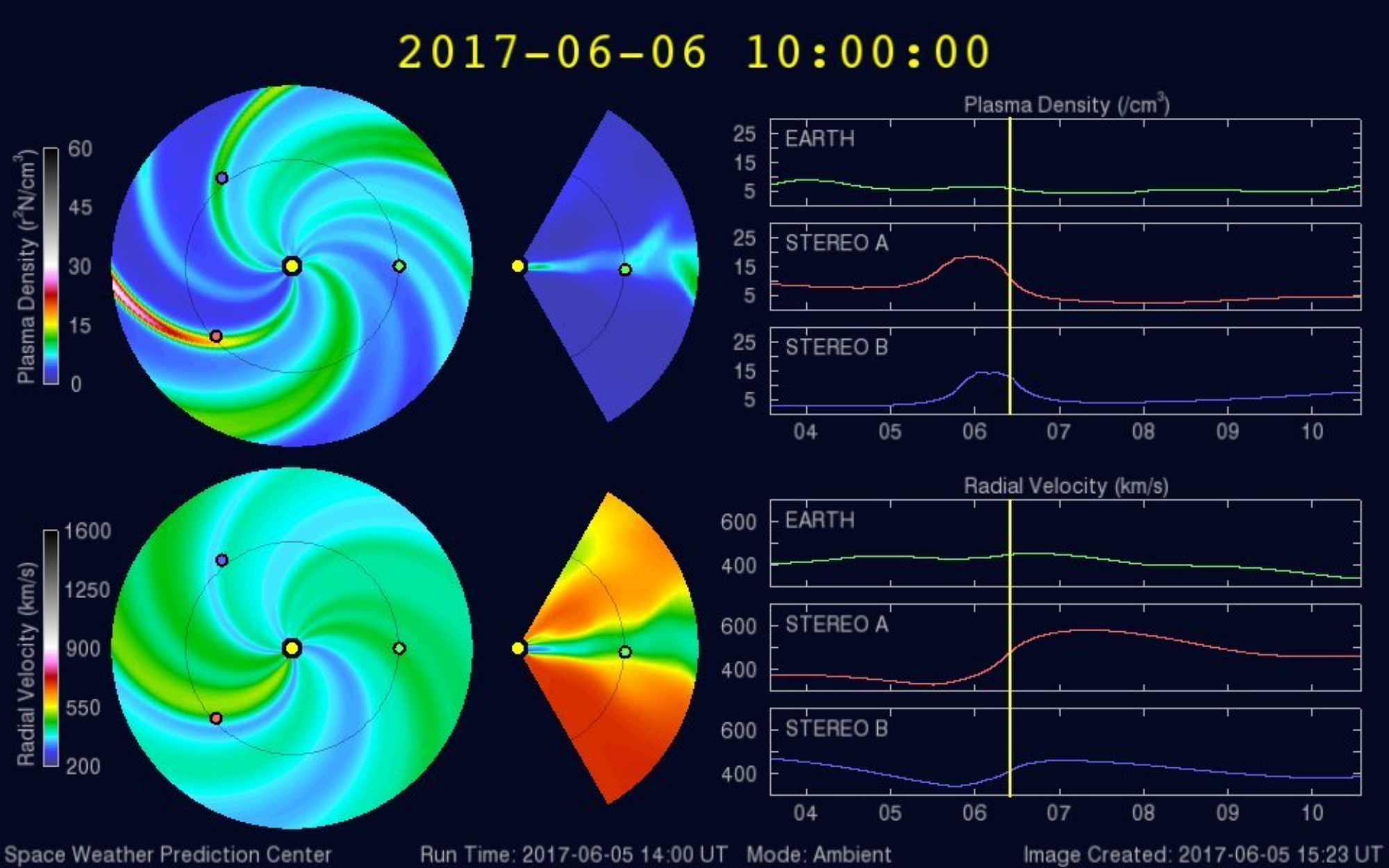}
\caption{Screenshot from the NOAA (National Oceanic and Atmospheric Administration) Space Weather Prediction Center website (\href{http://www.swpc.noaa.gov/products/wsa-enlil-solar-wind-prediction}{http://www.swpc.noaa.gov/products/wsa-enlil-solar-wind-prediction}) showing the density (top row) and solar wind speed (bottom row) predicted by the WSA--ENLIL model. The yellow, red, and blue dots indicate respectively the locations of Earth, STEREO-A, and STEREO-B at the time of the simulation.}
\label{fig:sec2:ENLIL}
\end{figure}

Figure~\ref{fig:sec2:ENLIL} shows an example of a frame from an ENLIL simulation of the solar wind, showing on the left-hand side the speed and density in the ecliptic and in a meridional cut at the location of Earth (yellow dot). Note the large-scale regions of slow and faster solar wind and their spiral configuration, as well as the CIRs indicated by density enhancements at the leading edges of the HSSs, similar to those in the schematic in Figure~\ref{fig:sec2:bdstreams}. The solar wind speed is also lower at low latitudes, resembling the Ulysses observations near solar minimum in Figure~\ref{fig:sec2:uly3orbit}. Time-series plots of the speed and density at Earth and the STEREO spacecraft are shown on the right-hand side, indicating the passage of (different) CIRs on days 5--6 at STEREO-A and -B, which can also be identified by the spiral density enhancements in the in-ecliptic density in the top left of the figure.    

The global structure of the solar wind in the inner heliosphere may also be inferred using remote-sensing observations such as IPS, which is driven by irregularities in the solar wind density \citep[e.g.,][]{Breen1998,Bisi2010,Bisietal2010a} and observations of variations in white light scattered from solar wind density enhancements \citep[e.g.,][]{Jackson2001, Rouillard2008, Eyles2009, Howard2013, Conlon2015, Plotnikov2016}.   However, inferring solar wind structures from such line-of-sight observations is complex.  Tomographic reconstructions of the global solar wind density have been derived from IPS and/or white-light observations \citep[e.g.][and references therein]{JacksonHick2002, Bisietal2010b, Jackson2011, Jacksonetal2020}, and solar wind velocity and density reconstructions using IPS are routinely provided by the \href{https://ips.ucsd.edu/}{University of California, San Diego (UCSD)}.  More details on IPS techniques for space weather and the implementation of IPS data in models are given in Section~\ref{sec:section4}.

Validating global solar wind models is a challenge: spacecraft observations only provide comparisons at widely-separated points in the heliosphere and, with the exception of Ulysses, near the ecliptic.  While a model may be “tuned” to agree with observations at a specific point, there is no guarantee that this tuning will also improve the agreement at other locations, where no observations may be available to provide validation.  Thus, the improved validation of global solar wind models ideally requires observations from as many spacecraft as possible.  Recently, \cite{Lang2021} have used DA to improve forecasts of the solar wind parameters at Earth by using observations from widely separated spacecraft to update model inner boundary conditions. \cite{Riley2021a} have discussed using PSP observations to constrain MHD heliospheric models with different coronal models as input.  Several studies have validated solar wind models using observations at Earth or other locations  \citep[e.g.,][]{Cohen2008,Owens2008, Gressl2014, Jian2015,Jian2016,Macneice2018,Reiss2020}.  For example, \citet{Gressl2014} and \citet{Jian2015} compared the validity of parameters derived from ENLIL simulations using different magnetograms and coronal field models as input. The validation of solar wind models is discussed further in the TI1 paper by \cite{ReissEtAl2022}. A validation of heliospheric modeling algorithms through pulsar observations is given in the TI1 paper by \cite{SHAIFULLAH2022}.

\subsection{Geomagnetic Effects from CMEs and SIRs/CIRs}

Geomagnetic effects are driven predominantly by the strength of the southward component of the solar wind magnetic field, $B_{\rm z}$, and the solar wind speed \citep[e.g.,][]{Newell2007}. Studies have shown that CMEs are the major drivers of strong geomagnetic storms, with a smaller fraction associated with CIRs \citep[e.g.,][and references therein]{Kilpua2017b}.  For example, \cite{Zhangetal2007} found that of 88 storms with $Dst\le -100$~nT in 1996-2005, only 13\% were associated with CIRs. Another 53\% were associated with single CMEs, and 24\% were produced by interactions of multiple CMEs. Although the southward fields driving CME-associated storms were generally in the ejecta, with the largest storms being associated with MCs/MOs with extended intervals of persistent southward field, 27\% of these strong storms were driven by sheath magnetic fields \citep[see also][]{Kilpua2017a, Kilpua2017b}. Also \cite{YermolaevEtAl2021b} highlighted that about 10\% of moderate to large geomagnetic storms are sheath-induced rather than driven by the ejecta. Geomagnetic activity associated with CIRs is largely driven by intermittent southward turnings of the magnetic field associated with Alfv\'enic fluctuations that results in extended enhanced activity as measured by the AE index persisting during passage of the HSS \citep[e.g.,][]{Tsurutani2006,Buresova2014}. Because of these differences in the storm drivers, the geomagnetic response as measured by magnetic indices such as Dst (Disturbance storm time), SYM-H (symmetric disturbance of horizontal geomagnetic fields), ASY-H (longitudinally asymmetric disturbance of horizontal geomagnetic fields), AE (auroral electrojet) and K\textsubscript{p} (\textit{planetarische Kennziffer}; global geomagnetic storm index), differs for CIR and CME-driven storms.  Further discussion of the geomagnetic effects of CIRs/SIRs can be found in Section~\ref{sec:section3}.

Since the coupling processes in the solar--terrestrial system during different kinds of solar wind are not fully understood, this may lead to discrepancies in models and forecasts of the effects of solar wind structures on Geospace.  Hence, the accurate prediction of the geoeffectiveness of space weather events (both large- and medium-scale) and the impacts on technological systems is a major challenge \citep[for more details, see the G Cluster TI2 papers by e.g.,][]{Opgenoorth2023,Bruinsma2023,Tsagouri2023,Zheng2023}.

Because of the close association between geomagnetic storms and CMEs, storm prediction often relies on the observation of a CME associated with frontside solar activity, perhaps combined with modeling to assess whether the related CME is likely to encounter Earth. However, a major challenge is to predict the strength and orientation of the CME magnetic field during Earth encounter as early as possible, ideally using observations of the related solar event \citep[e.g.,][]{SavaniEtAl2015,Savani2017}.   There is also evidence that stealth CMEs, without clear signatures of their solar source, may occasionally give rise to CMEs that produce significant geomagnetic activity. The circumstances of such so-called ``problem" geomagnetic storms were recently reviewed by \cite{Nitta2021}. 

\subsection{Summary}
In summary, this section has briefly described the main features of the solar wind in particular CIRs/SIRs, HSSs, and CMEs that are the major components of the solar wind that drive space weather. Observations from the recently launched PSP and SolO missions already have, and will continue to, provide valuable insights into the configuration and evolution of structures in the inner heliosphere far closer to the Sun than the 0.3~AU achieved by the Helios mission and, in the case of SolO, eventually to higher latitudes than previously attained at such distances from the Sun. New methods, such as ML, as well as new data sources, such as IPS, hold promise to better develop reliable solar wind structure classifications. Observations that extent into the upcoming cycle 25 will enable further studies of cycle-to-cycle variations of the characterisitcs of solar wind structures.

\section{SIRs/CIRs Formation and Propagation}
\label{sec:section3}

To properly forecast the arrival of transient events, we first need a reliable solar wind model which we do not have at the moment. For predicting the background solar wind structures in interplanetary space with higher accuracy, enhanced knowledge about the physics underlying the processes forming these structures is necessary. 
In this section, we explore the open questions \citep[see also][]{Viall2020} and ongoing scientific research specifically focusing on the generation and evolution of HSSs in the context of space weather, starting from their solar source regions, coronal holes, out to interplanetary space. For complementary ISWAT activities on solar wind generation and modeling we refer to S2 Cluster paper by \cite{Arge2023}.


\subsection{SIRs/CIRs and their Solar Sources}

From coronal observations, \cite{Waldmeier1956} was the first to associate dark regions in the corona (M-regions) with the recurrent geomagnetic activity noted by \cite{MAUnder1904}. Later, such geomagnetic activity would be clearly related to HSSs emanating from the dark coronal regions that are now known as coronal holes \citep[e.g.,][]{Newkirk1967,Wilcox1968}. Hence, HSSs are deeply linked to the presence and evolution of coronal holes on the Sun. In particular, low-latitude coronal holes are most relevant as sources of streams impacting planets in the ecliptic plane. The equatorward extensions of polar coronal holes start to form shortly after solar maximum \citep[][]{HarveyRecely2002}, leading to the appearance of the periodic geomagnetic storms that modulate planetary atmospheres and occur at a higher frequency close to solar minimum \citep[e.g.,][]{Temmeretal2007,LeiEtAl2008}.
With that, the number of SIRs/CIRs, and as such the heliospheric structure in general, varies strongly depending on the solar cycle and the coronal magnetic field configuration. To reliably forecast the solar wind structure, the number of streams per rotation and their properties need be sufficiently well known and modeled. 

\begin{figure} 
\centering
\includegraphics[width=1.\linewidth]{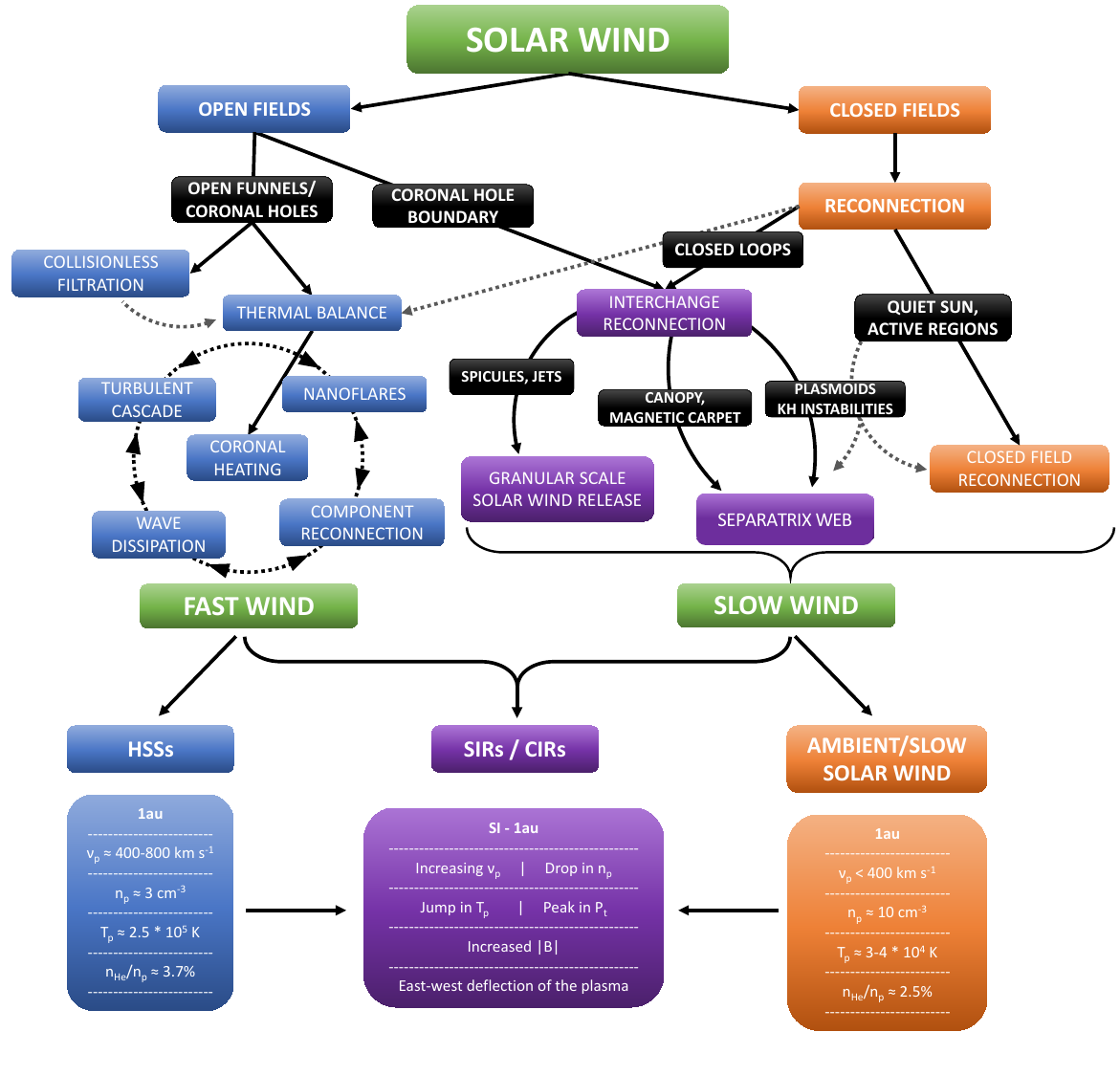}
\caption{Proposed pathways of the solar wind from origin to heliosphere and release mechanisms. A complex interaction of many different processes may finally produce the slow and fast solar wind that lead to the formation of SIRs. Solar wind values (proton speed, $v_p$, proton density, $n_p$, proton temperature, $T_p$, and charge states, $n_{\rm He}/n_p$) are taken from \cite{Schwenn2006} and stream interface (SI) criteria by \cite{Jian2006}.}
\label{fig:sec3:solar_wind}
\end{figure}

In many space weather forecasting models, the large-scale solar wind structures in the heliosphere are usually regarded as `quasi-time-stationary' and evolutionary aspects occurring during a solar rotation or on longer time scales, are often neglected. However, \cite{Heinemann2018a,Heinemann2020} showed with STEREO data that the evolution of coronal holes causes variations in the resulting HSSs as measured in situ. The more variable denser and slow solar wind, which is also found within coronal hole regions \citep{Bale2019}, plays a role that is not well established in the formation of SIRs. The parameters of the solar wind upstream and downstream of the stream interface, hence, the boundary separating the predominantly fast and predominantly slow wind regimes, have been well studied \citep[e.g.,][]{Crooker2012} and depend on the interplay of slow and fast solar wind (see Section~\ref{sec:section2}); however, short-term variations in both solar wind components have not been considered yet in determining the properties of the resulting SIR. Therefore, a more detailed understanding of the solar wind, heliospheric magnetic field, and their sources is vital for refining and validating space weather forecasting efforts. The processes leading to the formation of SIRs are manifold. Figure~\ref{fig:sec3:solar_wind} depicts several of them and shows how they might interrelate with each other. But it is still not well understood how the conditions in both slow and fast wind influence the formation of the SIR and the resulting space weather effects.  

\subsubsection{Fast Solar Wind}\label{sec3:fastSW}

\begin{figure}
\centering
\includegraphics[width=1.\linewidth]{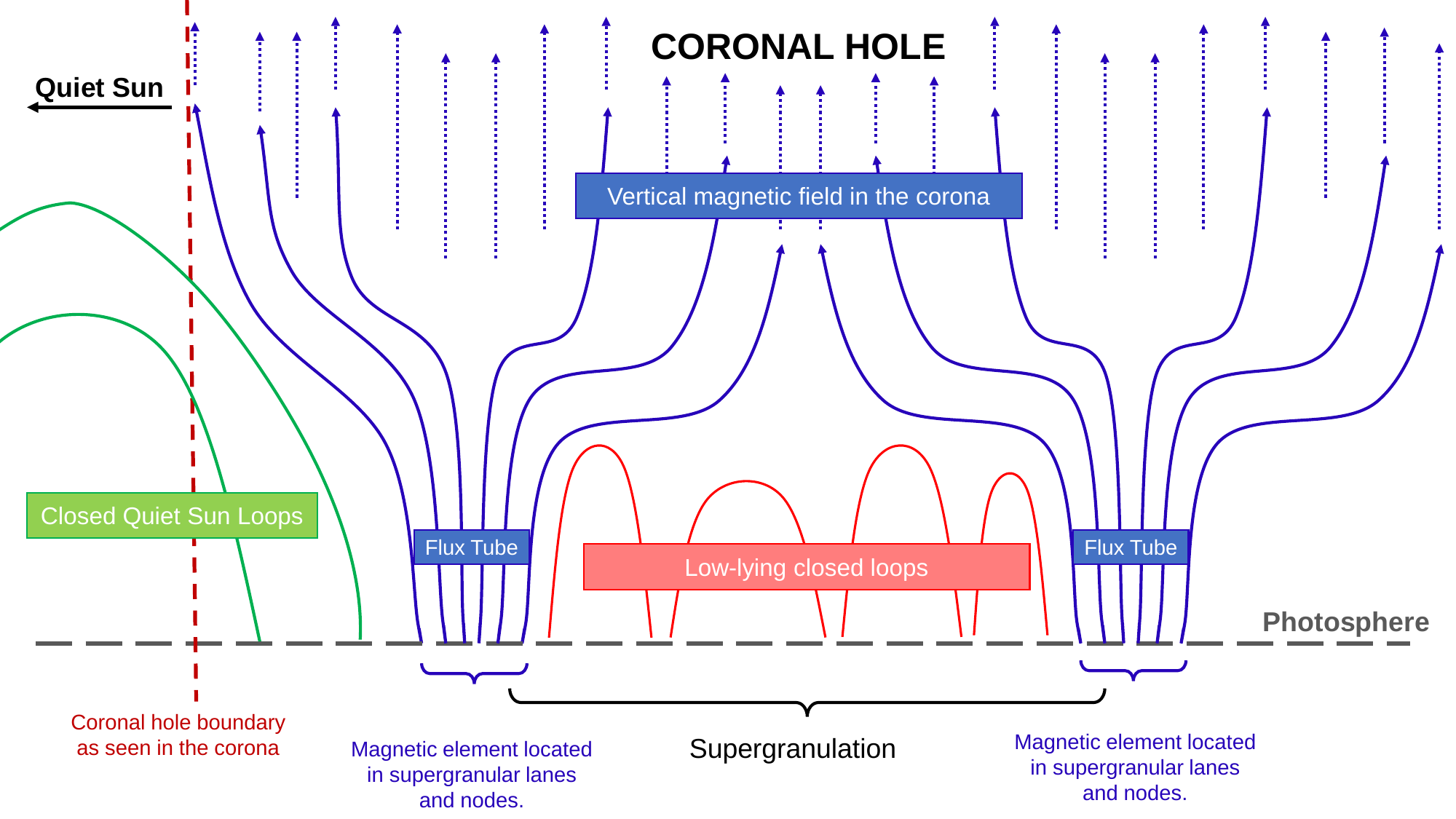}
\caption{Simplified depiction of the magnetic field configuration of a coronal hole from the photosphere to the corona. Figure by S.~G.~Heinemann, based on drawings by \cite{Cranmer2005} and \cite{Wedemeyer-Boehm2009}. Not to scale.}
\label{fig:sec3:magnetic_coronal_hole}
\end{figure}

Coronal holes are often regarded as coherent, rigid structures that evolve slowly. However, close inspection has revealed that the magnetic structure and substructure within coronal holes is highly complex. According to the standard model of the magnetic field configuration of coronal holes, open magnetic funnels \citep[e.g.,][]{Tu2005} that originate in small scale unipolar photospheric magnetic elements \citep[][]{Heinemann2018b,Hofmeister2019} located in the lanes and nodes of the magnetic network \citep[][and references therein]{Cranmer2005}, expand to fill the coronal space with an approximately uniform vertical magnetic field. This expansion is most likely modulated by low-lying closed loops existing in the space between the open fields \citep[][]{Wiegelmann2005}. Figure~\ref{fig:sec3:magnetic_coronal_hole} summarizes in a cartoon the mix of open and closed magnetic field structures reaching different heights in the corona and that subsequently extend to interplanetary space.

The magnetic funnels or flux tubes, that are the sources of the fast solar wind outflow, are the subject of many observational and modeling studies \citep[e.g.,][and more]{Wojcik2019,Tripathi2021,Bale2021}. However, \textit{it is still unclear how the funnel properties are linked to the properties of the outflowing solar wind.} The vertical expansion profile may depend on the height and, as such, on the field strength, of the low lying coronal loops in coronal holes which inhibit lateral expansion  \citep[for simulations see][]{Wiegelmann2005}. Detailed knowledge about the funnels can help to constrain the parameters of the resulting solar wind and improve understanding of the subsequent formation of SIRs.

Often, in situ plasma velocity profiles of HSSs near 1~AU show double or multiple  peaks, which suggest that there are multiple centres of solar wind outflow in individual coronal holes \citep[][]{Heinemann2018a, garton2018}. Knowledge about the source locations of the observed solar wind could increase the chances of observing the actual outflows, thereby improving not only solar wind backmapping methods (e.g., ballistic backmapping, \citealt{Peleikis2017,Macneil2022} or slip backmapping, \citealt{Lionello2020}) but also the modeling  of the solar wind release. However, investigation of the magnetic and plasma structure of coronal holes, especially magnetic funnels, is an arduous task due to the sparse availability of observations at low field strengths. It is questionable whether the commonly-used potential field source surface extrapolation (PFSS), which assumes a zero-current approximation that leads to a potential field, plus a prescribed source surface to open magnetic field lines \citep[][]{Altschuler1969,Schatten1971}, is valid at low heights in coronal holes. The community may want to go forward introducing a more realistic coronal source surface to better estimate solar wind boundary conditions \citep[see also e.g.,][]{Asvestari2019}. Promising approaches could be to examine high-resolution spectroscopy in coronal hole outflow regions using e.g., DKIST. However, it is not straightforward to relate solar surface parameters with solar wind parameters measured in situ (e.g., at 1~AU) as interaction processes (such as solar wind acceleration, slow--fast wind interaction, switchbacks, turbulences) may mask any correlation. In situ measurements close to the Sun, such as with PSP, where the slow and fast solar winds have had less time to interact, can help in revealing possible relations.

It is known that, in general, larger coronal holes produce HSSs of higher speed. From this relationship, 1D methods of forecasting the solar wind at 1~AU have been developed using empirical models relating the coronal hole area with the in situ measured solar wind peak speed \citep[][]{Nolte1976,vrsnak2007_I, Temmer2018,Heinemann2018a,Bu2019,Akhtemov2018,Heinemann2020} and the related geomagnetic activity \citep[][]{vrsnak2007_II,Nakagawa2019}. The relations for peak velocity hold well for defined coronal holes near disk center and a correction may be applied for different latitudes \citep[][]{Hofmeister2018}. \textit{However, the physical principles behind the relation between solar wind peak velocity and coronal hole area are not yet fully understood}. It has been suggested, and analytically shown, that the coronal area-HSS speed relation may be entirely a propagation effect (associated with slow--fast wind interaction) in interplanetary space caused by a discrete bimodal velocity distribution \citep{Hofmeister2022}. In contrast, the empirical relation ($v_{\textsc{sw}} \sim 1/f$) between the solar wind speed $v_{\textsc{sw}}$ and the flux tube expansion factor $f$ is often used to explain the connection between coronal holes and solar wind speed \citep[][]{Wang1990,Wang2010} that produce the observed bimodal distribution (see also Section~\ref{sec:section2:SWmodel}). Closed fields can influence the behavior of open fields and vice versa. In particular, magnetic field gradients along the vertical coronal hole boundary can influence the magnetic field expansion behavior and the resulting plasma outflow. Precise in situ solar wind measurements at different radial distances \citep[with PSP and SolO, see e.g.,][]{PerroneEtAl22}, as well as more advanced measurements of the first ionisation potential (FIP) effect \citep[][]{Pottasch1964a,Pottasch1964b} and  heavy ion charge states \citep[e.g.,][]{Lepri2013}, which help to connect the solar wind with the regions of origin \citep[][]{Brooks2011,Zambrana-Prado2019,Parenti2021}, will help to shed light on the origin of the relation between coronal hole area and HSS peak velocity.

\begin{figure}
\centering
\includegraphics[width=1.\linewidth]{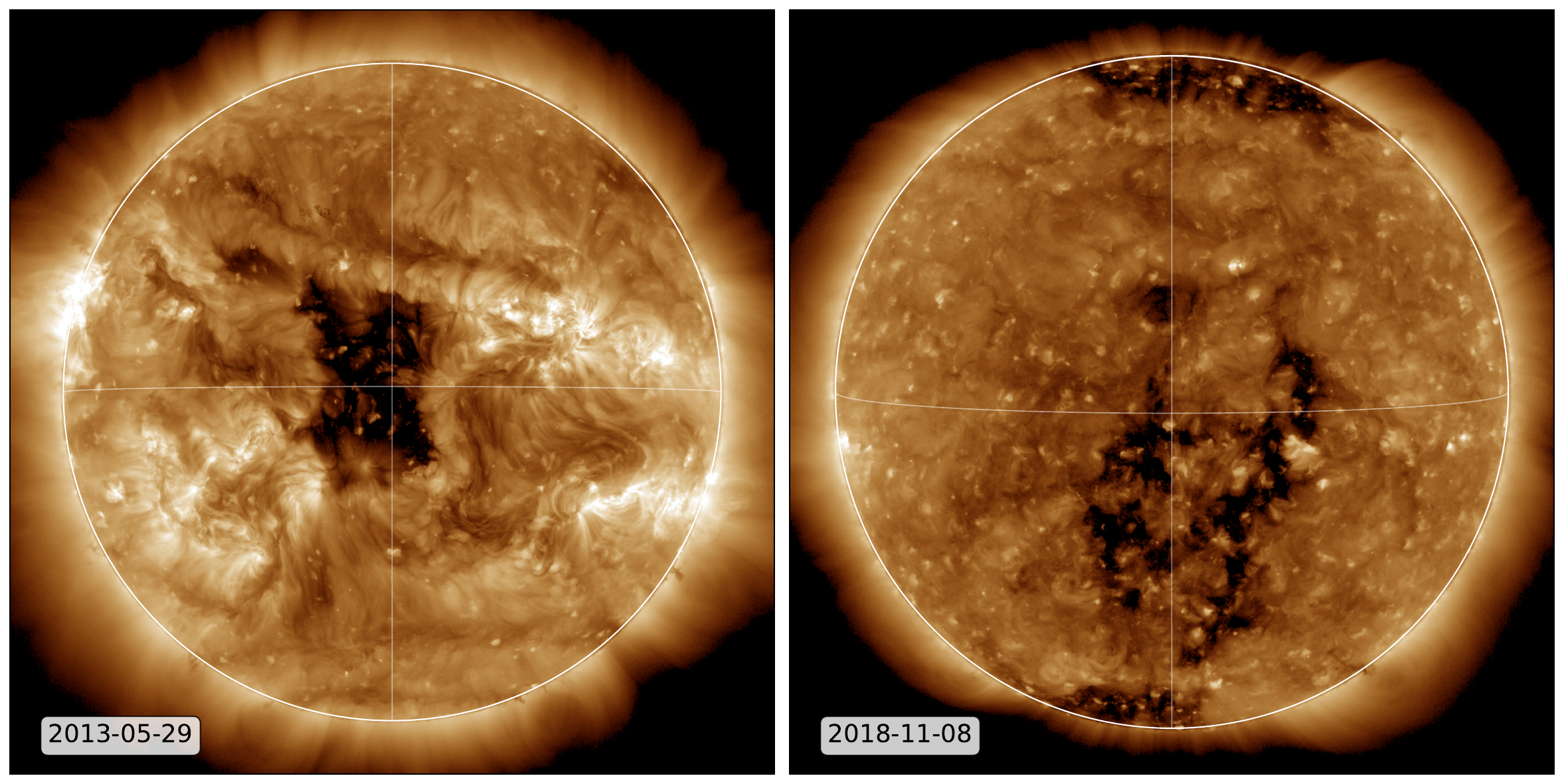}
\caption{Images of two SDO/AIA (Atmospheric Imaging Assembly) coronal holes observed in the 193~{\AA} filter. The left panel shows a clearly defined, compact coronal hole during solar maximum (May 29\textsuperscript{th}, 2013) and the right panel shows a very large but \textit{patchy} coronal hole during solar minimum (November 8\textsuperscript{th}, 2018). }
\label{fig:sec3:ch_comparison}
\end{figure}

It has been shown that for larger coronal holes without a clearly-defined boundary, the area to HSS peak velocity relation breaks down \citep[][]{garton2018,Geyer2021}. These coronal holes without clear boundaries have been observed preferentially during solar minimum (especially in the 2018--2020 minimum, e.g., see Fig.~\ref{fig:sec3:ch_comparison}). They may contain multiple brighter closed-field regions, and may stretch over large areas at low latitudes. The observed mean magnetic field strengths in such regions are around $\pm$1~G, with only a slight flux imbalance suggesting a low open flux. Because they resemble loosely-connected darker patches in EUV observations, these coronal holes have been dubbed \textit{patchy} coronal holes \citep[][]{Heinemann2020,Samara2022b}. The observed peak velocities of the solar wind emitted by such patchy coronal holes with areas larger than $10^{11}$~km$^2$ usually range from 450 to 600~km~s$^{-1}$, which does not follow the usual empirical relation. Due to their many differences from clearly-defined coronal holes, \textit{patchy} coronal holes need to be treated separately in terms of their space weather effects. We still do not know how HSS plasma emanating from a coronal hole is influenced by the presence or absence of closed magnetic field within the coronal hole and/or nearby it.

\subsubsection{Slow Solar Wind in the Frame of Solar Wind Interaction} \label{subsubsec:S3:slow_wind}
When discussing SIRs and CIRs, the contribution of the slow solar wind to the stream--stream interaction cannot be neglected. In contrast to fast solar wind streams, there is no full agreement on the source of the slow solar wind. Typically, slow solar wind has a composition resembling that of closed fields in the corona, but a closed-field source would appear to be inconsistent with the large angular widths of the slow solar wind. The strongest consensus is that reconnection is responsible for the slow solar wind outflow. This includes interchange reconnection of closed and open fields, typically, but not only, at coronal hole boundaries, or closed field reconnection, for example in active region cusps. The magnetic carpet of the Sun, resulting in a separatrix and quasi-separatrix web (dubbed the ``S-web"), is often proposed as the source of the ambient solar wind \citep[][]{2011Antiochos}. Pseudo-streamers \citep[][]{Riley2012}, coronal streamers \citep[][]{Habbal1997,Ofman2004}, coronal hole -- active region boundaries \citep[][]{Ko2006} and quiet-Sun regions \citep[][]{Fisk1998} have also been suggested as the origin of the slow wind. It has also been reported that the slow solar wind may not only originate in closed field regions but also in small equatorial coronal holes \citep[][]{Ohmi2004,Stansby2020}. Recent PSP observations clearly indicate slow and fast solar wind from an equatorial coronal hole \citep[][]{Bale2019}. There is also evidence for different types of slow solar wind (based on, e.g., FIP abundances, charge states and Alfv\'{e}nicity), further supporting that there are multiple sources for the slow wind. An open question is whether slow solar wind flows from different sources interact with HSSs in a different way, and how that affects the formation of SIRs and their 1~AU characteristics.

\subsection{Solar Wind Properties at 0.1~AU}\label{sec3.2:sw_model}

The solar wind properties at 0.1~AU, where it is assumed that most solar wind acceleration has ceased \citep[e.g.,][]{Cranmer2002,Bemporad2017}, are commonly used as input for heliospheric models. There are still open questions on the solar wind acceleration process itself that will not be discussed here; the interested reader is referred to \cite{Viall2020}. The 0.1~AU properties are usually inferred from coronal models based on photospheric magnetograms and empirical relations. However, the different assumptions and input data used can result in large variations of the inferred properties \citep[][]{Macneice2018,Samara2021,Riley2021b}. For the parameters relevant to SIRs/CIRs, their accuracy relies on how coronal holes are represented and on the assumptions made to estimate/derive the plasma and magnetic field parameters.

Most commonly, empirical relations between solar magnetic field quantities and the solar wind speed at 0.1~AU are used. These relations may be expressed as $v_{0.1AU} = v(f,d)$, and depend on the flux tube expansion factor $f$ and the distance from coronal hole boundary $d$. The exact form of the relation varies between different authors, studies and models \citep[e.g., see][]{Arge2000,Riley2001,Owens2008,McGregor2011,Wiengarten2014,Pinto2017,Pomoell2018}. The outer boundary conditions derived from a coronal model at 0.1~AU are typically used as the inner boundary conditions for heliospheric models. Current state-of-the-art solar wind models include the MHD models ENLIL \citep{OdstrcilPizzo1999b, Odstrcil2003}, EUHFORIA \citep{Pomoell2018,Poedts2020b}, ICARUS \citep{Verbekeetal2022}, hydrodynamic approaches \citep{Riley2011a, owensetal2020}, or kinematic models such as the WSA Inner Heliosphere model \citep[WSA-IH;][]{Arge2000}.

Although these models and methods are widely used in solar and heliospheric physics and space weather research, assumptions need to be made for plasma and magnetic field properties that cannot be observed directly. The solar wind speed, density, and temperature, as well as the magnetic field strength and structure, are not well constrained. As already noted, it is believed that a large proportion of the solar wind acceleration takes place below 0.1~AU, and so this distance corresponds roughly to the transition between the solar and heliospheric regimes. More precise knowledge about the environment at 0.1~AU would lead to a better representation of the  heliosphere through modeling. New missions that venture into the close proximity of the Sun, e.g., PSP that has already passed through the Alv\'{e}n point into the solar corona, will provide new in situ measurements of the environment close to and below 0.1~AU \citep[][]{Kasperetal2021}. In addition, new IPS observations could potentially be used to reconstruct solar wind maps at 0.1~AU \citep[similar to those in][]{Sokol2015,Jacksonetal2020}. Ideally, a ``universal" relation that successfully links solar surface properties to 0.1~AU should be established, making as few assumptions as possible, that will lead to  a community consensus on constraining solar wind parameters as input for heliospheric models for research and space weather prediction.

At 0.1~AU and beyond, plasma motion dominates the heliosphere ($\beta_\mathrm{plasma} \gg 1$) but the magnetic field structure cannot be neglected. Most heliospheric models produce a mostly smooth bipolar heliosphere (especially during solar minimum) separated by the HCS. However, recent PSP observations provide evidence of a much more complex magnetic field structure close to the Sun \citep[][]{Bale2019} including changing and mixed polarities due to different origins (e.g., open funnels, closed fields, coronal jets) as well as kinks and twists in the magnetic field \citep[switchbacks;][]{Mozer2020,DudokdeWit2020,Squire2020,Tenerani2020}. 
On larger scales, observed variations in the field (e.g., B$_{r}$) can be reproduced by combined PFSS and MHD models. However, such models cannot reproduce the fine structure. Although well-established and commonly used, it is not clear whether the flux tube expansion factor and distance to coronal hole boundary are optimal parameters for deriving magnetic field and plasma properties at 0.1~AU. A better knowledge of that would lead to improved, and more realistic models of the fractured structure of the open fields at 0.1~AU, and might also show what role these structures play in larger-scale heliospheric dynamics.

\subsection{Solar Wind Evolution in Interplanetary Space}\label{sec3.3}
Many observational and modeling studies have investigated the evolution of HSSs and SIR/CIRs with heliocentric distance, in particular during the Helios/Pioneer/Voyager and Ulysses eras \citep[e.g.,][]{Gosling1999,Whang1990,Burlaga1990,Burlaga1995,Burlaga1997,Gazis1999,Forsyth2001} as well as more recently \citep{Allen2021}. In particular, Helios observations at 0.3--1~AU showed that the velocity shear between slow and fast solar wind is largest closest to the Sun (consistent with different sources for slow and fast solar wind) and declines rapidly at 0.3--0.5~AU, before becoming approximately constant out to at least 1~AU \citep{Schwenn1990}. Beyond $\sim$1~AU, the expansion speed of a SIR may exceed the local magnetosonic speed, resulting in the formation of a forward shock at the SIR leading edge and a reverse shock at its trailing edge \citep[e.g.,][]{Smith1976,Gosling1976}. Such shocks are occasionally observed closer to the Sun. The increasing spiral field angle at larger heliocentric distances causes SIRs to become near tangential structures, almost perpendicular to the Sun-spacecraft line, leading to an increase in shock formation from $26\%$ at $1$~AU to $91\%$ at $5.4$~AU \citep[e.g.,][]{Jian2008_PhDT,Geyer2021}.  Expansion of the SIR with increasing heliocentric distance tends to erode the difference between the slow and fast solar wind speeds, leading to a weakening of the HSSs. In this respect, the wake of a HSS, i.e., where the fast wind merges with the slow wind, might be of interest for future studies. In addition, these streams can interact and merge, leading to a simplification of the stream structure further from the Sun \citep[e.g.,][]{Burlaga1990}. It also has been found that the tilt of a SIR does not necessarily match the tilt of the solar source coronal hole \citep{Broiles2012}. This implies that the shape of the solar source region might not be the dominant factor determining the SIR geometry.  Rather, the IMF configuration plays a role. The different space weather impacts resulting from variations in SIR/CIR geometry including the spiral angle, tilt and possible substructures due to local speed variations, need to be further investigated. This will be especially important for future exploration in interplanetary space requiring more detailed knowledge of space weather hazards at distances beyond Earth \citep[e.g.,][]{Kajdic2021}.     

This evolutionary behavior with heliocentric distance also influences the relations between different solar wind parameters. For example, the relations between solar wind density and velocity or proton temperature and velocity at 1~AU have been studied as far back as the 1970s \citep[][]{Burlaga1973,Eyni1980,Geranios1982}, while \cite{Lopez1986} used Helios data to study the radial dependence of the speed-temperature relation at 0.3 to 1~AU. It is found that the relations change with radial distance, suggesting that it is not possible to interpolate the solar wind properties measured in situ back to their source regions \citep[][]{Perrone2019b}. It has been shown that the relations found for HSSs and SIRs/CIRs may deviate from those found in slow solar wind. \cite{Wang2010} and later \cite{Fujiki2015} showed that the solar wind velocity is inversely proportional to the flux tube expansion factor and that the velocity increases linearly with the strength of the open field footpoints (see also Section~\ref{sec3:fastSW}). However, this relation cannot be used to improve prediction of the solar wind velocity without additional information about the mass flux, which is a fundamentally important physical parameter for solar wind acceleration. It was suggested that whereas the mass flux close to the Sun is proportional to the field strength, near 1~AU the mass flux is  latitudinally and longitudinally constant on average. This may imply  that interaction processes in the solar wind can break or smooth the proposed relations during propagation. PSP data might help to resolve these discrepancies.

The proton temperature of the solar wind might be expected to drop adiabatically with increasing radial distance, but it is found that it drops more slowly, implying that additional heating is required  \citep[e.g.,][]{Hellinger2011,Hellinger2013} while  \cite{Perrone2019b,Perrone2019a} noted that pure fast wind seems to follow an adiabatic cooling law as expected from radial expansion. The density decreases as function of radial distance as expected, but the magnetic field deviates from Parker's model.  The v--T relation for solar wind at 1~AU is usually described by a single linear fit for both slow and fast wind.  However, it has been shown that: (1) different solar wind may exhibit different relations, and (2) the relation evolves with radial distance \citep[][]{Elliott2012}. The behavior of the temperature near and within SIRs as function of the radial distance from the Sun is less well studied.

\subsection{Specific Challenges for Modeling SIRs/CIRs}
As described above, many physical processes related to the slow and fast solar wind acceleration are not fully understood. The successful modeling of the background solar wind is still a huge challenge especially keeping in mind that the observational input for numerical models comes from the photosphere (i.e., magnetograms) and/or EUV observations in case of analytical/empirical models.

The simulation of SIR/CIR formation relies mostly on the ability of the solar wind model used to produce a bimodal distribution to induce interaction between the fast and slow streams. Certain models, such as the Parker analytical model or the basic polytropic heating used in MHD, do not meet this requirement, although some modifications can be made by varying the adiabatic index or introducing Alfv\'{e}n waves or ad-hoc heating terms. Most models that can simulate SIRs/CIRs in 3-D space are heliospheric models driven by empirical coronal models (e.g., WSA--ENLIL and EUHFORIA). While such models produce rapid and robust results, they may not describe the source regions of the fast and slow wind very accurately. Studies discussing such model results and comparing them with observations include: \cite{Owens2008}, using  WSA--ENLIL simulations, \cite{Hinterreiter2019} using EUHFORIA,  \cite{Samara2021}, who compare HSSs modeled by EUHFORIA with observations and results of other models, and \citet{Samara2022b}, again using EUHFORIA. In the future, instead of empirical coronal models, MHD coronal codes optimized for space weather may be used to provide improved 0.1 AU input boundary conditions for heliospheric codes \citep[e.g., the Virtual Space Weather Modelling Centre VSWMC;][]{Poedts2020a}. The time-evolution of coronal codes may also become an important issue, as most of the current models are quasi-static. \textit{How can the time-evolution of solar source regions be incorporated in models to improve the modeling of SIRs/CIRs?} This might require a number of time-dependent extrapolations such as magneto-frictional \citep[MF;][]{Pomoell2019} or non-linear force-free \citep[NLFF;][]{Wiegelmann2012} modeling.  

As already discussed, predicting the solar wind at 1~AU and beyond is generally performed by combining models for different regimes \citep[][]{Macneice2018}, usually in the coronal and heliospheric domains \citep[e.g., solar wind models such as ENLIL or EUHFORIA which combine the coronal WSA model and a heliospheric MHD model;][]{OdstrcilPizzo1999b,Pomoell2018}. Identifying the source of discrepancies when comparing model results to those of other models and/or observations is often a challenge, leading to the question whether the coronal model, the input data or the heliospheric model is the least reliable part. For example, \cite{Linker2021} and \cite{Wangetal2022} showed that there are significant differences in the estimated open flux when different input magnetograms are used. \citep[For more details about open questions related to the global solar magnetic field, see the S2 Cluster TI2 paper by][]{Arge2023}. \cite{Asvestari2019} and \cite{Caplan2021} highlighted model--model and model--observation differences for several coronal models that led to differences in the heliospheric domain predictions. The results are strongly depended on which model combination was used and how the transition between the models was performed \citep[e.g.,][]{Jian2015,Jian2016}.  An objective evaluation of the performance of different models \citep[see e.g.,][]{Wagner2022} and model combinations is necessary to advance space weather modeling, which requires model developers to be transparent about their (often hidden) model parameters and how they are tuned \citep[see more details from the H1-01 team in][]{ReissEtAl2022}. Without constraints on how models are adjusted for various conditions, events and utilization, reliable comparison of models and estimation of uncertainties will continue to be challenging.

With the recent increase in available computational power, computer-based methods, such as DA, ML, and neural networks (NN), have become viable and widely available. Such techniques have been applied to, e.g., solar feature detection \citep[][]{Jarolim2021,Mackovjak2021}, solar wind forecasting \citep[][]{Wang2020,Upendran2020,Raju2021}, and the prediction of recurrent geomagnetic effects \citep[][]{Zhelavskaya2019,Haines2021}. These models are however still in their infancy; \cite{Camporeale2019} describes in detail some of the major challenges these models and methods face. Although there will be greater reliance on such computational methods in the future, it is important not to neglect the underlying physical principles and physics. 

\subsection{Geomagnetic Activity Associated with CIRs/SIRs}

The solar wind during the passage of a CIR is in itself a sufficiently strong driver for a magnetospheric storm \citep[][]{Koskinen2011}, and well-developed SIRs and faster HSSs can impact Earth's magnetosphere sufficiently to induce minor to moderate magnetic storms. During the passage of SIRs/CIRs and HSSs, typically $B_z$ fluctuates, and AE is relatively large for an extended interval, whereas the effect in Dst is relatively small, with the positive phase due to compression of the magnetosphere often larger than the negative phase. The geoeffectiveness of SIRs/CIRs has for a long time been underestimated by the space weather community; a significant impetus was provided by the deep solar minimum at the end of solar cycle 23 which was characterized by a large number of SIR/CIR events, and also by the 2005 Chapman Conference ``Recurrent Magnetic Storms: Co-rotating Solar Wind Streams" \citep[][]{Tsurutani2006}. An investigation carried out by \cite{Zhang2008} showed that about $50\%$ of $157$ “pure” SIRs/CIRs produced interplanetary shocks and $89\%$ of the shocks were followed by magnetic storms. Although the storm recovery phase is characterized by an abatement of perturbations and a gradual return to the ``ground state”, observations of the disturbed ionosphere show significant departures from climatology within this phase of a storm. For SIR/CIR events, the recovery phase is longer than is typical for the recovery of CME-induced storms (including both sheath or magnetic ejecta- driven storms) because of different method of energy input \citep[][]{Buresova2014}.   Statistical analyses of SIR/HSS-related events have revealed that their ionospheric effects may be comparable to the effects of strong CME-induced magnetic storms under higher solar activity conditions but are less dependent on the season \citep[Buresova and Lastovicka, pp.41--48 in][]{Fuller-Rowell2016}. 

\subsection{Summary}
In this section, we have explored several open and debated questions relating to SIRs and their space weather effects, ranging from their solar sources, formation and interaction processes to radial evolution and modeling challenges. The magnetic structure and plasma properties of the solar wind source regions, as well as solar wind acceleration processes close to the Sun, are major concerns. In particular: How can the solar wind parameters be constrained at small radial distances when the majority of the acceleration processes has ceased at 0.1~AU, and how can the constrained parameters be used to improve model input?

\section{CME Propagation Behavior}
\label{sec:section4}



As CMEs have the largest influence on space weather, CME forecasting is an important and wide field of research. The analysis and forecasting of  CME propagation can be divided into ``pre-event" (using model input from signatures/diagnostics occurring before the onset of the CME) and ``post-event" (after the onset of the CME). In this section we discuss the open scientific questions related to post-event forecasting, mainly focusing on CME propagation and interaction in the inner heliosphere starting from 0.1~AU. For the latest developments and future prospects for pre-event forecasting, see the TI2 paper by \cite{Georgoulis2023} from Cluster S. For a review on the relation between CMEs and flares as well as early CME evolution we refer to e.g., \cite{Temmer2021b,MishraTeriaca2023}.

\subsection{CME Propagation Behavior and Uncertainties}
\begin{figure}[ht]
\centering
\includegraphics[width=1.\linewidth]{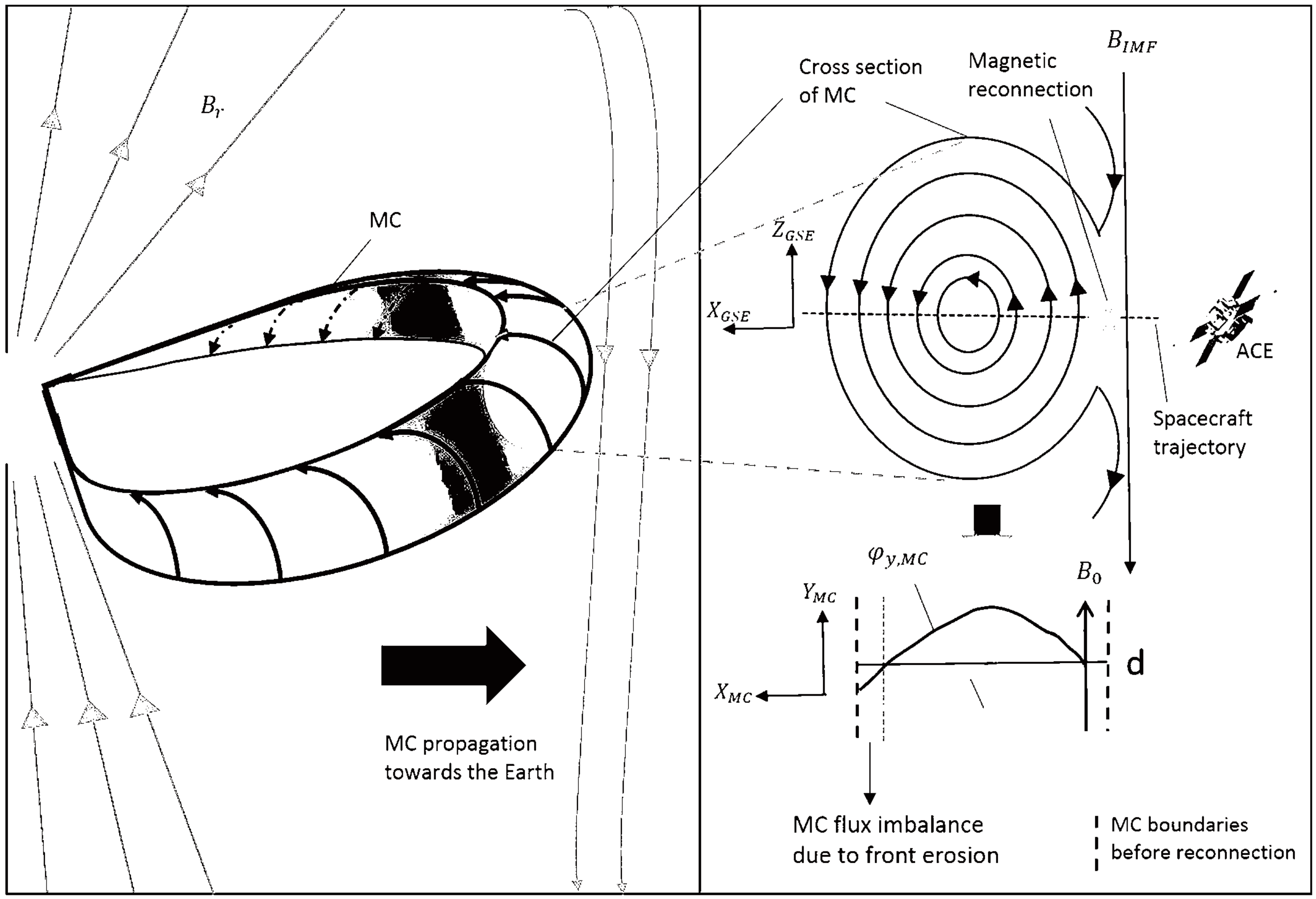}
\caption{Schematics of a magnetic flux rope (here referred to as MC) of a CME interacting with the IMF leading to physical processes affecting its propagation behavior and making it difficult to forecast the CME characteristics, especially its magnetic field component, at a target \citep[taken from][]{PalEtAl22}.}
\label{fig:sec4:erosion}
\end{figure}

Strong geoeffectiveness mainly results from the combination of a dynamic pressure enhancement (primarily associated with the sheath/compression region generated by the CME through compression of the preceding solar wind during propagation), and the local southward interplanetary field ($B_z$) component (primarily within the CME ejecta). Geomagnetic storm forecast modeling is therefore a double challenge, as it requires both --- well constrained  CME initial properties to feed the model together with a reliable ambient solar wind simulation. Current state-of-the-art CME forecasts have significant uncertainties for predicting the CME ToA, SoA, and its magnetic properties. This comes on the one hand from the uncertainties in the initial observational parameters used as model input, and on the other hand from the poorly-understood interaction processes between the different CME structures and the ambient solar wind (Section~\ref{sec:section3}).
The former is associated primarily with projection effects, as the features observed in coronagraphs are 2-D projections on the plane-of-sky (POS) of the actual 3-D structures, leading to an underestimation of the speed and overestimation of the CME angular width \citep[see e.g.][and references therein]{Burkepileetal2004, Vrsnaketal2007_pr, Temmeretal2009, Paourisetal2021_pr}. Many of the model input parameters are the result of modeling and fitting techniques where the observer plays a decisive but not objective role on the final CME parameters \citep[human-in-the-loop effect; see][]{Verbekeetal2022}. Moreover, the observed magnetic structures on the Sun related to the eruption may undergo significant development (cf.\,Figure~\ref{fig:sec4:erosion}), hence, predicting their properties at 1~AU distance is a major challenge \citep[e.g.,][]{PalEtAl22}. 

\subsubsection{CME Arrival Properties}\label{sec:section4:ToA}

A range of sources contribute to the total uncertainty, including the specification of accurate boundary conditions, the physical approximations used to model CME propagation (e.g., in drag-based or MHD models), the geometric representation used to parameterize the CME  (e.g.\ a cone or a FR model) and the interaction of the CME with the ambient solar wind. \cite{Lee_Arge_etal_2013} using ensemble modeling found that the accuracy of the modeled ToA not only depends on the initial input CME geometry, but also on the accuracy of the modeled solar wind background, which is driven by the input maps of the photospheric field.
\cite{Maysetal2015a} suggested that an ensemble of ambient solar wind WSA–ENLIL model outputs (an improved ensemble forecast of the maps of the photospheric field) would produce predictions that also reflect the uncertainties in the WSA–ENLIL modeled background solar wind in addition to the uncertainties in CME input parameters.
\cite{Pizzo2015} investigated CME ToA uncertainty in the WSA--ENLIL+Cone model and demonstrated that, for this model, the most important source of uncertainty was the correct specification of the CME initial conditions at the typical inner boundary distance for the heliospheric model of 0.1~AU, as well as the ambient solar wind structure. Accurate estimation of the ambient solar wind structure is the most challenging problem (see Section~\ref{sec:section2} and Section~\ref{sec:section3}) and depends sensitively on the nature of the coronal model and the observations used to drive this model \citep{Rileyetal2015, Gonzi2021}.

\cite{Riley2018} reviewed the performance of CME ToA forecasts for a range of models, within the Community Coordinated Modeling Center (CCMC) CME Scoreboard. They concluded that, on average, CME ToA forecasts were accurate to within about $\pm$10 hours, whilst the best performing models had a mean absolute error of 13~hours and a standard deviation of 15~hours. \cite{Vourlidasetal2019} presented a comprehensive analysis of the current status, open issues and path forward for the prediction of the geoeffective properties of CMEs. Taking into account many published works using different CME propagation models, they concluded that the current state of forecasting the ToA has an error of $9.8 \pm2$~hours. In addition, the authors stress that currently, it is not possible to predict $B_z$ reliably beyond a 40--60 minutes time window determined by the upstream solar wind in situ measurements from L1. What role additional magnetograms from a different viewpoint might play, will be answered with the upcoming ESA/\textit{Vigil} mission (to be launched 2029; see also Section~\ref{Data Availability}). The L5 view covers a larger portion of the solar surface, hence, gives more up-to-date magnetic field information to global numerical models. Advanced IPS techniques may give further insights into the CME magnetic-field rotation as it passes through interplanetary space \citep[see TI1 paper by][]{FALLOWS2022}. Additional views from the solar surface perspectives on the $B_z$ issue is given in the S Cluster TI2 paper by \cite{Arge2023}.

\cite{Riley2021b} explored the sources of uncertainty in CME ToA forecasts using a set of numerical MHD simulations of cone CMEs in ambient solar wind backgrounds. They concluded that uncertainty in each component of the CME initial parameters, such as longitude, latitude, width, and speed, introduces between 2.5 and 7.5~hours of uncertainty into the total ToA uncertainty. Furthermore, they concluded that the ambient solar wind structure was the largest source of uncertainty, and that without better constraints on the initial conditions of the heliospheric simulations, it is likely that the CME ToA error will remain close to $\pm$10~hours. For benchmarking and objective tracking of development improvements of background solar wind models, the H1-01 team has created a validation scheme \citep[see TI1 paper by][]{ReissEtAl2022}. This scheme and platform will be also used to test new models and to derive uncertainty estimates combining different model results in order to more accurately assess the magnitude and source of errors in the ToA.


During their evolution, CMEs are influenced and dominated by different forces such as the Lorentz-force close to the Sun and the drag force when propagating within the ambient solar wind. The latter force leads to the deceleration of fast CMEs, i.e., faster than the ambient wind, and to the acceleration of CMEs slower than the solar wind \citep[e.g.,][]{VrsnakGopalswamy2002}. In recent years, drag-based CME propagation models have attracted increased attention from the community. Despite their simple assumptions, including neglecting any other physical parameters besides the drag force, their ability to predict the ToA and SoA of CMEs is not necessarily worse than those of more sophisticated approaches \citep[e.g.,][]{Vrsnaketal2014}. The basic requirement is to feed the model with CME parameters derived from distances further out from the Sun (on average beyond 20 solar radii), i.e., where the driving Lorentz-force due to magnetic reconnection has ceased. Furthermore, these models are computationally inexpensive and can handle ensemble approaches faster than some other models can manage single runs.

While the speed of the CME relative to that of the ambient solar wind, is the most important factor when describing a drag-based motion, the drag parameter $\gamma$ includes information on other important parameters and is given by 
$$\gamma = C_{\rm{D}} \frac{A_{\rm{CME}} \rho_{\rm{sw}}}{m_{\rm{CME}}},$$ 
where $C_{\rm{D}}$ is the dimensionless drag coefficient (set to unity and therefore assuming an aerodynamic behavior), $A_{\rm{CME}}$ is the CME cross section the drag is acting on, $\rho_{\rm{sw}}$ is the solar wind density, and $m_{\rm{CME}}$ is the CME mass. Besides the effect on their overall behavior, deformations of CMEs can also occur locally on small scales due to the presence of preceding high-speed solar wind streams or other CMEs, which can lead to a change of the conditions in the preceding medium and influence the drag force on the CME.
Also the reconnection and/or the magnetic field erosion of the ejecta in interplanetary space or the driver may be part of the physics covered by the ``drag" phenomenon. There is a poor understanding of what constitutes that drag from a physical perspective but it can be interpreted with MHD waves \citep[][]{CargillEtAl96}. We can rely on such empirical treatments but it would be beneficial to understand the nature of this phenomenon better \citep[see e.g.,][]{ruffenach2012,ruffenach2015,PalEtAl22}.

\subsection{CME Propagation Model Input Parameters}

CME forecasting using MHD simulations usually introduces the CME at heights above 0.1~AU. However, using CME parameters derived at lower heights to introduce a CME into a model at a larger height is a potential source of uncertainty given that CMEs can undergo deflection and rotation while traveling through the corona \citep[e.g.,][]{Yurchyshynetal2009,Isavnin2014,Kay2015b}. Coupled solar-heliospheric models would be able to simulate the CME from the eruption site up to the arrival at a target in interplanetary space \citep[e.g.,][]{Torok2018}, but usually this is related to very high computational costs which is not practical for real-time forecasting.


CME forecast models require several data-driven or assumed initial parameters. Table~\ref{tab:section4_cme_input_parameters} lists these parameters along with the data sources and techniques used to estimate them. The parameters that are most often required by the models are the CME initiation time, initial height, latitude, longitude, and speed. If the CME model has a geometry that allows for a standard CME shape consisting of two legs and a curved front, or even one without legs (e.g., a spheromak), the model requires the CME tilt and an angular width. These parameters are usually derived via multi-viewpoint coronagraph observations and forward modeling techniques such as the Graduated Cylindrical Shell \citep[GCS;][]{Thernisien_etal_2006, Thernisien_etal_2009}, the FR in 3-D \citep[Fri3D;][]{Isavnin2016}, and the Stereoscopic CME Analysis Tool \citep[StereoCAT;][]{Maysetal2015a}. The CME parameters can typically be derived until the CME reaches the edge of the field-of-view of the observing coronagraphs. Obviously, these modeling techniques, especially for events which appeared very complex in white light data, might be very demanding for the observer. At such cases the human-in-the-loop plays a decisive role on the final CME parameters obtained by fitting process. The H2-01 team has made a thorough comparison of the skill of different GCS reconstructions to assess the bias and uncertainty in the derived parameters \citep[see TI1 paper by][]{Verbekeetal2022b}. To quantify the uncertainties of the CME parameters the team designed two different synthetic scenarios (ray-tracing from GCS and MHD simulations) where the “true” geometric parameters are known in order to quantify such uncertainties for the first time. From this effort interesting results occurred. CME reconstructions using a single viewpoint had the largest errors and error ranges overall for both synthetic GCS and simulated MHD white-light data. As the number of viewpoints increased from one to two, the errors decreased by approximately $4^\circ$ in latitude, $22^\circ$ in longitude, $14^\circ$ in tilt, and $10^\circ$ in half-angle. These results quantitatively show the critical need for at least two viewpoints to be able to reduce the uncertainty in deriving CME parameters.
\citet{Singh2022} performed a similar quantification of the uncertainty in GCS fits by comparing GCS parameters reported in multiple studies and catalogs. They determined that GCS estimates of the CME latitude, longitude, tilt, and speed have average uncertainties of about $6^\circ$, $11^\circ$, $25^\circ$, and 11.4\%, respectively.

Magnetized CME models are being developed to improve $B_z$ forecasting at Earth. These models have to be initialized with the correct magnetic field poloidal and toroidal fluxes and with the correct handedness (helicity sign). This requires expertise from Cluster S and knowledge about the solar surface structures related to the eruption. The poloidal flux is usually estimated via the reconnected flux in the post-eruption arcade (PEA) of the CME source region \citep{Gopalswamyetal2017} or the flare ribbons \citep{Kazachenkoetal2017}. The toroidal flux can be estimated from the flux in the coronal dimming regions near the CME source \citep[e.g.,][]{Dissaueretal2018}. The helicity sign can be estimated from EUV and magnetogram observations of the active regions \citep[][and references therein]{Palmerio2017, Palmerio2018}, or more simply via the hemispheric helicity rule \citep{Pevtsovetal2014,SavaniEtAl2015}.
Not all magnetized CME models have the capability to be initialized with the desired poloidal and toroidal fluxes, i.e.,\ the twist of the magnetic field lines may not be a free parameter in all models. For example, the spheromak model \citep{Shiota2016, Verbekeetal2019b} and the Gibson--Low model \citep{GibsonLow1998, Singhetal2019} use only one parameter to control the CME magnetic flux, making the poloidal and toroidal fluxes proportional to each other and the twist of the magnetic field lines a non-constant but fixed value. However, the removal of force-free assumptions in models such as the modified spheromak model \citep{Singhetal2020a, Singhetal2020b} and the constant turn FR model \citep{Singh2022} allows for the separate input of poloidal and toroidal fluxes, making the twist a free parameter that can be controlled by the model user. See also investigations about the magnetic morphology of CMEs from multi-spacecraft data \citep[e.g.,][]{Mstl2009,AlHaddadEtAl2013}.

MHD models also require the CME density or the CME total mass as inputs when introducing CMEs into the simulation domain. The total mass of the CME can be calculated from the total brightness of coronagraph images \citep{ColaninnoVourlidas2009, Beinetal2013} on an event-by-event basis. When such an analysis is not feasible (e.g.\ due to time constraints in forecasting/nowcasting conditions, or observational limitations), default values for the initial CME density may be used as input for propagation models \citep[e.g.][]{Odstrcil2003, Maysetal2015a, Pomoell2018}. Additionally, efforts towards defining a range of realistic CME densities to be used routinely as inputs into CME propagation models and ensemble realizations have been undertaken in recent years \citep{Temmer2021a}. Especially when including heliospheric imager (HI) observations, recent studies have shown that the CME kinematics beyond the coronagraphic field-of-view can be used to estimate the CME mass \citep{Amerstorferetal2018,Hinterreiteretal2021b}.
Cone and spheromak CME models may consider the density distribution inside the structure by considering pressure gradients.  However, since these models do not have solutions for more complex density distributions, usually, the mass is distributed uniformly throughout the CME volume. The validity of this assumption is supported by the recent study of \cite{Temmer2021a}, but needs to be further tested. Models such as the Gibson--Low model have an analytic solution for the mass density resembling the three-part structure of CMEs.

The thermodynamic evolution (e.g., pressure, temperature, heat, entropy) of CMEs is not well understood and it is one of the most challenging problems of space plasma physics \citep[e.g.,][]{Liuetal2005}. The combination of the density, temperature, and ionization states of CMEs constrains their thermal history and can be used to understand the physical processes within the CME plasma. The thermodynamics of the solar wind has been studied extensively since the seminal work of \cite{Parker1960}; however, such efforts are limited for the case of CMEs. The heating of plasma in the closed magnetic field configuration of a CME is expected to be different from that in the open magnetic field configuration of the background solar wind and needs to be examined over the different phases of heliospheric propagation. The CME is also an inhomogeneous structure, including substructures with different plasma characteristics. The thermodynamic evolution of a CME is often modeled using a polytropic approximation \citep[e.g.,][]{ChenGarren1993}. Although different values of the polytropic index might be used to imply different rates of heating, the ideal MHD models used for CME evolution often assume a fixed value of the polytropic index without any justification \citep{Pomoell2018}. Therefore, developing methods and models for estimating the radial gradient in kinematic, thermodynamic, plasma, and magnetic properties inside and outside CMEs is a major requirement for improving understanding of space weather. 

Earlier studies addressing the thermodynamic state of CMEs often estimated the thermodynamic properties of an expanding CME at a certain position or time \citep{Raymond2002, Ciaravellaetal2003}. The temperature of the plasma in the pre-and post-shock regions has been estimated using white-light, EUV, and radio observations of a fast CME \citep{BemporadMancuso2010}. The polytropic index of CMEs can be estimated by comparing in situ observations of the same CME observed by multiple radially-aligned spacecraft.  However, this situation is extremely rare due to the sparse distribution of spacecraft and the difficulty in identifying CMEs in the solar wind. Using observations of several CMEs made by spacecraft located over a range of radial distances, the polytropic index for CME plasma is inferred to be around 1.1 to 1.3 from 0.3 to 20 AU, and nearly constant over the solar cycle \citep{WangRichardson2004, Liuetal2005, Liuetal2006}. Thus, the expansion of an CME behaves more like an isothermal, rather than an adiabatic, process. It has also been shown that the magnetic field and density decrease faster in CMEs than in the solar wind, but the temperature decreases more slowly in CMEs than in the solar wind \citep{Tottenetal1995, Liuetal2006}. This implies that either the plasma in CMEs has to be  heated or that these analyses used oversimplified assumptions. 

A pioneering attempt to understand the thermodynamic evolution of an individual CME during its propagation from the inner to outer corona was made by \citet{Wangetal2009} by developing the Flux Rope Internal State Model (FRIS). This model was recently modified by \citet{MishraWang2018} so that the evolution of the CME's thermodynamic state is expressed in terms of its kinematics, which are governed by the Lorentz and thermal pressure forces. Although this simplified MHD model has not been used statistically for understanding the general thermodynamic behavior of CMEs, it has been applied to a few case studies giving different results \citep{MishraWang2018, Mishraetal2020, Nieves-Chinchillaetal2020}. Future studies should focus on investigating whether there is a critical height where CMEs turn from a heat releasing to a heat absorbing state and whether this has any dependence on CME characteristics. Such study would be feasible using CME kinematics derived from the Metis coronagraph and SoloHI on SolO, as well as from WISPR onboard PSP, in the FRIS model. The performance and reliability of such models need to be examined by comparing in situ observations by SolO and PSP with the results of fully three-dimensional numerical MHD modeling. 

The solar wind ion charge states in a CME are considered to be frozen-in in the lower corona, and the in situ charge state abundances  can provide information on the thermodynamic state of the CME \citep{Leprietal2001,Gruesbeck2011}. Future studies using such charge state compositions measured by spacecraft traveling to previously unexplored regions of the heliosphere are imperative for a better understanding of the heating and acceleration of CMEs as well as the solar wind in general. Observations in the Lyman-alpha line of hydrogen from the Metis coronagraph onboard SolO would help to link the solar atmosphere and inner heliosphere. There may also be variations in the thermodynamics of CMEs in different solar cycles, that future studies should explore (see also Section~\ref{sec2:solarCycle}). 
Better understanding of the thermodynamics of CMEs would improve modeling of CME expansion speeds which is crucial for improving CME ToA estimates.

\subsection{CME Propagation Models}\label{CMEpropModel}

The routine availability of data from spacecraft coronagraphs (e.g., LASCO, COR1, and COR2) and HI during the last two decades has triggered the development of new CME propagation models. These models utilize various CME characteristics to forecast CME kinematics and properties and address fundamental questions about CME propagation, such as CME ToA and impacts. They use a number of different approaches and may be  categorized as: empirical models \citep{Gopalswamyetal2001, Gopalswamyetal2005, Schwennetal2005, Nunezetal2016, PaourisMavromi2017_EAM, Paourisetal2021}, analytical and drag-based models \citep{Cargill2004, Vrsnaketal2013, Shietal2015, Amerstorferetal2018, MoestlEtAl2018,Dumbovicetal2018, Kayetal2022,Napoletano2022}, MHD models \citep{Odstrcil2003, Shiota2016, Jin2017, Pomoell2018, Torok2018}, heliospheric reconstruction approaches \citep{Sheeley1999,Kahler2007,Howard2006,Lugaz2009b,Davies2012,Davies2013,Rollettetal2016, Amerstorferetal2018,PaourisVourlidas2022}, and ML models \citep{Sudaretal2016, Liuetal2018}. These models and other related references are presented in Table~\ref{tab:section4_cme_models}. 

With so many models available, the comparison of the performance of these models is necessary. However, because of the different principles on which the models are based, model to model comparison is not straightforward. So far, most researchers have performed their own verification and validation studies \citep[see, e.g.,][]{Vrsnaketal2014, Maysetal2015a, PaourisMavromi2017_EAM, Dumbovicetal2018, Riley2018, Woldetal2018, Amerstorferetal2021, Paourisetal2021}. Typically, these validation studies each use different sets of CME events, CME parameters and metrics. However, some efforts have been made to compare models such as in \cite{Dumbovicetal2018} and \cite{Paourisetal2021}, where the performance of the Drag-Based Ensemble Model (DBEM) and Effective Acceleration Model (EAMv3) is compared with WSA--ENLIL by using the same set of events. The necessity of establishing a benchmark dataset that may be used for all validation analyses is clearly apparent. This benchmarking dataset will serve as a validation tool both for new models and for updated versions of already-existing models, where it will be possible to determine the difference in performance between the two versions of the model. 
With this in mind, the CME Arrival Time and Impact Working Team (H2-01) has been formed within the ISWAT H Cluster.  This team was originally founded in 2017 as part of the ``International Forum for Space Weather Capabilities Assessment". It aims to develop a data set with a statistical significant sample of 100 or more associated CME and CME arrivals covering different periods within the solar cycle \citep{Verbekeetal2019a}. However, this task requires considerable preparation and community coordination.  The first steps towards this goal were taken by the International Space Science Institute (ISSI) Team on \href{http://www.issibern.ch/teams/understandcormasseject/}{``Understanding Our Capabilities In Observing and Modeling Coronal Mass Ejections"} (formed of a subset of the H2-01 team). 

As can be seen from the \href{https://kauai.ccmc.gsfc.nasa.gov/CMEscoreboard/}{CME scoreboard}, even the same model may produce different outputs if different CME input parameters and model settings are chosen. Human bias also plays a role as different forecasters may generate different forecasts owing to their level of experience or skill (human-in-the-loop effect). As such, when benchmarking CME arrival time models, it is important to collect accurate information about the CME and solar wind inputs selected for each model. The CCMC scoreboard acts in a very similar way as the solar wind benchmarking scheme given in \cite{ReissEtAl2022}.  With this approach, it may be possible to at least reduce the ambiguity coming from observational data.  

In addition to CME dataset required for benchmarking, developing a community-agreed, unified set of metrics is of high importance. \citet{Verbekeetal2019a} made a first effort towards this goal. To assess CME arrival predictions, they used two categories of metrics: a) event detection performance metrics (from contingency tables) that aim to determine whether an event was correctly predicted, and b) ToA and SoA metrics (i.e., hit performance metrics) that assess the performance of the model's predicted events. As part of the event detection performance metrics, the observed arrival and/or non-arrival of a CME and the corresponding CME forecast were used to create a contingency table containing information about `hits' (observed and predicted arrival), `false alarms' (predicted arrival but not observed), `misses' (observed arrival but not predicted) and finally, `correct rejections' (arrival not observed nor predicted). Note though that the definition of a hit is dependent on the chosen time interval within which a forecast arrival is assumed to be correctly predicted. See  \cite{Verbekeetal2019a} for more details about the skill scores that can be derived from the contingency table. 

Hit performance metrics focus on the predicted hit arrivals and assess how well the model predicts CME ToA, SoA, or DoA as well as other arrival parameters such as, magnetic field, and temperature (see also Section~\ref{method}). Different metrics can be used for the ToA error, such as the mean error, mean absolute error, the root mean squared error, and the standard deviation. Each of these metrics provides different information on the accuracy of CME propagation model.  For example,  the mean error quantifies the bias of the model in terms of whether the predictions are early or late on average, while the mean absolute error quantifies the absolute time difference irrespective of whether it is early or late. It remains a difficult and ongoing task to determine how prediction errors originating from the ambient solar wind modeling (see Section~\ref{sec:section3}) and from the chosen CME model can be separated and determined. See more details of the H2-01 and H2-03 team efforts at \url{https://www.iswat-cospar.org/h2}. 

Forecasting the $B_z$ component is one of the key challenges in space weather forecasting. Currently, the most reliable estimates use measurements of the magnetic structures at L1 propagated to Earth, giving a lead time of 40--60 minutes \citep[][]{Vourlidasetal2019}. Recent attempts have applied new methodologies, such as deep learning (DL) and ML, to remote sensing image data and in situ measurements with the aim of increasing this lead time \citep[e.g.,][]{dosSantosEtAl2020,ReissEtAl2021}. Statistical and analytical methods using information from the solar surface, e.g., the helicity rule \citep[][]{BothmerSchwenn1998}, as described in \cite{SavaniEtAl2015}, or using a combination of several forecasting tools, such as the Open Solar Physics Rapid Ensemble Information \citep[OSPREI; see][]{Kayetal2022}, also show promise.

Although models and methods utilizing heliospheric imaging data have successfully tracked CMEs and estimated their kinematics away from the Sun, particularly within the large space between the Sun and Earth, they have their limitations.  These are mainly due to the line-of-sight integration of the visible-light signal, and the interaction of CMEs with the background solar wind, CIRs/SIRs, and, most importantly, other CMEs, which complicates the use of such observations in space weather forecasting (see Section~\ref{sec:section5} for more details on interaction processes). Attempts to address limitations in the localization of large-scale solar wind features have led to the development of a plethora of techniques to aid in the interpretation of HI observations.  Several approaches to derive the kinematic properties of CMEs from HI observations are based on the analysis of their time--elongation profiles combined with assumptions about the CME cross-section \citep{Liuetal2010, Lugazetal2010, Davies2012, Davies2013, Rollettetal2016, Amerstorferetal2018, Bauer2021, Hinterreiteretal2021b, PaourisVourlidas2022}. Such approaches have often used just the manually-extracted time--elongation profile along a single position angle corresponding to the CME leading edge in the ecliptic plane. Future reconstruction methods need to be developed that track CME features along different heliolatitudes \citep[see e.g.,][SATPLOT tool]{MostlEtAl14} while considering the time-varying geometry of CMEs.

Some recently developed techniques are ``scientifically rich" \citep{Rollettetal2012}, but show limitations in terms of their operational potential. Some HI-based methods incorporate inputs from other methods, such as the 3-D CME propagation direction from the analysis of coronagraphic observations, to reduce the number of free parameters in the HI-based analysis \citep{Mishraetal2014,Mishraetal2015a}. Furthermore, some well-established HI-based techniques include aerodynamic drag to extrapolate the CME speed profile beyond the field-of-view of the HI observations \citep{MishraSrivastava2013, Rollettetal2016}. Recently, such a method was enhanced by including the deformation of the CME during propagation arising from interaction with the ambient solar wind \citep{Hinterreiteretal2021b}. This approach of using external information in HI-based analysis including drag forces may surpass the performance of other methods based only on single and multiple viewpoint observations. 

\cite{PaourisVourlidas2022} adopted a slightly more realistic approach for CMEs propagation using HI data. They replaced the common assumption of constant speed in the inner heliosphere with a two-phase behavior consisting of a decelerating (or accelerating) phase from 20 Rs to some distance, followed by a coasting phase to Earth. This new approach improved the ToA of CMEs in some cases. For example, the difference between predicted and observed ToA was below  52 minutes for 21 of the cases considered. The analysis indicates that reasonable forecasts may be attainable with CME HI measurements up to 0.5 AU and with a (mean) lead time of 31 hours \citep[see also][]{Colaninnoetal2013}.

Furthermore, because interactions with other large-scale structures can lead to a significant change in CME speed and direction, the accuracy of HI-based techniques used for CME ToA prediction will severely be reduced if post-interaction kinematics are not taken into account \citep{Shen2012, MishraSrivastava2014, Rollettetal2014, Temmeretal2014, Mishraetal2016}. Several studies have demonstrated that CME--CME interactions are poorly understood \citep{Temmeretal2014,Shenetal2016,Mishraetal2017}. However, since interacting CMEs may give rise to enhanced space weather effects \citep{Farrugiaetal2006, Mishraetal2015b}, future research to understand the nature of such interactions at different heliocentric distances is imperative. Even if two CMEs do not physically interact, the preceding CME can ``pre-condition" the background solar wind \citep{TemmerNitta2015}. More details about interacting CMEs and pre-conditioning effects can be found in Section~\ref{sec:section5}.  Predicting CME arrival at Earth when CME--CME interactions occur remains challenging using observations as well as MHD models. A time-dependent modeling of interplanetary space is needed.  

The limited cadence and resolution of the HI onboard STEREO has prevented their full use for monitoring solar wind structures. The next generation of HI making observations from a vantage point off the Sun--Earth line, onboard NASA/PUNCH, to be launched in 2025, and ESA/\textit{Vigil}, to be launched 2029, have carefully tailored instrument specifications (field-of-view, cadence, exposure time, and resolution) and may be expected to track CMEs in the heliosphere more accurately. This will help to make further refinements in HI-based reconstruction techniques \citep{Davies2012, Davies2013, Rollettetal2014, PaourisVourlidas2022}, and in the models that combine these techniques with drag-based motion \citep{Zicetal2015, Rollettetal2016, Amerstorferetal2018, Hinterreiteretal2021b}.

Above we mention the ESA \textit{Vigil} mission, planned to be launched in 2029 to a location at L5 where it will view the solar surface and active regions 4--5 days before they rotate to central meridian with respect to Earth. With that, \textit{Vigil} will give us advance warning of how the solar surface behaves, hence, we gain more time to protect vulnerable space equipment and exploration as well as vital infrastructure on the ground. {\it Vigil} observations will be used as valuable input to improve heliospheric models and will help to estimate the probability of solar eruptions \citep[see also S Cluster TI2 paper by][]{Georgoulis2023}. Adding a complementary L4 mission \citep[][]{Posner2021} will provide information about those active regions on the western hemisphere and beyond the west limb that are the sources of the most intense SEP events observed at Earth \citep[see Cluster H3][]{Guo2023}. Combining observations from the vantage points of Earth, L4 and L5 will cover more than 80\% of the solar surface, significantly improving modeling inputs and both short- and long-term forecasting abilities.

\begin{table*}
\caption{CME parameters commonly used to initialize CME propagation models}
\label{tab:section4_cme_input_parameters}

\centering
\begin{tabular}{|p{4.5cm}|p{5cm}|p{4cm}|}
\hline
Parameter & Source & Useful references \\
\hline\hline

CME start time, start height  & Model-dependent; stereoscopic coronal observations + forward modeling techniques 
& \cite{Thernisien_etal_2006, Thernisien_etal_2009}; \cite{Isavnin2016}; \cite{Maysetal2015b}; \cite{WoodHoward2009} \\ 
CME longitude, latitude & Stereoscopic coronal observations + forward modeling techniques 
& \cite{Thernisien_etal_2006, Thernisien_etal_2009}; \cite{Isavnin2016}; \cite{Maysetal2015b}; \cite{WoodHoward2009} \\ 
CME volume, geometry (e.g. angular width, aspect ratio)  & Model-dependent; stereoscopic coronal observations + forward modeling techniques  
& \cite{Thernisien_etal_2006, Thernisien_etal_2009}; \cite{Isavnin2016}; \cite{Maysetal2015b}; \cite{WoodHoward2009} \\ 
CME total, translational speeds              & Model-dependent; stereoscopic coronal observations + forward modeling techniques 
& \cite{Thernisien_etal_2006, Thernisien_etal_2009}; \cite{Isavnin2016}; \cite{Maysetal2015b}; \cite{WoodHoward2009} \\ 
CME-driven shock speed  &  Model-dependent; stereoscopic coronal observations + forward modeling techniques; associated-flare location + SXR peak & \cite{Thernisien_etal_2006, Thernisien_etal_2009}; \cite{Isavnin2016}; \cite{Maysetal2015b}; \cite{WoodHoward2009}; \cite{Nunezetal2016} \\ 
CME HI time-elongation profile & Heliospheric images
& \cite{Zicetal2015, Rollettetal2016} \\ 
CME total mass          & Geometry-dependent, linked to CME volume and mass density; stereoscopic coronal observations 
& \citet{ColaninnoVourlidas2009, Beinetal2013, Temmer2021a} \\
CME mass density        & Single view-point coronal observations; stereoscopic coronal observations 
& \cite{Falkenbergetal2010, Maysetal2015b, Werneretal2019, Temmer2021a} \\
CME drag parameter      & Model-dependent, linked to CME speed and solar wind pre-conditioning & \citet{Vrsnaketal2014, Calogovic2021} \\

CME temperature         & Model-dependent &  ad-hoc parameter\\

FR handedness           & EUV and/or X-ray estimates; hemispheric helicity rule 
& \cite{BothmerSchwenn1998, Palmerio2017, Palmerio2018, Pevtsovetal2014} \\
FR axial orientation    & Stereoscopic coronal observations + forward modeling techniques; EUV and photospheric magnetic field estimates
& \cite{Palmerio2018}; \cite{Yurchyshynetal2001, Marubashietal2015}; \cite{Yurchyshyn2008} \\
FR axial magnetic field strength; FR total, toroidal, poloidal magnetic fluxes; FR magnetic field twist 
& EUV and photospheric magnetic field estimates of reconnected flux based on different eruptive signatures
& \cite{Gopalswamyetal2017, Dissaueretal2018, Kazachenkoetal2017} \\

%
\hline\hline

\end{tabular}
\end{table*}

\begin{table*}
\caption{Most known and widely used CME propagation models}
\label{tab:section4_cme_models}

\centering
\begin{tabular}{|p{6cm}|p{5.5cm}|p{5.5cm}|}
\hline
Model Category/Model name & Input data & Useful references \\
\hline\hline

Empirical Models & & \\
\hline
Effective Acceleration Model (EAM) & Coronagraph data & \cite{PaourisMavromi2017_EAM, Paourisetal2021} \\ 
Empirical Shock Arrival model (ESA) & Coronagraph data & \cite{Gopalswamyetal2001, Gopalswamyetal2005, Manoharanetal2004} \\ 
Shock ARrival Model (SARM) & Coronagraph and soft X-Rays data  & \cite{Nunezetal2016} \\
\hline\hline

Drag-based Models & & \\
\hline
Drag Based Model (DBM) & Coronagraph data & \cite{Vrsnaketal2013, Cargill2004} \\ 
Drag Based Ensemble Model (DBEM) & Coronagraph data & \cite{Dumbovicetal2018, Calogovic2021} \\
Drag-based Model Fitting (DBMF) & Coronagraph data & \cite{Zicetal2015} \\
ELlipse Evolution model based on Heliospheric Imaging (ELEvoHI) & HI data & \cite{Rollettetal2016, Amerstorferetal2018} \\
\hline\hline
Reduced-physics Models & & \\
\hline
Heliospheric Upwind eXtrapolation with time dependence (HUXt) & Magnetograms and coronagraph data
& \cite{owensetal2020} \\ 
Open Solar Physics Rapid Ensemble Information (OSPREI) & Magnetograms and coronagraph data & \citet{Kayetal2022} \\
\hline\hline

MHD Models & & \\
\hline
ENLIL + Cone & Magnetograms and coronagraph data & \cite{OdstrcilPizzo1999b, Odstrcil2003, Odstrciletal2005} \\ 
CORona-HELiosphere (CORHEL)/Magnetohydrodynamic Algorithm outside a Sphere (MAS) + modified Titov-Demoulin (TDm) & Magnetograms and coronagraph data & \cite{Riley2012b, Lionello2013, Torok2018} \\ 
Alfv\'en Wave Solar Model (AWSoM) & Magnetograms and coronagraph data & \cite{vanderHolstetal2014, Jin2017} \\ 
MSFLUKSS + Gibson-Low   & Magnetograms and coronagraph data & \cite{Singhetal2019} \\
MSFLUKSS + modified spheromak   & Magnetograms and coronagraph data & \cite{Singhetal2020a} \\
EUropean Heliospheric FORcasting Information Asset (EUHFORIA) + Cone & Magnetograms and coronagraph data & \cite{Pomoell2018} \\
EUHFORIA + Linear Force-Free Spheromak (LFFS) & Magnetograms and coronagraph data & \cite{Verbekeetal2019b} \\
ICARUS + Cone & Magnetograms and coronagraph data & \cite{Verbekeetal2022}\\
Space-weather-forecast-Usable System Anchored by Numerical Operations and Observations (SUSANOO)-CME & Magnetograms and coronagraph data & \cite{Shiota2014, Shiota2016} \\

\hline\hline

Heliospheric Reconstruction Approach & &  \\
\hline
Fixed-Phi Fitting (FPF) & HI data & \cite{Rouillard2008} \\ 
Harmonic Mean Fitting (HMF) & HI data & \cite{Lugaz2009a} \\
Self-Similar Expansion Fitting (SSEF) & HI data & \cite{MoestlDavies2013} \\
ELlipse Evolution model based on Heliospheric Imaging (ELEvoHI) & HI data & \cite{Rollettetal2016,Amerstorferetal2018} \\
Drag-based Fitting (DBMF) & HI data & \cite{Zicetal2015} \\
Heliospheric Reconstruction and Propagation Algorithm (HeRPA) & HI data & \cite{PaourisVourlidas2022} \\
\hline\hline

ML Models & & \\
\hline
CME Arrival Time Prediction Using ML Algorithms (CAT-PUMA) & Coronagraph and solar wind data & \cite{Liuetal2018} \\ 
\hline\hline

\end{tabular}
\end{table*}

\subsubsection{The Heliosphere Observed in Radio}

Since the 1950’s, there have been attempts to relate solar observations to heliospheric structures. 
Early analyses used metric \citep{WildMcCready1950} and later kilometric \citep{Bougeretetal1998} radio observations to track shocks moving outward from the Sun and predict their arrival at Earth \citep{Fryetal2001}. IPS and Thomson-scattering observations have been utilized to provide the near-Earth morphology of outward-flowing heliospheric structures. Some of the best early studies of this type used IPS data from the Cambridge IPS array \citep{Hewishetal1964, Houminer1971} to fit both remotely-sensed and in situ observations with modeled co-rotating and transient structures \citep[][]{BehannonEtAl1991}.  These “by eye” model fits to data were followed by more sophisticated analyses of the IPS observations \citep[made at the UCSD, USA, and Nagoya University, Japan; see][]{Jacksonetal1998, Kojimaetal1998} employing iterative 3-D tomographic reconstruction techniques that used no preconceived notion of heliospheric structures present other than assuming outward radial expansion of the solar wind. Since the IPS observations were available with delays of only $\sim$12 hours, these analyses were also developed to forecast the arrival of heliospheric structures at Earth. To improve the 3-D reconstruction of CMEs, an even more sophisticated, time-dependent model was developed that conserved mass and mass flux, and could also incorporate Thomson-scattering observations \citep{Jackson2001, Jacksonetal2008, JacksonHick2002}. Results from this model have been fit to in situ data at Earth, usually through least squares Pearson’s “R” correlation procedures, in a way that helps refine the remote-sensing analyses and certify forecast performance. With the more abundant Thomson-scattering brightness data available over most of the sky from the Solar Mass Ejection Imager \citep[SMEI;][]{Jacksonetal2004} launched in early 2003, far higher time-dependent 3-D reconstruction resolutions of heliospheric density became possible \citep{Jacksonetal2006, Jacksonetal2008}. These have led to a capability to use Thomson-scattering analyses to iteratively reconstruct 3-D densities that match in situ measurements near the observer with cadences of about one-hour \citep{Jacksonetal2020}. This technique has also recently been used with well-calibrated STEREO HI data \citep{Harrisonetal2008, Eyles2009} to provide high-resolution 3-D reconstructions \citep[][]{Jacksonetal2020} throughout the region of the heliosphere viewed by these instruments.

Other variations of these iterative data-fitting techniques have been developed using IPS and Thomson-scattering observations. The Japanese IPS iterative technique \citep[e.g.,][]{Hayashietal2003} has been used to provide boundary conditions for a 3-D MHD model, and more recently IPS data have been used to modify the spheromak-initiated 3-D MHD CME model (SUSANOO-CME) in a time-dependent way \citep{Iwaietal2019}. Results from the UCSD iterative technique are also currently extracted at 0.1~AU and used to drive 3-D MHD models \citep{Yuetal2015} in an analysis that determines CME structure and forecasts their velocity and density as well as the magnetic field components \citep{Jacksonetal2015}.  Additionally, the ENLIL model \citep{Odstrcil2003, Odstrciletal2005} can now be used as a kernel in the 3-D reconstruction analyses \citep{Jacksonetal2020}. The use of 3-D MHD modeling in the 3-D reconstruction analyses allows the incorporation of more sophisticated physical processes such as shocks and compressive structures, and, as a result, non-radial plasma transport, modifying the outward solar wind flow by temperature and magnetic fields. 

\subsection{Summary}
In this Section we have given an overview of state-of-the-art CME propagation models. Despite the plethora of these models as well as observed and modeled CME parameters, reliably simulating CME propagation has still many open questions for research due to the complex and rather poorly understood interplay between CME and solar wind characteristics. There is huge future potential for utilizing multiple observational data (e.g., from coronagraphs and HI from the Lagrange points L1, L5, and maybe L4) as input to CME propagation models (DA) in order to better investigate CME propagation behavior in interplanetary space. This will decisively improve the capability for producing more accurate space weather forecasts.

\section{Interaction Phenomena (HSSs--CMEs, CIRs/SIRs--CMEs, CME--CME) and Preconditioning}
\label{sec:section5}



At any instant of time, interplanetary space is filled with various large-scale solar wind structures. As already discussed in more detail in Sections~\ref{sec:section2}--\ref{sec:section4} the key players are transient events, i.e., CMEs, and SIRs/CIRs together with their related HSSs. Each of these structures generates a perturbation in the smooth outflow of the slow solar wind, and interactions between them cause complex processes that alter the characteristics of these structures and hence, the prevailing conditions in interplanetary space. This section gives an overview of the preconditioning effects and interaction processes between CMEs--SIRs and CMEs--CMEs, and how these relate to CME propagation models and space weather forecasting. For a more detailed review on CME-CME interaction we refer to \cite{Lugazetal2017} and on the nature of CME collisions to \cite{Zhang2021}.

\subsection{Variability of Space Environment on Short and Long Terms}\label{Interaction}

The evolution of CMEs during propagation through interplanetary space is strongly shaped by the interplay between the internal and external factors controlling their interaction with the surrounding solar wind and other transients \citep[][]{Manchester2017}. The magnetic structure of CMEs is therefore the result of a complex chain of physical processes, including: expansion due to differences in the internal plasma and magnetic pressure, as well as magnetic field magnitude, with respect to the ambient environment, which basically controls the size of the ejecta \citep[see e.g.,][]{DemoulinDasso2009,PalEtAl22}. CMEs can occur in sequence when successive releases of energy (primarily magnetic) occur in the parent source region. Interactions among multiple CMEs may involve a faster CME that ``overtakes'' a slower, preceding CME. A CME launched close to a coronal hole may interact with the associated HSS and SIR (see Section~\ref{sec:section3} for more details). Hence, other CMEs and SIRs present magnetic obstacles to the interacting CME. According to the frozen-in field theorem, interacting magnetic structures cannot easily penetrate each other, resulting in strong changes in the physical properties of CMEs such as: 

\begin{itemize}
    \item geometry and size (deformation, compression)
    \item propagation direction and orientation (rotation, deflection)
    \item kinematic properties 
    \item magnetic structure and field (amplification, distortion, reconnection -- erosion or flux injection, magnetic tension)
    \item plasma parameters, thermal properties 
\end{itemize}

The space weather impact at a target due to these changes might be larger by up to a factor of 2--3, especially due to compression and the enhancement of the pre-existing negative $B_{z}(t)$ to more negative values \citep[see e.g.,][]{Farrugiaetal2006,Zhangetal2007,Lugazetal2016,Lugazetal2017,Dumbovicetal2015, Shenetal2017, Shenetal2018,Kilpua2019b,Xuetal2019,Scolinietal2020,Koehnetal22}.

The presence of multiple transient structures also leads to a ``preconditioning" of the solar wind into which subsequent structures are propagating.  As consequence large uncertainties may be introduced into space weather forecasts based on simple (i.e., undisturbed) background solar wind flow simulations. One of the best examples for preconditioning of interplanetary space is the super-fast CME event observed in situ at STEREO-A on July 23, 2012 \citep[][]{RussellEtAl2013}. It propagated the Sun to 1~AU distance in less than 21~hours and would have caused major geomagnetic effects, if Earth directed \citep[][]{BakerEtAl2013}. The arguments and effects of CME propagation into a previously rarefied region from an earlier CME on July 19, 2012 were very clearly pointed out in the work by \cite{LiuYDetal2014}. Follow-up studies showed that the strong density depletion lowered the drag by a factor of 10 \citep[][]{TemmerNitta2015} making the July 23, 2012 event super-fast. The idea that extreme events can result from these combinations (and historical extreme events probably have) is important. Therefore, improving knowledge of the role of preconditioning, and implementing that into models, is a key goal of future research.

\begin{figure*}
\centering
\includegraphics[width=.85\linewidth]{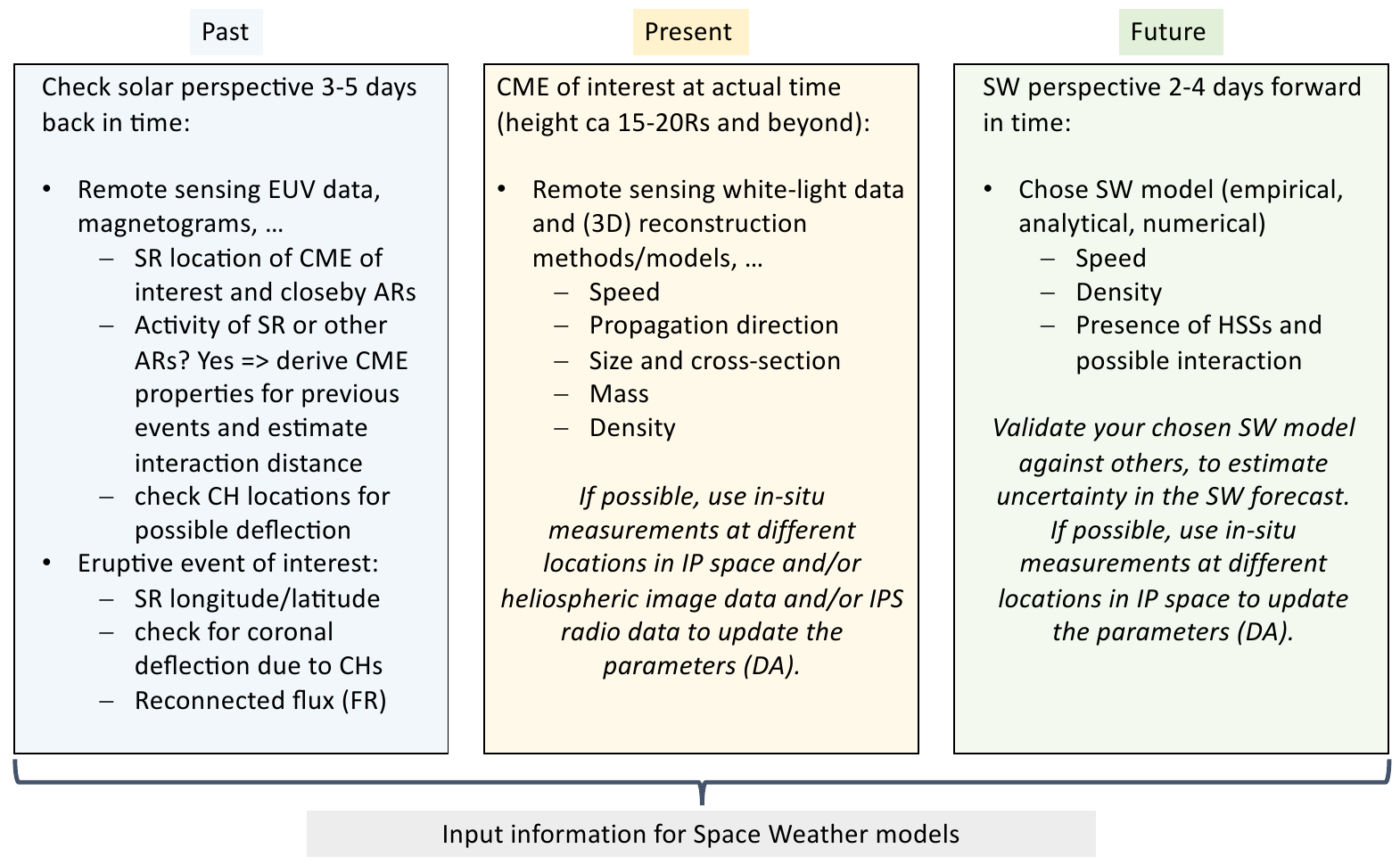}
\caption{For more accurate forecasts of a specific CME of interest, it is necessary to know the ``history'' of the erupting active region (AR), i.e., CME source region (SR). In addition to the actual CME properties, information is needed about the ambient environment in which the CME is embedded in, such as nearby coronal holes (CHs) and, hence, fast solar wind (SW) that has not arrived yet at any in situ measurement location. With that, three pillars of information feed forecasting models. The total time range to check covers a window of about 5--9 days. As the CME evolution in interplanetary space progresses, DA from in situ measurements, heliospheric images or radio data might be used to adjust the model input. The increase in accuracy gained due to DA is usually on the cost of a decrease in the forecast lead time.}
\label{fig:sec5:history}
\end{figure*}

The significance of preconditioning of interplanetary space and CME properties is clearly expected to be related to the solar cycle \citep[e.g.,][]{Cremades2006}. The CME occurrence rate (as viewed by coronagraphs) is only about 0.3/day during solar minimum but rises up to 4--5/day during solar maximum \citep[e.g.,][]{StCyretal2000,Gopalswamy2006}. With CME transit times from the Sun to 1~AU of about 1--4~days (average speeds reported are in the order of $\sim$500 to $\sim$3000~km~s$^{-1}$), there might be only a few CMEs at solar minimum, or as many as 20 at solar maximum, in the 4$\pi$ heliosphere between the Sun and 1~AU \citep[][]{Lugazetal2017}. Hence, during solar minimum, interactions between successive CMEs are rare but the occasional CME that is present is more likely to interact with CIRs/HSSs. Model evaluations confirm that during times of increased solar activity, preconditioning dominates and forecasts are more likely to fail \citep[see. e.g.,][]{Gressl2014}. It is found that disturbed solar wind conditions at a specific measurement location resulting from a sequence of interacting CMEs extend over 3--6 days after the CME start, which is much longer than the average duration of an individual CME disturbance \citep[][]{Temmeretal2017,Janvieretal2019}. To fully understand and to successfully simulate a specific CME, we need to know the history of the solar wind configuration in a wide analysis window extending back to several days before the time the event is observed \citep[see][]{Schrijver2015,Palmerio2021b}. Figure \ref{fig:sec5:history} gives an overview of suggested time windows and parameters that might be useful to estimate the history of solar activity related to a CME and to check for information and parameters required to feed the CME parameter into CME propagation models (see Section~\ref{tab:section4_cme_models}). Besides the number of transient events also the solar wind parameters themselves changes over the cycle (see also Section~\ref{sec2:solarCycle}). For cycle 24 a clear drop in the magnetic field and heliospheric pressure led to stronger CME expansion throughout the heliosphere that changed their propagation behavior and the build-up of shocks \citep[e.g.,][]{Gopalswamyetal2015,Jianetal2018,Lugazetal2020}. There is still more to learn about the solar cycle influence on solar wind parameters and how this knowledge can be fed into models. It is important to take into account whether a CME forecast is made in a weak or strong solar cycle, and how active the Sun is in the specific forecast window \citep[][]{owensetal2021}. In that respect we may use long-term averages of the solar wind pressure, density, and speed to compare the characteristics of individual cycles (see Cluster S1 TI2 paper by \cite{Pevtsov2023}).

\subsection{Interaction Processes with Large-Scale Field Structures}

Knowing the propagation direction and orientation of a CME event is key to a) interpret observational data and b) properly feed models. Changes in the initial propagation direction have manifold reasons. During different phases of the solar cycle CMEs are launched from different latitudes, which is related to the global solar magnetic field configuration \citep[see also the S2 Cluster TI2 paper by][]{Arge2023}. Even high latitude CMEs from active regions may cause intense geomagnetic storms \citep[e.g.,][]{Zhou2006}, suggesting that CMEs are deflected, in this case, towards the ecliptic, during propagation. CME deflection is related to magnetic pressure gradients that are stronger in the corona \citep[][]{MacQueen1986,Shen2011,Moestl2015,Wang2015,Kilpua2019a,Wangetal2020} than in interplanetary space \citep[][]{Wang2004,Siscoe2008}. Thus, CMEs tend to be deflected towards regions of weaker magnetic field \citep[][]{Gui2011}. The majority of CMEs are deflected in latitude towards the equator \citep[][]{MacQueen1986,Kilpua2009,Wang2011}. Longitudinal deflections may be either towards or away from the Sun-Earth line. Slow CMEs are found to be deflected more easily than fast ones, and usually an E-W asymmetry is observed such that fast CMEs are deflected to the East and slow ones to the West \citep[][]{Wang2004}.  

\subsubsection{Interaction between CMEs and SIRs/CIRs/HSSs}

Most strongly CMEs are affected by the presence of open magnetic fields, in particular, coronal holes close to the eruption site \citep[][]{Gopalswamy2009,Heinemann2019,Sahade2020}. Being large-scale magnetic structures, HSSs and related SIRs/CIRs (see more details in Section~\ref{sec:section3}), may cause significant changes to the intrinsic physical properties of a CME. It has been shown that, due to interactions between CMEs with HSSs and SIRs, the FR structure of a CME may deform, kink or rotate  \citep{manchester2004,riley2004,wangY2006,Yurchyshyn2008,Isavnin2013}, or erode due to reconnection \citep{Dasso2006,ruffenach2012,lavraud2014,ruffenach2015,wangY2018,PalEtAl22} and be deflected \citep{Wang2004,manchester2005,kay2013,wangY2014,Kay2015a,Kay2015b,wangY2016,zhuang2019, Heinemann2019}. In addition, SIR/CIR/HSS--CME interactions can alter the magnetic field complexity inside CMEs \citep[][]{Winslowetal2021b,Scolinietal2022}. The specific effects of the interaction depend on whether the SIR/CIR and related HSS is ahead of or behind the CME. If behind and catching up with the CME, the interaction processes are always associated with the deformation, compression, and acceleration of the CME \citep{Winslowetal2016,Winslowetal2021a,Heetal2018}.  This can also enhance the geoeffectiveness of the CME and the ability of the CME to form a shock, especially for a slow CME. If ahead of the CME, \cite{Heinemann2019} found that the CME may be deflected through more than 30° due to the SIR/CIR/HSS--CME interaction. Recently, \cite{LugazEtAl2022} studied an extended CME simultaneously observed in situ by STEREO-A and Wind. They found that in the part of the CME facing Earth and propagating inside the preceding HSS, a shock and sheath were absent at Wind, whereas a shock structure and a sheath region were found to be associated with the part of the CME observed by STEREO-A that had not interacted with the HSS.

\subsubsection{HCS Crossing and Connection to Source Region Location}

The HCS separates regions of open magnetic fields with opposite polarities originating (in the simple case of a dipolar solar magnetic field), in opposite solar hemispheres, and maps down to the streamer belt. During solar minimum CMEs tend to occur in or near the streamer belt and HCS, while near solar maximum, streamers occur all over the Sun, and the connection between a CME and the HCS is less obvious  \citep[e.g.,][]{Smith2001}. Pre-existing helmet streamers that are disrupted or blown out by CMEs generally reform in a time interval much shorter than the lifetime of the HCS \citep{Zhao1996}, while the HCS exists throughout the solar cycle. Hence, the location of the HCS relative to the source region of a CME and the observation target is important. \cite{Henning1985} first noted the “same-opposite side effect”, that disturbances (CMEs and related shocks) associated with flares located on the same side of the current sheet as Earth were of larger magnitude than those associated with flares located on the opposite side. This effect was later confirmed by several other studies. For example, based on observations of hundreds of events over five years, \cite{Zhaoetal2007} found that (1) shocks with the associated flares located near the HCS had a lower probability of reaching Earth, (2) the initial speeds of shocks that encountered Earth were noticeably faster when the associated flares were located near the HCS, (3) shocks associated with flares on the same side of the HCS as Earth were more prone to arrive at Earth than those with their associated flares on the opposite side.

The HCS can also serve as a boundary that affects CME expansion and propagation. Several recent in-depth studies of CMEs and HCSs have used multipoint observations. For example, \cite{Winslowetal2016} attributed a highly turbulent region with distinct properties observed within a FR at STEREO-A (but not at MESSENGER, which was in longitudinal alignment with STEREO-A) to the interaction between the CME and the HCS and the surrounding heliospheric plasma sheet during propagation of the CME. To better understand the physical processes involved in interactions between CMEs and the HCSs, more coordinated remote-sensing and in situ observations, as well as multi-scale modeling, are needed. 

CME deflections in latitude are constrained by the location of the streamer belt or HCS, and the deflection occurs mostly close to the Sun near the streamers \citep[see e.g., TI1 paper by][]{WANG2022}. Based on coordinated remote-sensing and in situ observations, \cite{Yurchyshyn2008} speculated that the axis of an ejecta might be rotated in such a way that it aligns with the local orientation of the HCS. See also more recent studies using observations and (space weather) models \citep[see e.g.,][]{Isavnin2013,wangY2014,Kay2015a,AsvestariEtAl2022}.  

\textit{The degree of influence on the evolution of large-scale CME properties depends on the ambient solar wind conditions. All of the aforementioned evolutionary aspects of CMEs are found to be amplified by interactions with HSSs, CIRs/SIRs, as well as the HCS and/or heliospheric plasma sheet \citep[see more in e.g.,][]{OdstrcilPizzo1999, Rodriguezetal2016, ZhouFeng2017, Liuetal2019, Daviesetal2020, Scolinietal2021}.}

\subsection{CME--CME Interaction}  

A variety of magnetic structures resulting from CME--CME interactions have been classified based on 1~AU observations. These include: ``multiple ejecta'' \citep[][]{WangEtAl2002}, in which a single dense sheath precedes two (or more) distinct ejecta. In such cases, the ejecta are separated by a short period of large plasma beta, which may indicate magnetic reconnection taking place between the structures. It is relatively easy to distinguish individual ejecta in magnetic field time series, especially if simultaneous plasma data are also available at the target location. ``Complex ejecta'' \citep[][]{BurlagaEtAl2002,FarrugiaBerdichevsky2004} are events where the two (or more) original ejecta cannot be distinguished anymore based on magnetic field observations. Such structures often exhibit the decreasing speed profiles typical of individual CMEs, but have a long duration compared to average ejecta. The magnetic field profile can range from smoothly-rotating magnetic field components to complex magnetic fields. In the former case, it is easy to be misled that such structures are the counterparts of individual CMEs, even when plasma data are available; their interpretation requires information on the broader context (e.g., remote-sensing observations, multi-point in situ observations at different heliocentric distances).

Progress on understanding the complex CME interaction processes was not really possible until heliospheric imaging became routine with STEREO/HI (see also Section~\ref{sec:section4}). One of the first CMEs observed by STEREO was in fact a series of two interacting CMEs in January 2007 \citep[][]{OdstrcilPizzo2009}. The energy transfer between the two CMEs was investigated by \citet{Lugaz2009a}, who found clear indications that the leading CME was accelerated due to its interaction with the overtaking, initially faster, CME. As an CME shock interacts with and propagates through a preceding ejecta, it can cause radial compression, amplification of the magnetic field, a change in the CME aspect ratio, acceleration and heating in the region downstream of the shock within the preceding ejecta \citep[][]{Vandas1997, SchmidtCargill2004, Lugaz2005, Xiong2006}. Observational and numerical studies have also shown that the preceding ejecta might quickly over-expand during this later phase of interaction \citep[][]{Xiong2006,Gulisano2010,Lugaz2012}, such that, as the ejecta continues to propagate away from the Sun, the space weather impact may progressively return to pre-interaction levels.

Assuming that CMEs are magnetically coherent structures, which is debatable \citep[][]{Owens2017,Lugazeetal2018}, elastic or super-elastic collisions may occur \citep[][]{Shen2012, Temmeretal2014, Mishraetal2015a, Mishraetal2015b, Mishraetal2016, Mishraetal2017, Lugazetal2017} by converting the magnetic or thermal energy of the CMEs to kinetic energy by some process. In particular, magnetic reconnection plays a crucial role in CME--CME collision \citep[][]{Lugaz2005}. It may lead, as in the case of the interaction with the ambient magnetic field, to magnetic erosion and flux injection occurring at the CME boundaries \citep[][]{Dasso2006,ruffenach2012,PalEtAl22} or in their interiors \citep[][]{Crooker1998}, fundamental topological changes of the magnetic structures \citep{Winslowetal2016, Winslowetal2021a, Scolinietal2021, Scolinietal2022}, as well as local magnetic field distortions \citep[][]{Torok2018}. This can alter the magnetic connectivity, topology, and size of CME magnetic structures and causes the formation of magnetically complex structures leading to strong geomagnetic effects \citep[][]{Gopalswamy2001,Wangetal2003,Goslingetal2005}. In the most extreme cases, this may result in the full coalescence of the two original structures \citep[][]{Odstrcil2003,SchmidtCargill2004,ChatterjeeFen2013}. \cite{MishraSrivastava2014} and \cite{MaricicEtAl2014} have shown possible signatures of magnetic reconnection in in situ observations at 1~AU as a result of CME--CME interaction. Hence, knowledge of the relative orientation of the MFR in the interacting CMEs is important \citep[e.g.,][]{Xiong2009,Lugaz2012,Shen2012,Shenetal2017}. 

Magnetic tension associated with interactions between FRs and ambient magnetic fields has been widely discussed \citep[e.g.,][]{Kay2015a,MyersEtAl2015,Vrsnak2016}. It is worth mentioning the work by \cite{MyersEtAl2015}, in which they concluded experimentally that the magnetic tension force resulting from the interaction between the background field and current sheet in the FR would halt the eruption process. A general consequence of FR-FR interaction is a change in the  magnetic field inside the FR \citep[][]{Shenetal2017,Lugazetal2017} which leads to  a change of magnetic tension force arising from the change in the toroidal magnetic field component associated with compression of the FR cross-section and thus an enhancement of the tension force. The enhanced magnetic tension force then restricts further deformation of the FR \citep[][]{Suess1988,manchester2004}. More detailed 3-D modeling of CMEs and observational constraints from white-light and IPS data would shed more light on this topic \citep[see e.g., TI1 paper by][]{FALLOWS2022}. 
 
The heliospheric distance where the interaction takes place can vary from the low corona to interplanetary space and determines the degree of impact at a specific target. This distance has been termed the ``helioeffectiveness"  \citep[][]{Scolinietal2020} and means that the time interval between the CME eruptions and their relative speeds are critical factors in determining the resulting impact of complex CMEs at various heliocentric distances.

\subsection{Simulations}

It is fair to say that CME--CME interactions are complex, acting on different spatial and temporal scales with respect to, for example, energy transfer, momentum exchange, magnetic reconnection, heating, compression, and over-/under-expansion. Sophisticated numerical modeling will help to improve understanding of the processes involved in CME--CME interactions, and recent efforts have been reported in a number of studies \citep[e.g.,][]{LugazEtAl2013,Lugaz2015,Shenetal2016,zhuang2019,Scolinietal2020}. More details of single CME propagation models, their observational input requirements and limitations, are given in Section~\ref{sec:section4}. These models may also be used in a simple approach to simulate multiple events by considering the distance where the CMEs interact and where it is necessary to change the model parameters \citep[e.g., when using DBEM;][]{Zicetal2015,Dumbovicetal2019}.

\subsection{Summary}
In conclusion, the reliability of CME space weather forecasts is especially complicated at times when other large-scale solar wind structures lie in the path of the CME. In such cases, neither estimates of the speed at 1~AU, nor in situ magnetic field data upstream of Earth might be sufficient to accurately estimate the magnetic field strength and orientation at the impact location, and more sophisticated modeling tools capable of describing interactions in a physically-consistent manner, in combination with reliable remote-sensing CME observations, are required. Such studies may be possible with the help of in situ observations from PSP, SolO, and other missions sampling heliospheric plasma at different distances from the Sun.

\section{Improving Heliospheric Modeling/Forecasts}
\label{sec:section6}


The previous Sections have reviewed the current state of modeling of heliospheric transients, i.e., CMEs (see Section~\ref{sec:section4}) and identified the issues that impact the accuracy of forecasts of their impacts on Geospace (see Sections~\ref{sec:section4} and \ref{sec:section5}). This section  identifies paths forward to improving the physical understanding, modeling, and consequently, forecasting of heliospheric transients. The section starts with a short overview of the current state of forecasting the key physical parameters of transients, and the performance required by various space weather users  (Section~\ref{sec:sec7_current_state}). We then outline the top-level gaps in physical knowledge and data availability (Section~\ref{sec:sec7_gaps}), setting the stage for suggestions for closing these gaps and moving the field forward in Section~\ref{sec:sec7_pathforward}. 

\subsection{The Current State of Modeling and Forecasting of Heliospheric Transients Properties} \label{sec:sec7_current_state}
A concise method to identify the state of space weather forecasting of heliospheric transients is to compare the current and desired performance of predictions of the key physical parameters used in space weather forecasting. Some of these parameters are identified in Table~\ref{tab:section4_cme_input_parameters}. We also use information from a recent NASA-sponsored  \href{https://science.nasa.gov/science-pink/s3fs-public/atoms/files/GapAnalysisReport_full_final.pdf}{Gap Analysis} \citep[see also][]{Vourlidas2021} that examined a wider range of space weather-related phenomena. Table~\ref{tab:state_summary} presents the resulting summary of the current and desired state of the forecasting of heliospheric transients, which is the focus of the H1+H2 Cluster. The table lists the key parameters and their current forecasting accuracy. The desired state is based on space weather user requirements (see Sec. 5.1 in the  Gap Analysis, for details). The last column lists the high-level issues that prevent current forecasts from meeting users’ expectations. These issues are derived from the literature and discussions in the previous sections and within the H1+H2 Cluster groups. 

   \begin{sidewaystable*}
    \centering
    \caption{State of forecasting key CME and SIR properties at 1 AU. Each row describes (from left to right): observations required to initiate a model (usually at 20 Rs); to forecast a key parameter (usually at 1 AU);  the current state of forecasting (if known); the accuracy required by space weather users (Gap Analysis and references therein); issues that restrict the accuracy of forecasting; SEL (Sun-Earth-line); IP (interplanetary); MSE (mean squared error); MAE (mean absolute error); LOS (line-of-sight);}
    \label{tab:state_summary}
        \begin{tabular}{*{3}{|p{4cm}}|p{4cm}|p{5cm}|}
    \hline
    Model Inputs (at 0.1 AU) & Model Output (at 1 AU) & Forecasting Status (Current) & Forecasting Status (Desired) & Issues \\
    \hline
      Source region location; CME direction, width, speed & 'All Clear' & 97\% CME detection rate from SEL & CME detection rate: 90\% of user-directed events within 2~h of eruption & ’stealth’ events (from Sun-user viewpoint); LOS effects; insufficient off-SEL coverage \\
         \hline
      Source region location; CME direction & CME Arrival (hit/miss) & 80\% (from off-SEL visible imaging) & 24-h lead-time of a hit & “stealth” events; IP evolution; LOS effects; latency \\
         \hline
      source region location, CME/shock direction, size, speed & Time of Arrival (ToA) & 10 ±2 h* (based on off-SEL imaging). Reduces to $>17$ h for single point imaging & 24-h lead-time and $0\pm2$~h ToA (for actionable electric grid countermeasures) & IP propagation; CME/shock structure at 1 AU; insufficient off-SEL coverage \\
         \hline
      Same as above & Speed on Arrival & $> 30\%$ & $< 10\%$  & Same as above \\
         \hline
    CME/SIR mass & CME mass density (to estimate flux transfer to magnetosphere) & 2-3x overestimate & 24-h lead-time of a hit but duration accuracy requirement is undefined & small-scale density structure of CME/SIR front; IP evolution \\
       \hline
    CME volume & CME size along SEL (to estimates duration of interaction & $\sim3x$ & same as above & IP propagation; LOS effects; CME/shock shape only measured at L1 (~30 min lead time)\\
       \hline
MFR strength, helicity, size based on source region proxies (polarity inversion line, post-eruption arcades); CME volume, shape & CME Magnetic configuration (strength, orientation, duration) & $\sim30$ min (from L1) & strength: 24~h; orientation/duration: 2–-3h, both lead-times & 
     unknown at-birth CME magnetic properties; unknown sub-Alfvénic corona properties; remote sensing of CME magnetic field in corona/heliosphere; CME/shock IP propagation; Earth trajectory through magnetic structure\\
   \hline
Photospheric magnetic field map & SIR ToA/Duration & ToA: $24\pm6$h; Duration: $<$12~h & Same as CME requirements & uncertain coronal boundary specification (magnetic field, plasma) \\
   \hline
Background properties (velocity, temperature, density, magnetic field) & SIR Plasma/Magnetic Properties & B strength: MSE $\pm30\%$; B orientation; density: MSE $\pm80\%$; speed: MAE 80 km/s & undefined & same as above; radial evolution; off-ecliptic compression effects on $B_z$ \\
    \hline
    \end{tabular}

    \end{sidewaystable*}
\subsection{Knowledge and Capability Gaps} \label{sec:sec7_gaps}
The issues listed in the last column of Table~\ref{tab:state_summary} can be  broadly classified into two categories: issues arising from gaps in observational coverage, including latency and spatial coverage, and issues arising from limited knowledge of the physics involved in the formation and evolution of the transients in the inner heliosphere. 

\subsubsection{Observational Gaps}
\label{subsubsec:sec7_observational_gaps}
\noindent\textsl{Sparse coverage of the Sun-Earth space.} At present, solar activity is remotely monitored from just two viewpoints--from Earth/Lagrange L1 and from STEREO-A (the time-varying spacecraft positions can be viewed under \url{https://stereo-ssc.nascom.nasa.gov/where.shtml}).  In the next two years (assuming that STEREO-A continues to operate), the two viewpoints will effectively be reduced to one, as STEREO-A orbits at small angular separations from Earth. The incomplete coverage of the photospheric magnetic field, the coronal layers where activity originates and the Sun-Earth line affects all aspects of forecasting (e.g., CME source properties, line-of-sight confusion, interplanetary propagation). The issue is discussed in more detail in \citet{Vourlidasetal2019} and the NASA Gap Analysis. On the in situ side, consistent measurements upstream of Earth are only available from L1, providing 15 to 60-minute advanced warnings of the arrival of CMEs and interplanetary shocks at Earth. Numerous events have been measured by two or more spacecraft in radial alignment. However, these were serendipitous cases, mostly captured by spacecraft orbiting the inner planets, with only magnetic field measurements available. As a result, changes in the CME properties such as size, expansion, and velocity were difficult to interpret, and the study of shock speed and strength was impossible. Some CMEs exhibited drastic changes in their properties, associated with interactions with ambient structures, while rapid geometric expansion may lead to distortion of CMEs by the ambient solar wind and a resulting lack of coherence in CME structure at different heliospheric locations. \citep{Owens2017}.  It is hardly surprising, therefore, that even two or three in situ measurements are insufficient to constrain the properties of transients \citep{Lugazeetal2018}. The event-to-event variability means that any highly-accurate forecast, especially of the magnetic field strength and direction, will need to rely on plasma and field measurements made within 0.02 to 0.25 AU upstream of Earth (between L1 and Venus), providing a few hours up to a day of advanced warning.\\
    
\noindent\textsl{Low sensitivity of heliospheric imaging.} The STEREO HI achieved breakthrough observations of CMEs and SIRs to 1 AU. Yet, the faintness of the structures, the long lines of sight, and the required long exposures to detect those emissions reduce the structure contrast, particularly of the transient fronts. As a result, ToA prediction have improved only modestly \citep{Woldetal2018,Vourlidasetal2019}.\\
    
\noindent\textsl{High latency of near-Sun observations.} SDO/AIA provides real-time imaging of coronal activity but lacks the field-of-view (and viewpoint) coverage to enable robust detection of CME eruptions and their kinematics for model initialization. This information comes from coronagraphic measurements beyond 2 Rs, at least. However, real-time coronagraphic imaging is not always available from the LASCO or STEREO coronagraphs, even though the latter provide a continuous stream of low-resolution EUV and white-light images (known as ‘space weather beacon’). Lack of ground-based downlink availability is usually the reason.\\
    
\noindent\textsl{Inability to measure the coronal magnetic field.} Routine spatially-resolved coronal magnetic fields measurements across the solar disk/limb are currently beyond our reach due to the high demands in instrument throughput \citep{casinietal2017}. Yet, it is precisely the evolution of this field that, through the accumulation of energy and helicity and their subsequent release, powers flares and CMEs. Our inability to measure the spatio-temporal evolution of key parameters, such as free energy, helicity or currents, in the corona is the biggest impediment in predicting eruptions \citep[see][for details and path forward suggestions]{Patsourakosetal2020}.\\

\noindent\textsl{Small event samples.} Comprehensive ‘sun-to-mud’ analyses of solar transients became available only in the last cycle thanks to the triple-viewpoint capability of STEREO + SOHO/Earth observations. The larger number of datasets, however, requires more complex analyses, which, in turn, results in small sample studies. Such studies cannot easily avoid selection biases and may have inconsistent criteria for, say, ToA (see \citet{Vourlidasetal2019} for discussion).

\subsubsection{Knowledge Gaps}\label{sec:section6:gap}
\noindent \textsl{Incomplete description of the state of the ambient inner heliosphere.}  The structure of the ambient heliosphere (background solar wind) plays a critical role in the modeling of transient evolution in the inner heliosphere (discussed in Sections 1--5). CME interaction with HSSs or with other CME en route to Earth can influence the extent and kinematics of the event significantly (Section~\ref{sec:section5}). This is primarily a concern for medium-speed events ($\sim600-900$ km/s within 20 Rs) as their speeds are close to the typical ambient solar wind speeds in the inner heliosphere and they appear to evolve kinematically well beyond the typical coronagraph field-of-views \citep[e.g.][]{Colaninnoetal2013,Sachdevaetal2017}. Yet, the current heliospheric modeling performance is insufficient, primarily due to a single reason, its ‘Achilles heel’ \citep{Vourlidasetal2019}---incomplete boundary conditions. It is a two-fold weakness: (1) the background photospheric field is measured only across the Earth-facing part of the disk (corresponding to about 1/3 of the total surface), requiring strong assumptions about the far-side and polar field distributions \citep[e.g.][]{Linker2017, Temmer2021b}, and (2) the sub-Alfv\'enic corona is poorly understood due to the lack of consistent measurements of its state (temperature, density, composition, kinematic profiles, etc). Expanding the coverage of photospheric magnetic field measurements, from, say the L4/L5 Lagrangian points and the poles, and bringing in long-term off-limb spectroscopic coronal measurements, will go a long way towards closing this knowledge gap.\\
    
\noindent\textsl{Poor knowledge of internal CME structure} We, presently, lack knowledge about the initial configuration of CMEs in the corona (especially the amount of twist) and how to incorporate more realistic CME initiation models into space weather models. Most current space weather models either assume a very highly twisted FR initiated in the upper corona (EUHFORIA), or a non-magnetized eruption, also in the high corona (ENLIL), or a highly-twisted FR initiated in the low corona (SWMF, SUSANOO). While we have some insight from more complex simulations and non-linear force-free reconstructions, these are not yet adapted for real-time space weather forecasting. CME--CME interaction and energetic particles associated with a series of CMEs are especially problematic and these cases are common during solar maximum. In addition to the initial conditions, we still do not understand well how the CME internal magnetic field evolves as the CME propagates and interacts with the solar wind and other transients. This knowledge gap arises from (1) lack of data about the CMEs as described in Section~\ref{subsubsec:sec7_observational_gaps}, (2) lack of detailed simulations of the background solar wind with small and intermediate scale features (turbulence, more complex density, magnetic field and velocity profiles), (3)  lack of numerical studies focusing on complex and realistic CME topologies with propagation to 1 AU \citep[an exception is the work of][]{Torok2018} and (4) overly simplified models to reconstruct single-spacecraft measurements at 1 AU. There has been some progress on this last point in the past few years \citep[e.g.][]{Nieves-Chinchillaetal2020} but we still rely primarily on fitting models of a force-free FR with a circular cross-section for space weather applications.\\
    
\noindent\textsl{Incomplete knowledge of transient mesoscales.} While imaging has probed the large (tens of degrees) scales and in situ observations have measured the small (sub-degree) scales of transients, the results remain far from satisfactory for space weather users. Key constraints on the structure of transients seem to reside in mesoscales (roughly $\sim 1^\circ$, \citet{Lugazeetal2018}), which are almost totally unexplored due to the lack of closely-space in situ measurements and/or high spatial resolution imaging. For example, the uncertainties in the CME internal magnetic field can be as high as 60\% at 1 AU, when considering the limits in the  drop-off rate of the magnetic field with distance (between $r^{-1}$ to $r^{-2.5}$).\\

\noindent \textsl{Inefficient use of available assets and capabilities.} Although not a knowledge gap, the sub-optimal use of available data is certainly hindering progress in space weather forecasting. We tend to under-utilize individual data streams and to under-exploit their synergies. For example, (i) reconstructions of the solar magnetic field beyond potential field are typically not integrated into space weather models, (ii) remote-sensing observations are typically used to constrain only the CME direction and speed but not its 3-D shape (iii) three-dimensional plasma flows, composition, charge states and pitch-angle distributions of suprathermal electrons are often not integrated consistently in the discussion of CMEs (for example, to check whether the measured flow speed is consistent with the assumed CME shape).

\subsection{Moving the Field Forward} \label{sec:sec7_pathforward}
Advancing the capability of heliospheric modeling and forecasting requires the closure of the knowledge and capability gaps identified in Section~\ref{sec:sec7_gaps}. Here, we outline a strategy for making effective progress on this issue that identifies challenges that can be tackled on short-term, near-term, medium-term and long-term horizons.

\subsubsection{Short Term (Leverage Existing Knowledge and Assets)}
 
\noindent\textsl{Think ‘Outside-the-Box’.} We offer two suggestions:
    \begin{itemize}
        \item Observing System Simulation Experiments (OSSEs) have long been used in the terrestrial weather arena to inform measurement strategies, to design space-based architectures to acquire those measurements and to fine-tune DA schemes to ingest the resulting data products \citep{zengetal2020}. Since we face very similar challenges, investment in leveraging terrestrial weather experience and in developing OSSEs to address the issues in Table 1, seems the most beneficial path forward.
        \item Developing the capability to obtain a missing measurement may not always be the most practical solution for improving space-weather forecasting.  What if modeling could provide a sufficient substitute for a missing/incomplete measurement or perhaps an alternative observation/measurement that is available by some different means?  For example, could models based on photospheric magnetic-field measurements replace direct solar wind or EUV (or other wavelength) irradiance measurements for some niche space weather-applications or users?  Also, could more work be done in using observations of IPS and improving ground-based networks as an alternative way of driving ENLIL as already been explored by \cite{Gonzi2021,JACKSON2022}?  Such models, methodologies, alternative observations already exist in early forms and/or can be solicited with targeted funding opportunities bridging into heliophysics expertise from other communities; a prime example here for the modeling since would be from the fields of ML or data analytics.
        \end{itemize}

\noindent \textsl{Standardize data quality and analysis approaches.} The cross-calibration of magnetograph data is a well-known problem that impacts the reliability and validation of MHD models \citep{Rileyetal2014,Wangetal2022}. Various image processing and measurement techniques are applied to imaging data for kinematic or dynamic measurements of CMEs, using samples that are not always vetted for selection bias or data quality, resulting in statistics for ToA (or other space weather-relevant quantities) that cannot be properly assessed. The development of standard data products for space weather analysis (similar to, say, the creation of ML/AI-ready data sets) would greatly improve the assessment of model and forecasting performance and, perhaps more importantly, enhance peer-review validations and data distribution across the community.\\
        
\noindent\textsl{Standardize Performance Metrics.} It is currently challenging to assess and compare the performance of heliospheric modeling frameworks \citep[e.g.,][also the TI1 paper by \cite{ReissEtAl2022}]{Verbekeetal2019a}. Developing a set of common performance metrics with wide community acceptance would provide better insight into the physical realism of different heliospheric models, as well as their performance for operational forecasting. \\
        
\noindent \textsl{Improve DA Workflows.} The objective of DA is to provide an optimal estimate of the state of a dynamical system by combining knowledge of the system's state derived from both a physical model and observations. In practice, DA incorporates a wide range of mathematical techniques whose use depends upon the specifics of a model (e.g., linear and non-linear) and observations (e.g., in situ or remote sensing) of a particular system. DA techniques have revolutionized the performance of terrestrial weather and climate modeling and it is reasonable to assume DA will return similar benefits to heliospheric modeling. 
        
Currently, heliospheric modeling constrained by DA is implemented primarily as a research tool only, although there are examples where these techniques are being configured for operational purposes. For example, ADAPT assimilates magnetogram observations of the photosphere into a flux-transport model, returning improved estimates of the state of the photosphere \citep{Argeetal2010}. Recent works have pursued assimilating both in situ and remote sensing observations of the solar wind and CMEs into heliospheric models, incorporating a range of complexities of both the DA scheme and heliospheric model. For example, \citet{Langetal2017} demonstrated a proof of concept sequential DA scheme for the assimilation of in situ plasma observations in the ENLIL 3-D MHD model. Similarly, \citet{Langetal2017} implemented a more advanced variational DA scheme into the reduced physics HUX solar wind model. \citet{Barnardetal2020} presented a method for constraining an ensemble of solar wind simulations with HI observations of CMEs, demonstrating that these could lead to improved hindcasts of CME arrival times, and providing a first step towards the formal DA of HI data in solar wind models. Similarly, \citet{Iwaietal2021} successfully constrained the SUSANOO‑CME 3-D MHD solar wind model with IPS observations, resulting in improved CME ToA forecasts by constraining an ensemble of simulations with the IPS data.
One immediate issue is the currently disconnected nature of these efforts. In terrestrial meteorology, forecasts typically rely on coupled DA schemes, which facilitate the self-consistent assimilation of a range of different observables across coupled models \citep{Leaetal2015}. Heliospheric simulation and prediction could be improved by the development of a coupled DA system that can simultaneously assimilate a range of in situ and remote sensing data. The existing archives of magnetogram, coronagraph, HI, IPS, and in situ plasma data, provide an excellent test bed for establishing the potential of such a coupled DA modeling scheme for use with future assets such as ESA's \textit{Vigil} and NASA's PUNCH missions.\\
        
\noindent \textsl{Introduce new/enhance potential data streams.}  Observations of IPS can provide important data for improving the forecasting output of MHD simulations \citep[e.g ][and references therein]{Iwaietal2019,Jacksonetal2020,Gonzi2021}.  They are used to improve background solar wind distributions \citep[e.g.][see Sec. 2 for details]{Jacksonetal2020}.  They can also follow CMEs propagating from ~0.1 AU to ~1 AU, a range dictated by the metric to deci-metric wavelength range of current IPS stations, and offer the potential for bringing out confirmed CME features and/or indications of the orientation (but not the sign) of CME magnetic fields \citep[e.g.][and references therein]{Bisietal2010a,FALLOWS2022}.  Thus, IPS data can be used to validate and/or drive MHD simulations across the inner heliosphere.  For example, \cite{Iwaietal2022} observed a CME using both LOFAR and ISEE (Nagoya University, Japan) arrays and included those data into the SUSANOO-CME MHD simulation, which successfully improved the reconstruction of the CME.  

Observations of IPS, being ground-based in nature, have the advantage of easily obtainable real-time data.  On the other hand, these observations are available only during daytime (and just before sunrise/after sunset) of each observing station.  This limitation can be overcome by coordinated observations across multiple IPS stations in different time zones, known as the Worldwide IPS Stations Network (WIPSS) \citep{bisietal2016,BisiWIPSS2016a,BisiWIPSS2016b}.  So far, only ISEE in Japan provides real-time IPS data but several other stations (e.g., LOFAR, MEXART - Mexican Array Radio Telescope) have the potential to do so.  Finally, we note that current IPS-based forecasts have only a 0.5--1 day lead-time. Another important new addition to radio-based space weather capabilities is LOFAR4SW \citep[e.g.][]{carleyetal2020} with stations spread around Europe and observing capabilities across the Solar-Heliosphere-Geospace regimes.  Each station can form a two-dimensional steerable beam to track a single radio source.  With these new capabilities, observations of IPS can help improve CME modeling, heliosperic reconstructions, and their accuracies \citep[see TI 1 papers by][]{FALLOWS2022,Iwaietal2022,JACKSON2022,TIBURZI2022,SHAIFULLAH2022}.  A key recommendation here would be for the proposed LOFAR4SW upgrades to be undertaken, thus making the LOFAR system a comprehensive space-weather observatory on the ground, and alongside LOFAR4SW implementation, for other WIPSS Network IPS observatories to make their data available in real time (which would include the full real-time implementation of the ISEE IPS data which are, as noted in earlier sections, only available in one-day intervals lagged by almost a day).  This recommendation links into the next subsection of the $<$10-year horizon also.

\subsubsection{Near-term (Within the Next 10 Years)}

\begin{figure*}
\centering
\includegraphics[width=.85\linewidth]{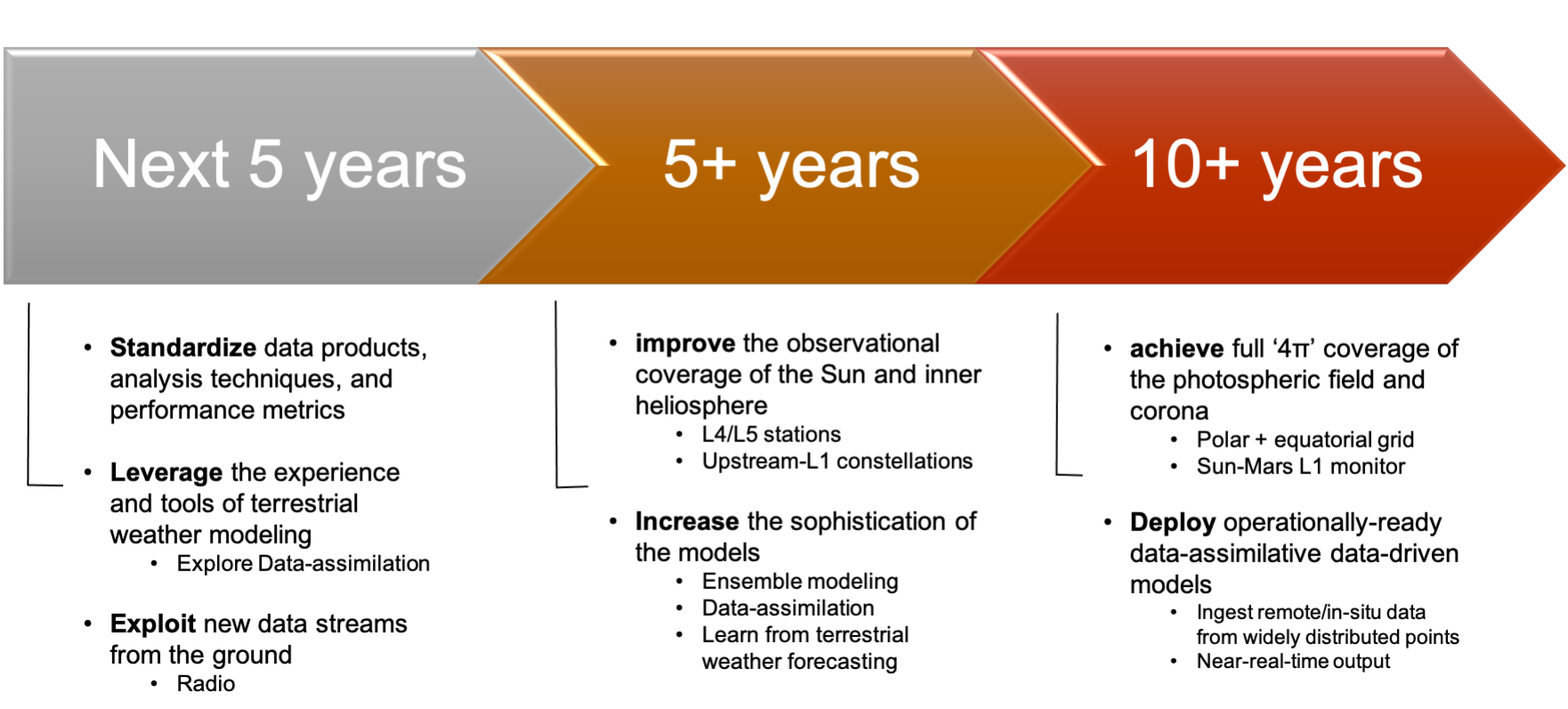}
\caption{A \textsl{multifaceted strategy\/} is required to significantly increase the accuracy of CME and SIR propagation models within the next ten years}
\label{fig:sec7:findings}
\end{figure*}

\noindent\textsl{Maintain existing off-SEL coverage.} Transient measurements from off-SEL viewpoints are now a vital input for many, if not most, models used in research or operational space weather forecasting \citep[e.g.,][]{dekoningetal2009,rodriguezetal2020,Bauer2021}. STEREO-A is the sole provider of off-SEL imaging into the near future, followed (potentially) by the ESA L5 Lagrange mission in the 2028-30 time frame. However, it is not certain that STEREO-A, launched in 2007, will continue to operate until then. It is, therefore, urgent to consider smaller missions with shorter development schedules as ‘gapfillers’ for off-SEL coverage.\\ 
            
\noindent \textsl{Improve coverage of the inner heliosphere.} As discussed in the previous section, improving space weather models requires data from more places in the heliosphere; namely, distributed in situ measurements, primarily between Venus and L1,  in tandem with multi-viewpoint coronal/SEL imaging from the L1, L4, and L5 Lagrange points and wider synchronous coverage of the photospheric magnetic fields. Such measurements are achievable with current technologies and specific implementations have been discussed in the literature, such as the Space Weather diamond \citep{stcyretal2000b}, the L5 pathfinder \citep{Vourlidas2015} or the Heliospheric Research Grid \citep{Vourlidasetal2018}. The fact that several other concepts are currently under design for NASA’s Heliophysics Concept Mission Studies and the Living With a Star (LWS) Architecture Study is encouraging.\\
            
\noindent\textsl{Improve model sophistication.} This is another area where heliospheric modeling can further benefit from the terrestrial weather experience. Terrestrial weather forecasting uses data-assimilative ensemble modeling extensively. Ensemble modeling approaches are being developed \citep[e.g.,][]{Maysetal2015a,Amerstorferetal2018, Barnardetal2020, Weissetal2021} but the DA aspect is still in its infancy and needs to be developed to extract the maximum benefit from the widely distributed  measurements discussed above. The proposed increase in spatial coverage will produce a corresponding increase in the available data for assimilations, required integrated data mining and ML/AI workflows.
    
\subsubsection{Long-term (10+ Years)}

\noindent \textsl{Close the ‘coverage’ gap.} Ultimately we need complete, so-called ‘$4\pi$’ coverage of the solar surface and atmosphere to achieve robust boundary conditions for heliospheric and space weather models \citep{kleimann2012,Gibson2018,Vourlidasetal2020}. A system of 3-4 spacecraft in both near-polar and ecliptic orbits can provide this coverage within realistic cost, schedule and technology constraints while establishing the cornerstone of a long-term systems approach to Solar-Heliosphere-Geospace observations. To enable the human exploration of Mars, the addition of a Sun-Mars L1 monitor to the Lagrangian and Sun-Earth stations will  ensure actionable forecasting for both Earth-Mars transits and Martian outposts.\\
            
\noindent \textsl{Deploy Next Generation operational models.}  Any long-term modeling development strategy should aspire to the smooth transition of research-grade models (developed during the 'near-term' steps above) into the operational theater, thus providing the space weather community with data-assimilative data-driven models that both meet the performance requirements of space weather users and continue to push the boundaries of our physical knowledge of the inner heliosphere.

 \subsection{Summary} \label{sec:sec7_summary}
 We present a set of ideas for improving the accuracy of modeling, and subsequently of forecasting space weather-relevant parameters of solar transients (CMEs and SIRs). The ideas are based on the current research status as discussed in Sections 1--5, and are focused specifically on the issues surrounding the modeling of transient propagation in the inner heliosphere. Figure~\ref{fig:sec7:findings} summarizes the key findings for moving the field forward.

\section{Closing Thoughts}
\label{sec:section7}


In the years since the last COSPAR Roadmap \citep{Schrijver2015} novel methodologies and increasingly sophisticated methodologies have been developed. We attempted to review the status of the field regarding a specific, but highly important, component of Space Weather forecasting chain---CME propagation. Sections 2--5 expanded on the various aspects of CME propagation and relevant background solar wind structures (SIR/CIR).  In Section~\ref{sec:section6}, we offered ideas on moving forward with our current gaps in observing, modeling, and physical understanding.  We close this effort with an outline of the near-future exciting prospects in observations and modeling and a final summary of our top-level findings.

\subsection{Novel Observing Capabilities}\label{sec8:novel_obs}

The recent launches of PSP and SolO (in 2018 and 2020, respectively), constitute a major leap forward for the solar and heliospheric physics communities. The two missions investigate the solar wind in the corona and inner heliosphere from heliocentric distances far closer than the 0.3~AU achieved by the Helios mission.  SolO, in particular, will obtain off-ecliptic imaging of the near-polar regions for the first time and will bring new insight about the magnetic field characteristics at high latitudes (see also Section~\ref{sec:section2:SWmodel} and Section~\ref{sec:section6:gap}). 

The upcoming ESA L5 (\textit{Vigil}; estimated launch date 2029) operational mission will provide valuable data for further improving operational forecasting. In additional to the Sun-Earth line coverage, Vigil will obtain photospheric magnetic field observations over the East solar limb, thus providing, for the first time, accurate information on the magnetic conditions of the regions rotating towards Earth. The \textit{Vigil} mission will nicely complement the NOAA SWFO-L1 (Space Weather Follow On-Lagrange 1) and GOES-U (Geostationary Operational Environmental Satellite-U) observatories (early 2025 estimated launch) that will provide operational coronagraphic imaging from the Sun-Earth line.


\subsection{Novel Computing Capabilities}\label{sec8:novel_comp}

Novel methodologies, such as ML and AI, have increased in sophistication and gained lots of momentum in the recent years, mostly thanks to the impressive improvements in computational power and investments from the commercial sectors. ML methods have shown considerable promise in addressing the CME heliospheric propagation (see Section~\ref{sec:section4}). ML/AI is an accessible and powerful methodology to investigate large amounts of data, on a statistical basis. We briefly outline some perspectives of ML/AI for space weather forecasting.

\begin{itemize}
    \item Current methods make limited use of the high resolution/cadence solar data (when available)  using extracted parameters or single images as input to the models \citep{Camporeale2017}. NNs are able to extract complex relations from multi-dimensional data \citep{LeCun2015}. The increasing computational capabilities will enable the use of larger spatial-, spectral-, and temporal-resolution data, that should lead to novel prediction methods.
    \item The inner workings of NNs are opaque, preventing a clear interpretation of the results ('black-box' problem). Making ML/AI interpretable is a major challenge but promising approaches, such as the Grad-CAM algorithm and visual attention mechanisms \citep{xu2015show}, may address this challenge and hopefully lead to physical insights. 
    \item As discussed in Sections~\ref{sec:section1} and \ref{sec:section4}, forecasting models can be computationally expensive. DL enables the acceleration of existing methods by training a NN with the results of the simulation. Applications to fluid simulations have already demonstrated that comparable results can be achieved in a fraction of the time \citep{tompson2017accelerating,sanchez2020learning}.
    \item Extending this concept, NNs can be used to directly learn from physical equations. Physics-informed NNs integrate the information from a physical model (e.g., differential equations) and measured data \citep{karniadakis2021physics}.  The ability to handle noisy data and imperfect assumptions makes this method promising for future simulation methods that combine multi-instrument data.
\end{itemize}

\subsection{Path Forward}\label{sec8:path-forward}
Finally, we close this section with a top-level list of recommended actions for improving the modeling of the propagation of CMEs in the heliosphere.

\begin{enumerate}
    \item Improve background solar wind modeling. The outputs from the current background solar wind models have large uncertainties. It is not clear which model performs better under what conditions (see \#3 below). Permanent model evaluation will be able to react on the varying conditions in interplanetary space on different temporal scales (Cluster H2). In general, we recommend driving a variety of models to obtain uncertainty estimates since we do not yet know ``the'' most reliable one (this holds for CME propagation as well as background solar wind models). 
    \item Invest on sophisticated ensemble modeling using different models; see discussion in \#1 above.
    \item Standardize data analysis techniques and metrics; We need a way to intercompare model results and identify whether the problems arise from the inputs or the computations. Developing/adopting standards for data, analysis and performance metrics will greatly facilitate this effort.
    \item Facilitate data preparation and sharing to boost collaboration \citep[e.g., see concept by][]{RINGUETTE2022}.
    \item Establish regular off-Sun-Earth line observations (e.g. from L4/L5) with complementary instrumentation (following the STEREO paradigm); future mission for 4$\pi$ coverage of magnetic field to overcome the $B_z$ issue; 
    \item Exploit new data streams (e.g., IPS as well as other space-weather observations across the S, H, and G domains, e.g. by the implementation of \href{http://lofar4sw.eu}{LOFAR4SW upgrades}) and new forecast techniques (ML, DA, NN);
    \item Explore the eruption prediction capabilities from active regions in order to increase the lead time of space weather forecasts. Cluster S3 teams are investigating the maximum likelihood of a CME occurring together with its most likely speed and acceleration. Conceivably, if an estimate of the mass is available, the kinetic energy of a CME may also be predicted. The pre-eruption magnetic helicity may also be estimated (see TI2 papers by \cite{Georgoulis2023} and \cite{Linton2023}). These predicted values could be used for Sun-Earth modeling well before the event actually occurs.
    \item Improve communications with peer-users (Cluster G community) and end-users; emphasis for end-users must be placed on explaining the complexity of the system versus the terrestrial weather system \citep[e.g., see][]{MarshallEtAl2022} and on setting realistic expectations for the performance of Space Weather forecasting methods given, for example, the issues discussed in this paper. 
\end{enumerate}



\section{Acknowledgments}
We thank Janet Luhmann for critically reading the paper and giving valuable comments that helped to improve the manuscript. 
C.S.\ acknowledges the NASA Living With a Star Jack Eddy Postdoctoral Fellowship Program, administered by UCAR's Cooperative Programs for the Advancement of Earth System Science (CPAESS) under award No.\ NNX16AK22G, and NASA grants 80NSSC19K0914, 80NSSC20K0197, and 80NSSC20K0700.
I.G.R.\ acknowledges support from the ACE and STEREO missions and NASA program NNH17ZDA001N-LWS.
S.G.H.\ acknowledges funding from the Austrian Science Fund (FWF) Erwin-Schrödinger fellowship J-4560. 
E.Paouris acknowledges support from the NASA LWS Grant 80NSSC19K0069.
A.V.\ was supported by 80NSSC19K1261 and 80NSSC19K0069.
M.M.B.\ acknowledges support from UKRI-STFC in-house research funding and Space Weather Core funding used in part for the organisation and writing inputs to the paper. 
T.A.\ thanks the Austrian Science Fund (FWF): P-36093.
D.B.\ acknowledges support from the Horizon 2020 Framework Programme H2020-INFRAIA-2020-1 Project 101007599 — PITHIA-NRF.
L.K.J.\ thanks the support of the STEREO mission and NASA's LWS and Heliophysics Support Research (HSR) programs. 
J.A.L.\ and E.\ Palmerio acknowledge support from NASA's LWS-SC program (grant no.\ 80NSSC22K0893) and NSF's PREEVENTS program (grant no.\ ICER-1854790).
D.S.\ was supported by NASA LWS 80NSSC17K0718.
M.O.\ is part funded by Science and Technology Facilities Council (STFC) grant numbers ST/V000497/1


\bibliographystyle{jasr-model5-names}
\biboptions{authoryear}


\end{document}